\documentclass{aa}  

\usepackage{graphicx,subcaption}
\usepackage{multirow}
\usepackage{lscape}
\usepackage{pdfpages}
\usepackage{txfonts}
\newcommand{\logMHI}{$\text{log}(\text{M}_{\text{HI}})$}
\newcommand{\Amod}{$A_\mathrm{mod}$}
\newcommand{\AmodH}{$A^{15}_\mathrm{mod}$}
\newcommand{\AmodL}{$A^{5}_\mathrm{mod}$}
\newcommand{\Aflux}{$A_\mathrm{flux}$}
\newcommand{\cm}{$\mathrm{cm}^{-2}$}
\newcommand{\kms}{$\text{km}\text{s}^{-1}$}
\newcommand{\HI}{H\,\textsc{i}}
\newcommand{\degree}{$^{\circ}$}
\begin{document}

   \title{\HI\ asymmetries of galaxies in the Ursa Major and Perseus-Pisces environments}
   \titlerunning{Asymmetries in UMa and PP}

   \author{P.~V.~Bilimogga \inst{1}\fnmsep\thanks{Work done at the University of Groningen},
           E.~Busekool \inst{1}$^{,\star}$,
           M.A.W. ~Verheijen \inst{1}$^,$\thanks{Corresponding author: verheyen@astro.rug.nl} 
           \and 
           J.M. van der Hulst
           \inst{1}
          }
    \authorrunning{Bilimogga et al.}

   \institute{Kapteyn Astronomical Institute, University of Groningen, Postbus 800, 9700 AV Groningen, The Netherlands}

   \date{Received: 4 August, 2025 ; Accepted: }

 
  \abstract
   {The morphology and kinematics of atomic Hydrogen (\HI) gas in galaxies are influenced by both local and large-scale cosmic environments. Differences in global environment and galaxy interactions can leave distinct signatures in \HI\ asymmetry, offering insight into environmental effects on galaxy evolution.}
   {We investigate the role of environment on \HI\ asymmetries in galaxies located in two contrasting cosmic structures: the Ursa Major (UMa) group and the Perseus-Pisces (PP) filament. We quantify asymmetries in \HI\ global profiles and column density maps to assess their relation to local and large-scale environments.}
   {We analyze \HI\ 21cm spectral line imaging from the Westerbork Synthesis Radio Telescope (WSRT) and the Very Large Array (VLA), targeting the UMa and PP volumes. Data cubes are homogenised in resolution for fair comparison. Asymmetries are measured using observational constraints established in \cite{Bilimogga2022} and results are compared to asymmetries of simulated mock galaxies presented in \cite{Bilimogga2022}.}
   {The PP volume contains a significantly higher fraction of galaxies with asymmetric global \HI\ profiles (33 percent) compared to the UMa volume (9 percent). Similarly, the proportion of galaxies with \HI\ morphological asymmetries greater than 0.5, measured above a column density threshold of $15\times10^{19}$\cm , is also higher in the PP sample (46 percent) than in the UMa sample (13 percent). The higher column density sensitivity of the UMa data allows detection of lopsided features in interacting galaxies and close pairs down to $5\times10^{19}$\cm\, and allows for reliable measurement of \HI\ morphological asymmetries at this threshold. We identified a category of mock galaxies which are not representative of real galaxies due to unrealistic feedback models in the \textsc{EAGLE} simulations. In both volumes, we find stellar asymmetries and \HI\ morphological asymmetries are uncorrelated. Global profile asymmetries and morphological asymmetries are found to be uncorrelated, consistent with our previous results and external studies.}
  {}


   \maketitle
%

\section{Introduction}
\par It is widely known that the morphology, star formation rate, colour and gas content of galaxies depend on the environment in which they are located. From his study of 55 nearby galaxy clusters, \cite{Dressler1980} proposed the morphology-density relation, which shows that the fraction of star-forming spiral and irregular galaxies in a cluster decreases when the local galaxy density increases, while the fraction of early-type ellipticals and S0 galaxies increases. This relation was traced to higher redshift (z$\sim0.5$) by \cite{Dressler1997} and to intermediate redshift ($0.1\leq\text{z}\leq0.25$) by \cite{Fasano2000} who found that the morphology density relation is strong for centrally concentrated clusters. \cite{Postman1984} found that the morphology-density relation is also valid in less-extreme environments. \cite{Zabludoff1998} and \cite{Tran2001} studied a sample of X-ray-selected galaxy clusters and found that the fraction of bulge-dominated galaxies decreases with increasing distance from the cluster center.

\par The star formation rate of galaxies is suppressed near the cluster center, and remains so beyond the cluster's virial radius with a break at a projected density of 1 galaxy per Mpc$^{2}$ \citep{Balogh1997,LewisI2002,Gomez2003}. \cite{Martinez2002} studied star formation rates in galaxy groups and found a correlation between the relative fraction of star-forming galaxies in a group and the mass of the group. \cite{Kodama2001} studied galaxies surrounding the cluster Abell 851 out to 11 Mpc from the cluster center. They found a break in galaxy colour and luminosity at densities typical of galaxy groups and concluded that the quenching and morphological transformation of galaxies have terminated in groups well before they even enter the dense cluster environment. Similarly, \cite{Hogg2004} showed that galaxies in high-density environments are redder than their counterparts in low-density regions.   

\par The above trends show that it is important to study the gaseous component of galaxies to gain a more complete picture of the environmental processes on galaxy evolution. The atomic neutral hydrogen (\HI\ hereafter) disk of a galaxy is extended and very fragile thus making it an excellent tracer of the undergoing environmental processes. One of the first studies investigating the effect of the environment on \HI\ was performed by \cite{Davies1973}, who found galaxies in the Virgo cluster to be \HI\ poor compared to their counterparts in a field sample. Subsequently, several studies have established how the environment influences the \HI\ content of a galaxy.  The \HI\ content of galaxies in different nearby clusters has been studied in great detail, including Virgo (e.g. \cite{Cayatte1990, Cayatte1994, Chung2009, Taylor2012}), Coma \citep{Molnar2022, Healy2021, BravoAlfaro2000}, Abell 1367 \citep{Dickey1991, Cortese2008, Scott2018}, Hydra \citep{Reynolds2021, Wang2021}, and Fornax \citep{Serra2023}. These studies show that galaxies in cluster regions are \HI\ deficient and have truncated or asymmetric \HI\ disks due to ram-pressure stripping or tidal interactions. In their study of 18 nearby clusters, \cite{Solanes2000} found that about two-thirds of the cluster galaxies show a dearth of cold neutral gas and that in such clusters, the morphology of galaxies relates to their \HI\ deficiency. In groups of galaxies, signatures of frequent interactions such as tidal tails, common HI envelopes, and \HI\ features decoupled from the optical disk have been reported \citep{VMontenegro2001, Borthakur2010}. \cite{Hess2013} investigated the global \HI\ content of galaxies in many groups and found that larger groups are \HI\ deficient while it is more probable to find an \HI\ rich galaxy in the outskirts of a group than in the central regions. Galaxies found in the deepest of the voids have been studied with the Void Galaxy Survey (VGS, \cite{Kreckel2012}) in which many galaxies show kinematic and morphological asymmetries, which has been attributed to ongoing accretion.\citep{Hank2025} investigated the \HI\ asymmetries of simulated SIMBA \citep{Dave2019} galaxies in the context of their dynamical history over the past 2.5 Gyr and found that galaxies with a merger event are more asymmetric than galaxies that only experienced a close encounter, while galaxies that remained isolated are least asymmetric in the latest epoch (their Figure 9). They also found that lower mass galaxies tend to have more asymmetric \HI\ morphologies, presumable because they can be more easily disturbed.

\par To study the \HI\ features arising from environmental effects, spatially and spectrally resolved observations are required. To obtain an unbiased estimate of how often these \HI\ features occur, untargeted surveys are necessary. Some of the above-mentioned studies use targeted interferometric observations of optically selected galaxies. Other studies were performed with single-dish telescopes that only provide information about the global profile shape and the total \HI\ content of galaxies as they lack angular resolution. In this chapter, we present the results of two volume-limited, high-resolution \HI\ imaging surveys of the Ursa Major (UMa hereafter) and the Perseus-Pisces (PP hereafter) volumes, where we aim to investigate the environmental effects on the \HI\ disk of galaxies using the asymmetry index of the global \HI\ profiles (\Aflux) and the modified asymmetry (\Amod) index measured from the 2-dimensional \HI\ column density maps.

\par In \cite{Bilimogga2022}, we derived constraints on observational parameters to robustly measure the asymmetries in the \HI\ global profiles and column density maps of galaxies using the \Aflux\ and \Amod\ indices. We found that robust quantification of asymmetries in the global profile of galaxies requires a minimum average signal-to-noise ratio of 6. Similarly, to quantify asymmetries in the \HI\ disk of galaxies, a column density threshold of $5\times10^{19}$\cm\ or lower should be applied. Moreover, this \HI\ column density threshold should be at a signal-to-noise ratio of at least 3 to mitigate the effect of noise on the measured asymmetry values. We also studied how \Amod\ depends on resolution and concluded that the galaxy under consideration should have at least 11 beams or resolution elements across its major axis for the asymmetry values to be sufficiently unaffected by resolution. In this paper, these constraints will be considered when interpreting \HI\ asymmetries in the UMa and PP galaxies.

\section{Data samples} \label{sec:sample}

\subsection{The Ursa Major super-group}

\subsubsection{The volume}

\par The Ursa Major association of galaxies is located behind the nearby large-scale structure of the Coma-Sculptor cloud and is at a junction of filaments connecting to the Virgo cluster \citep{Tully1996}. Galaxies with radial velocities relative to the Local Group $700 < V_\mathrm{LG} < 1210$ \kms\ and within a projected radius of 7.5 \degree centered on $11^h56.9^m +49$\degree 22\arcmin\, (see Fig 1) are considered members of the Ursa Major association \citep{Tully1996}. According to Tully's catalogue \citep{Tully1988}, the `12-1' cloud of this collection of galaxies consists of 79 members, mostly late-type galaxies with a normal gas content, that lack a central concentration towards any core. Of these members, 62 galaxies form a complete sample with $\mathrm{M_B}<-16.5$. The Ursa Major association of galaxies has a low velocity dispersion of approximately 150 \kms\ , a virial radius of 880 kpc \citep{Tully1996}, and does not show any X-ray emitting intra-group gas \citep{Verheijen2001}. It has a total mass of $4.4 \times 10^{13} M_{\odot}$ and a total B-band luminosity of $5.5 \times 10^{11} L_{\odot}$ \citep{Tully1987}. The complete sub-sample of the Ursa Major association constitutes of only 9 early-type and 53 late-type galaxies (see \cite{Tully1996} Table 1) whereas the Virgo cluster consists of 66 early-type galaxies and 91 late-type galaxies \citep{Tully1987}. Compared to the Virgo cluster, Ursa Major has only 5 percent of its mass but 30 percent of its light \citep{Tully1987}. Considering the low abundance of early-type galaxies, absence of X-ray emitting intra-group gas, low velocity dispersion as well as total mass, it is evident that the Ursa Major association does not meet the criteria for a typical dense cluster like the Virgo cluster. Therefore, we consider Ursa Major as a super-group. 

\par More recently, \cite{Wolfinger2013} presented results from a blind \HI\ survey of the wider Ursa Major region, observed as part of the \HI\ Jodrell All Sky Survey (HIJASS). This single-dish survey covered a region of approximately 480 sq. degrees between $\text{10}^h\text{45}^m < \alpha < \text{13}^h\text{04}^m$ and $\text{22\degree} < \delta < \text{54\degree}$ and in the velocity range 300\,\kms\ $< V_{\mathrm{hel}} <$ 1900\,\kms encompassing the entire Ursa Major super-group mentioned above. They identified 166 \HI\ sources within this region. As the HIJASS survey was performed with a single-dish telescope, the galaxies detected are not spatially resolved. \cite{Wolfinger2013} provide only information about the global \HI\ profiles and gas content of the galaxies but did not consider profile asymmetries. Based on a sample of 1209 galaxies, combining the HIJASS survey with the optical SDSS redshift catalog, \cite{Wolfinger2016} identified 41 galaxy groups with four or more group members. They used the friends-of-friends algorithm with an adopted linking length of $D_{\mathrm{0}} = 0.3$ Mpc projected on the sky and a line-of-sight linking length of $V_{\mathrm{0}} = $150\,\kms .\cite{Wolfinger2016} defined this collection of 41 groups as a `super-group' that is gravitationally bound and will eventually merge with the Virgo cluster. They found that galaxy groups in the HIJASS volume are in an early evolutionary state and that the optical and \HI\ properties of the galaxies are similar to those in the field. Nevertheless, several \HI\ rich galaxies are found in the HIJASS volume that tend to reside in a lower density environment while \HI\ deficient galaxies are preferentially situated in the region closer to the Virgo cluster.

\par The Ursa Major region is also featured in a comparative asymmetry analysis performed by \cite{Angiras2007}. In this study, a Fourier decomposition analysis was performed on the HI column density distributions of a small sub-sample of 11 HI detections from the Ursa Major region. The average of the first harmonic coefficient ($\mathrm{A}_1$) was calculated along different radii, and these were used to infer asymmetry in the Ursa Major galaxies. In the results presented by \cite{Angiras2007}, the subsample of Ursa Major galaxies used in the analysis was shown to have small values of $\mathrm{A}_1$. The eleven averaged $\mathrm{A}_1$ values of the Ursa Major sample were compared against those of the sixteen Eridanus galaxies \citep{Angiras2006} and it was found that the Ursa Major galaxies had smaller $\mathrm{A}_1$ values than the Eridanus galaxies. 

\begin{figure*}
\centering
\includegraphics[width=\textwidth, trim=0.5cm 0.5cm 3cm 11cm, clip]{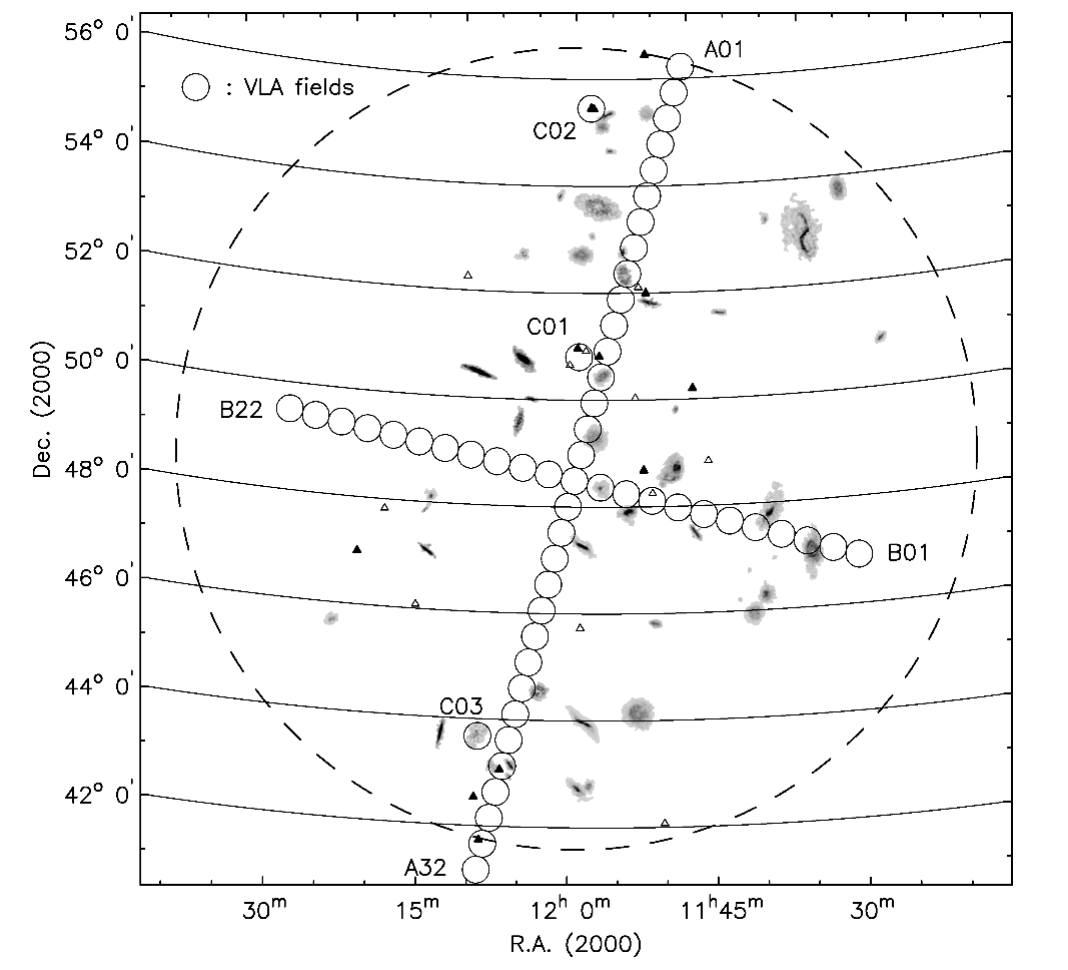}
\caption{This figure illustrates the layout of the 54 VLA-D survey pointings in the Ursa Major volume, indicated by the solid circles. The grayscale images show the \HI\ column density maps from the targeted WSRT observations, magnified by a factor of four. The large dashed circle outlines the adopted boundary of the Ursa Major environment. Solid triangles represent primarily early-type galaxies that are brighter than the optical completeness limit but lack sufficient \HI\ for meaningful WSRT synthesis imaging. Open triangles represent dwarf galaxies fainter than the optical completeness limit. Image taken from \cite{Verheijen2001}.}
\label{fig:UMa_Mosaic}
\end{figure*}

\subsubsection{HI Observations} \label{sec:UMa VLA details}

\par In this work we use results from both a volume-limited \HI\ survey of the UMa volume with the Very Large Array (VLA) in its D configuration \citep{Busekool2020}, as well as targeted \HI\ observations with the Westerbork Synthesis Radio Telescope (WSRT) \citep{Verheijen2001}. 

\par The VLA-D array \HI\ survey covers a 20 Mpc$^3$ volume, which constitutes 16 percent of the total volume within the 7.5\degree projected radius defined by \cite{Tully1996}. At an adopted distance of 17.1 Mpc \citep{Tully2008}, the 45\arcsec\ synthesised beam of the VLA-D array corresponds to 3.8 kpc. The Ursa Major volume is observed in a mosaic of 54 pointings arranged in a cross pattern, as shown in Figure \ref{fig:UMa_Mosaic}. The north-south direction of the cross pattern aligns with the Supergalactic plane. A bandwidth of 3.125 MHz with 128 channels covering a velocity range of 532\,\kms\ $< V_{\mathrm{hel}} <$ 1161\,\kms\ is used for the VLA-D survey. This yields a channel width of $\Delta V=$5.15\,\kms and, with an offline Hanning taper applied, a velocity resolution of 10.3\,\kms. Each pointing has a typical integration time of $2\times35$ minutes and a noise level of 0.76 $\mathrm{mJy}\,\mathrm{beam}^{-1}$ per channel. Observational details of the VLA-D survey are described by \cite{Busekool2020} and summarised here in Table \ref{tab:obsBlind_details}. 

\par In addition to the volume-limited VLA-D \HI\ survey, we also include in our analysis the targeted \HI\ observations of 50 Ursa Major galaxies presented by \cite{Verheijen2001}, comprising a complete, magnitude-limited sample of gas-bearing galaxies in this volume. These galaxies were observed with the WSRT with varying integration times ($1\times12$ to $5\times12$ hours) and various angular and velocity resolutions depending on the correlator restrictions. Observational details of galaxies observed with the WSRT are summarised in Table \ref{tab:obsWSRT_details} in Appendix \ref{sec:WSRTObs}. 

\subsubsection{Data reduction}

\par The data reduction steps for the VLA-D \HI\ survey are described in \cite{Busekool2020} and are summarised here. Calibrated visibilities for each pointing were Fourier transformed using \texttt{IMAGR} in the Astronomical Image Processing System (AIPS) \citep{Greisen2003} with robust weighting (R=1). Subsets of individual pointings were combined into 9 mosaics, four along the north-south direction, four along the east-west direction and one central mosaic. The mosaicing increased the sensitivity across the surveyed volume where adjacent pointings overlap. Thereafter, continuum sources were removed in an iterative way by fitting linear baselines until the residual noise converged. Separate continuum maps and \HI\ line emission data cubes for the 9 mosaics were generated. Frequency-dependent masks containing \HI\ line emission regions were made interactively for the mosaic data cubes. Thereafter, the dirty mosaic data cubes were cleaned in regions restricted to the masks generated in the previous step and restored using MIRIAD's \texttt{MOSSDI} and \texttt{RESTORE} algorithms respectively. Subsequently, a smooth-and-clip algorithm was used on the cleaned \HI\ mosaic data cubes to search for \HI\ sources. This resulted in the detection of 26 previously known galaxies and 16 new \HI\ sources.

\par Data reduction for the targeted WSRT observations was performed in a standard manner using the Groningen Image Processing SYstem (GIPSY) software package \citep{vdHulst1992, Vogelaar2001}, as described in \cite{Verheijen2001}.

\subsection{The Perseus-Pisces ridge}

\subsubsection{The volume}

\begin{figure*}
\centering
\includegraphics[width=0.85\textwidth]{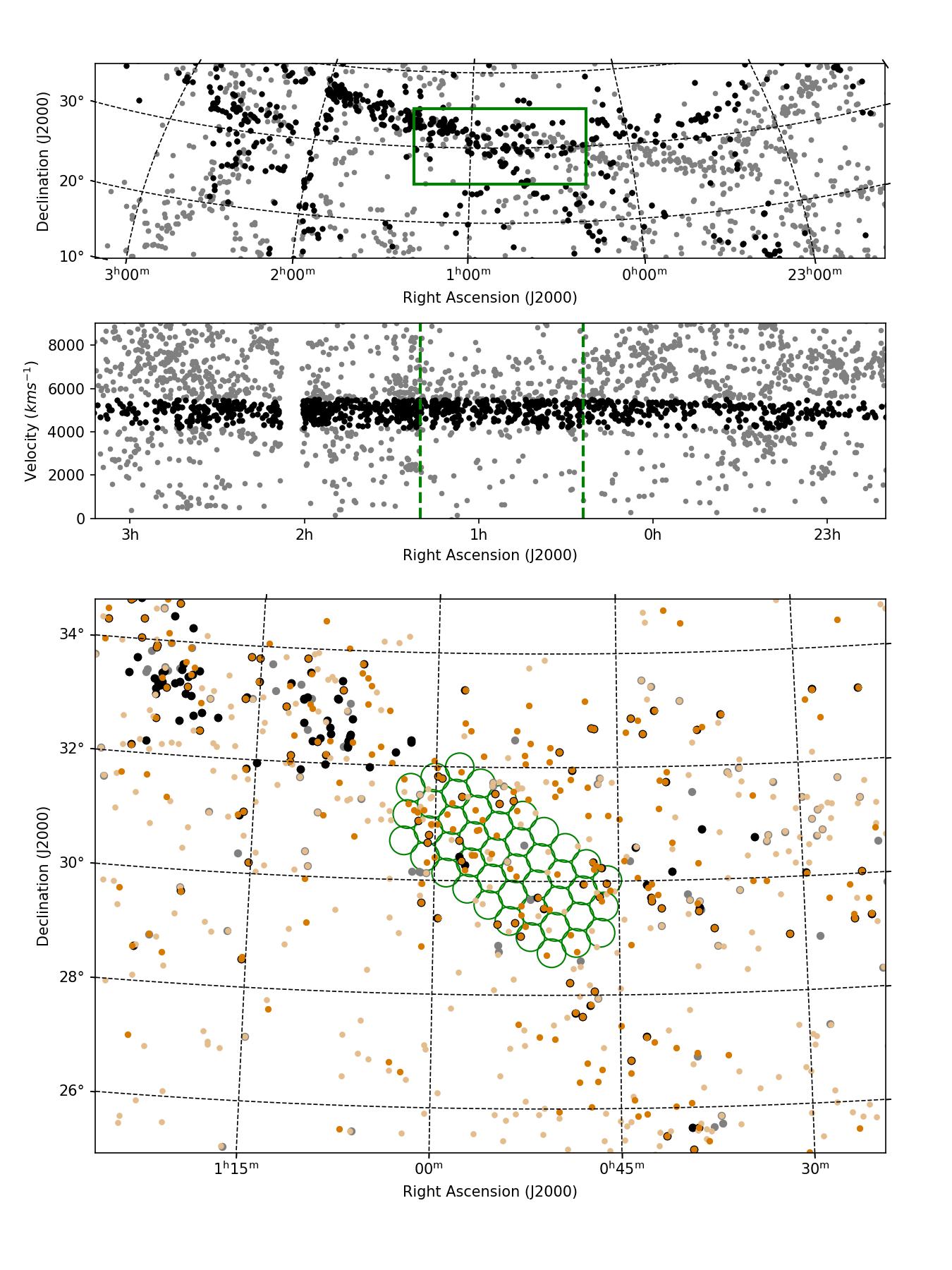}
\caption{The top panel of this figure shows the extended large-scale structure of the Perseus-Pisces Super-cluster. The green small rectangle contains the region where the VLA-C HI survey was conducted. The middle panel shows the distribution of galaxies in velocity space. The bottom panel shows the region outlined by the green rectangle and illustrates the layout of the volume-limited survey with the VLA-C array, where each green circle shows an individual pointing. In the three panels of this figure grey and black points indicate galaxies from the CfA redshift survey \citep{Huchra1999}. The orange and light-orange points in the bottom panel indicate galaxies from the ALFALFA survey \citep{Alfa100Cat}. Galaxies from the CfA redhsift survey and the ALFALFA \HI\ survey that lie within the bandwidth of the VLA-C survey are shown with black and orange points respectively.}
\label{fig:PP_Mosaic}
\end{figure*}

\cite{Bernheimer1932} observed `a metagalactic cloud' extending between the constellations of Perseus and Pegasus, in the region bound by $\text{4}^h > \alpha > \text{22}^h\text{, 0\degree} < \delta < \text{+50\degree}$, which is now identified as the Perseus-Pisces super-cluster. This super-cluster has been extensively studied by \cite{Gregory1981} using optical redshifts and by \cite{Giovanelli1983} using 21cm \HI\ line emission with the 1000ft Arecibo and the 300ft Green Bank radio telescopes. Both studies have shed light on the geometry of the Perseus-Pisces super-cluster and the overall filamentary large-scale structure in which it is embedded, shown in Figure \ref{fig:PP_Mosaic} based on the galaxies detected in the CfA redshift survey of the South Galactic cap \citep{Huchra1999}. The ridge of the Perseus-Pisces super-cluster, shown in the top panel of Figure \ref{fig:PP_Mosaic}, runs almost perpendicular to the line-of-sight and connects various clusters such as Pisces, Pegasus, Abell 262, Abell 347, and Abell 326, as well as rich groups associated with NGC 7626, NGC 383, and NGC 507 \citep{Gregory1981, Haynes1986_2}. The super-cluster is situated between two large voids on the near and far sides, making it an obvious feature in redshift space \citep{Haynes1986_2}. The void on the near side is centered at $\mathrm{0}^h \mathrm{30}^m$, $cz = 3200$\,\kms\ and extends over 50 degrees in declination (see Fig 2. of \cite{Haynes1986_2}). This prominent void, also seen in the middle panel of Figure \ref{fig:PP_Mosaic}, separates the Perseus-Pisces super-cluster from the Local super-cluster \citep{Haynes1986_2}, recently labelled as Laniakea \citep{Tully2014}.

\par \cite{Trasarti1998} used 21cm \HI\ line red-shifts from \cite{Giovanelli1983} and optical red-shifts from \cite{Wegner1993} to study loose groups in the Perseus-Pisces super-cluster. They selected a sample of 3014 galaxies in the region $22^h30^m\leq\alpha\leq03^h0^m $, $\text{0\degree}\leq\delta\leq\text{40\degree}$ and redshift $cz\leq27000$\,\kms. They applied the friends-of-friend algorithm to this sample with $D_{\mathrm{0}}=0.231\,h^{-1}$Mpc and $V_{\mathrm{0}}=350$\,\kms at $cz_0=1000$\,\kms and identified 188 loose groups with 3 or more members. They also found that the spatial distribution of the identified loose groups follows the large-scale structure of the Perseus-Pisces super-cluster.

\subsubsection{HI observations}

\par In this work, we utilise the results from a volume-limited \HI\ survey carried out with the VLA in its C configuration of a limited region in the extended Perseus-Pisces (PP) filament. An area of roughly $\text{2\degree} \times \text{4\degree}$ centered on $\alpha=0^h 52^m, \delta=\text{30\degree} \text{00\arcmin}$ is covered by a mosaic of 44 pointings. The top panel of Figure \ref{fig:PP_Mosaic} shows the location of this mosaic within the large-scale structure of the Perseus-Pisces ridge, where we observe an apparent bifurcation of the PP ridge. The bottom panel shows the layout of the 44 pointings of the VLA-C survey mosaic. A heliocentric velocity range covering 4200\,\kms\ to 5500\,\kms\ is observed with a channel width of $\Delta V = 10.3$\,\kms. Applying an offline Hanning taper results in a velocity resolution of 20.6\,\kms. Each pointing in the mosaic was observed for $2\times45$ minutes and the nearby calibrator 3C48 was used for complex gain and band-pass calibrations. Further observational details of the survey are summarised in Table \ref{tab:obsBlind_details}. The distance corresponding to the center of the velocity range is 66 Mpc, at which the 15\arcsec\ synthesised beam of the VLA-C array corresponds to 4.8 kpc. The surveyed volume comprises approximately $197\,\mathrm{Mpc}^3$, which is much larger than the volume observed in the Ursa Major VLA-D \HI\ survey as described in Section \ref{sec:UMa VLA details}.

\subsubsection{Data reduction}

\par The observed visibilities of the individual pointings were calibrated in the standard VLA fashion using AIPS \citep{Greisen2003}. The flux, phase and bandpass calibration were linearly interpolated in time. Radio Frequency Interference (RFI) was flagged automatically with the AIPS task \texttt{RFLAG} in an iterative procedure. For each baseline, the root-mean-squared (rms) values of the visibility amplitudes were calculated in a frequency-time window. Visibilities inside this window with amplitudes higher than a threshold determined by the local rms were flagged along with visibilities close in time and frequency. After the calibration and flagging, the visibilities for each individual pointing were Fourier transformed into a single data cube with 5\arcsec\ pixels using the \texttt{IMAGR} task, applying a `robust' weighting of 1. We also create a cube with the corresponding antenna patterns during this procedure. We do not perform continuum subtraction or cleaning at this step since it was not known beforehand where the HI emission would occur in the data cubes. The ‘dirty’ data cubes and antenna patterns were imported into GIPSY for further inspection and cleaning.

\par We verified the astrometry of the 44 individual data cubes by comparing the coordinates of known bright continuum sources to their coordinates in the NRAO VLA Sky Survey (NVSS) \citep{NVSS1998}. The brightest continuum sources and their sidelobes were removed from all channels in the `dirty' individual data cubes by cleaning them down to $3\sigma$, using the Högbom algorithm \citep{Hogbom1974}. This was required to properly remove the sidelobes, especially in the three channels that were heavily affected by RFI. The other less bright continuum sources were removed from the cubes through iterative baseline fitting and subtraction to retain unknown \HI\ emission above the zero-level baseline. After the first baseline fit and subtraction, the rms noise level was evaluated from the residual cube and pixels with values above or below a certain rms noise level were discarded from the new baseline fit. This procedure was repeated until the rms in the residual cube converged to a stable value. After this procedure, the quality of the `dirty' individual data cubes was checked. In the data cube of field C7 (central co-ordinates RA=$00^h56^m48.67^s$ Dec=+30\degree47\arcmin23.8\arcsec), RFI artifacts were still present in six channels. These artifacts were difficult to remove as there was an extended continuum source in this field NVSS J005748+302114, which had residuals similar to the RFI signatures. It was decided to set the affected channels to blank in a small area around the position of this continuum source.

\par Next, the individual cubes were smoothed spatially and in velocity to increase the sensitivity to extended emission at lower column densities and to line emission profiles wider than the original instrumental velocity resolution. The first step was to Hanning smooth the data cubes to a rest frame velocity resolution of 21.2\,\kms\ at the central redshift of the survey volume. This velocity resolution is referred to as `R2' hereafter. Afterwards, a Gaussian smoothing kernel was used to further smooth to velocity resolutions with a FWHM of 4, 6, and 8 channels which corresponds to 42.4, 63.6, and 84.8\,\kms and are referred as `R4', `R6', and `R8' hereafter. The data cubes at each velocity resolution were then spatially smoothed from a resolution of 15\arcsec\ to
 30\arcsec,\ 60\arcsec\ and 90\arcsec. This resulted in 16 low-resolution data cubes for each individual pointing.

\par We used the GIPSY task \texttt{OBJECTS} on each of the low-resolution data cubes to find \HI\ emitters. In Table \ref{tab:ObjectCriteria}, we quote different detection thresholds which depend on a combination of the velocity resolution and the number of adjacent channels. For example, a detection in a data cube with a `R6' velocity resolution (63.6\,\kms) requires a signal in excess of $4\sigma$ over 18 adjacent channels (190.8\,\kms) to be considered an \HI\ detection. \texttt{OBJECTS} provides a combined mask for all detections identified with the task which is then used to clean the individual data cubes with their corresponding antenna patterns, using the standard Högbom algorithm \citep{Hogbom1974}. Subsequently, the cleaned individual cubes were mosaiced together to increase the volume sensitivity of the survey. For this purpose, the individual cleaned cubes were imported back into AIPS and the AIPS routine \texttt{FLATN} was used for the mosaicing process. \texttt{FLATN} provides a Primary Beam Corrected (PBC) mosaic and a weights cube. This weights cube was used to obtain a mosaic without PBC.

\begin{table}
	\centering
	\caption{Observational details of the VLA observations.}
	\label{tab:obsBlind_details}
	\begin{tabular}{lcc}
                                     & Ursa Major          & Perseus-Pisces      \\
        \hline
        Configuration                & VLA-D               & VLA-C               \\
        Beam size [arcsec$^{\rm 2}$] & 45$^{\prime\prime}$ & 15$^{\prime\prime}$ \\
        Correlator mode              & Dual IF             & 4IF                 \\
        Bandwidth [MHz]              &  3.125              & 2$\times$3.125      \\
                                     &                     & (5.712 total)       \\
        Nr. pointings                &  54                 & 44                  \\
        T$_{\rm int}$ [min/pointing] & 2$\times$35         & 2$\times$45         \\
        Central Freq. [MHz]          & 1416.91 (2AC)       & 1399.21 (AC)        \\
                                     &                     & 1396.47 (BD)        \\
        Channel width [kHz]          & 24.414              &  48.828             \\
                                     &  5.15               &  10.3               \\
        noise [mJy/beam]             & 0.79                & 0.48                \\
        \hline
	  \end{tabular}
\end{table}

\subsubsection{Detection of sources in the mosaic}

\par The mosaic without PBC was smoothed spatially and in velocity, similar to the process adopted for the individual cubes. The GIPSY task `objects' is used again on the mosaic cube with the detection criteria stated in Table \ref{tab:ObjectCriteria}. A total of 113 possible \HI\ detections were identified in the mosaic. To verify if these are real detections, we searched for optical counterparts in the SDSS survey. Of the 113 detections, only 67 had optical counterparts and these will be used further in our asymmetry analysis. We list these 67 \HI\ sources in Table \ref{PPAsyTable} with their central coordinates. In the following section, we describe the process of extracting \HI\ data products after the PP mosaic cube was smoothed to the same spatial and velocity scale as the UMa cubes.

\begin{table}
    \centering
    \caption{Source detection criteria for \HI\ sources within the PP data cubes. The columns represent the four different velocity resolutions defined in the text, the rows represent the required signal strength for a detection and the table entries indicate the number of adjacent channels in which the signal should appear.}
    \label{tab:ObjectCriteria}
    \begin{tabular}{ccccc}
         &  R2 & R4 & R6 & R8\\
    \hline
    $8\sigma$ & 1 & 1 & 1 & 1 \\
    $5\sigma$ & 4 & 5 & 12 & 16 \\
    $4\sigma$ & 6 & 12 & \textbf{18} & 24 \\
    $3\sigma$ & 8 & 16 & 24 & 32 \\
    \hline
    \end{tabular}
\end{table}

\section{Homogenization of the HI samples}\label{sec:DataProds}

\begin{figure*}
    \centering
    \includegraphics[width=\textwidth]{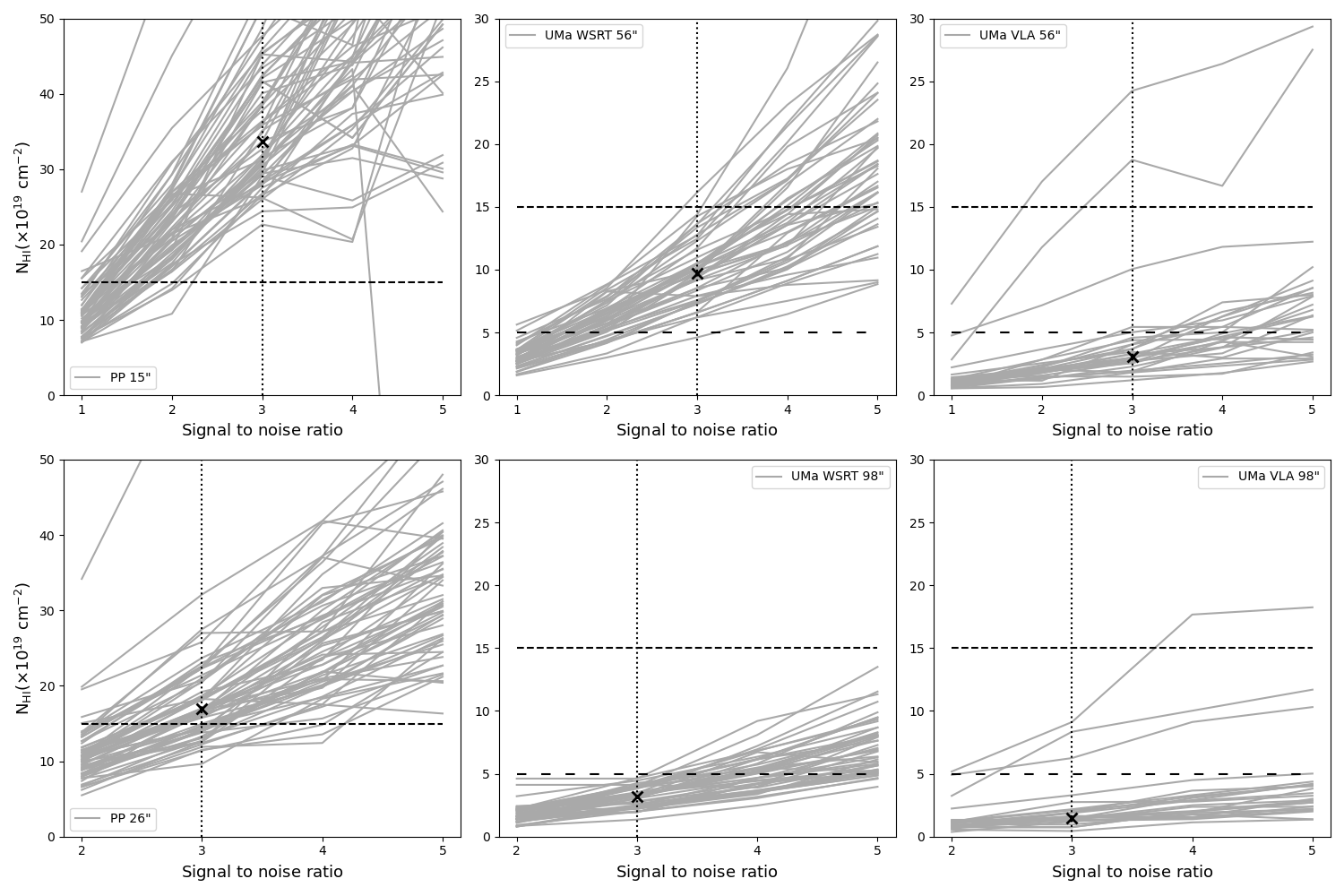}
    \caption{This figure illustrates the median column densities of the pixels with different S/N values in the column density maps of the UMa and PP galaxies. The top row shows $N_{\rm{HI}}$ values measured in maps at 4.8 kpc resolution, while the bottom row shows $N_{\rm{HI}}$ values at 8 kpc resolution. Each gray line in a panel illustrates how the median $N_{\rm{HI}}$ varies with S/N value in the column density map of the galaxy. A vertical dotted line is drawn at $\rm{S/N}=3$ in each panel, while a cross represents the median column density threshold for the sample at $\rm{S/N}=3$. These column density values are tabulated in Table \ref{tab:NoiseDiag}. The horizontal dashed line indicates the adopted column density threshold of $15\times10^{19}$\cm, used to measure morphological asymmetry in UMa and PP galaxies (see Section \ref{MorphAsym}). The long-dashed line represents an additional threshold of $5\times10^{19}$\cm\, at which morphological asymmetries are also measured in UMa galaxies.}
    \label{fig:CombinedND}
\end{figure*}

\par Individual WSRT, VLA-D and VLA-C data cubes and mosaics used in this work have diverse spatial and velocity resolutions, as is evident from Tables \ref{tab:obsBlind_details} and \ref{tab:obsWSRT_details}. To investigate the effect of the environment on the \HI\ disk of galaxies, we measure and compare the asymmetries in the \HI\ morphology of galaxies in the UMa and PP volumes. Therefore, to ensure a fair comparison of the measured asymmetries, it is necessary to bring all data cubes to the same resolution in terms of kpc and \kms. The VLA-C mosaic of the PP survey volume has the lowest spatial (4.8 kpc) and velocity (20\,\kms) resolution. Thus, we bring all the UMa data cubes from the VLA-D and WSRT observations to the same spatial (in kpc) and velocity resolution. At a distance of 17.1 Mpc, 4.8 kpc corresponds to 56\arcsec\ in angular resolution, to which the UMa WSRT data cubes and VLA-D mosaic were smoothed. Similarly, the velocity resolution of the UMa data was lowered to 20\,\kms\ to match the velocity resolution of the PP observations. Data cubes observed with the WSRT that have a channel width $\Delta V = 16.5$\,\kms\ were not smoothed to lower velocity resolutions as their native resolution is already close to or exceeding 20\,\kms. 

\par Additionally, to improve column density sensitivity, both the UMa and PP data cubes and mosaics were smoothed to a spatial resolution of 8 kpc (26\arcsec\ for PP and 98\arcsec\ for UMa). The data cubes at 8 kpc resolution were further smoothed to a velocity resolution of 40\,\kms\,. Frequency-dependent masks that define regions of \HI\ emissions were created by applying a $2.5\sigma$ cut-off to the 8 kpc and 40\,\kms\ data cubes. This allows the detection of \HI\ gas at lower column density in the outer regions of galaxies. Furthermore, the masks are visually inspected and unrelated noise peaks are interactively removed. In the aforementioned steps, both spatial and velocity smoothing are achieved by convolving with a Gaussian kernel of the appropriate size.

\subsection{HI global profiles}

\par An \HI\ global profile for a galaxy is created from the primary-beam corrected flux contained within the masked region of the homogenised data cube. From the global profiles, the width of the \HI\ emission lines at 20 per cent and 50 per cent of the peak flux are determined. The systemic velocity of each galaxy is calculated as follows:

\begin{align}
\label{Vsys}
\begin{split}
    V_\mathrm{sys} &= 0.5(V_{a,20}+V_{r,20})\\
    W_{20} &= V_{r,20}-V_{a,20}\\
    W_{50} &= V_{r,50}-V_{a,50}
\end{split}
\end{align}

\noindent where $a$ and $r$ in the subscript denote the approaching and receding sides of the global profile respectively; $20$ and $50$ in the subscript denote velocities corresponding to 20 percent and 50 percent of the peak value in the global profile. $W_{20}$ and $W_{50}$ denote the full measured widths of the global profile at 20 percent and 50 percent respectively.

\par Global profiles are also used to calculate the \HI\ mass of a galaxy, using the following equation:

\begin{equation}
    M_\mathrm{HI} = 2.36 \times 10^5 D^2 \int S_{\nu}dv\ \ \text{M}_{\odot}
\end{equation}

\noindent For the PP galaxies we estimate the distance $D$ in Mpc according to $D = \frac{V_\mathrm{sys}}{H_\mathrm{o}}$ where $H_\mathrm{o}$\, is the Hubble constant ($H_\mathrm{o}=\text{73\,\kms}\mathrm{Mpc}^{-1}$). For the more nearby UMa galaxies, we adopt a common distance of $D=17.1\,\mathrm{Mpc}$. Furthermore, $\int S_{\nu}dv$ is the total integrated \HI\ flux in $\mathrm{Jy}$\,\kms derived by integrating the global profile.  For galaxies in the UMa and PP volumes, measured values of $V_\mathrm{sys},\,W_\mathrm{20}$ and $W_\mathrm{50}$, the integrated flux and the estimated \HI\ mass are tabulated in tables \ref{UMaAsyTable} and \ref{PPAsyTable}.

\par In the \HI Atlas pages of PP galaxies shown in Figure \ref{fig:PPHIAtlas}, \HI\ global profiles are illustrated in the third panel of the bottom row. The systemic velocity calculated using Equation \ref{Vsys} is indicated by an arrow in this panel. In each channel of the profile, the noise is estimated as the product of the RMS flux value and the square root of the number of pixels in the frequency-dependent mask of that channel and is denoted with error bars.

\subsection{HI column density maps}\label{coldenmap}

\par A \HI\ column density map for a galaxy is created by applying frequency-dependent masks to the homogenised data cube, applying a primary beam correction, and subsequently integrating along the velocity/frequency axis of the data cube. The pixel values are converted from units of flux density to units of column density by using the following equation:

\begin{equation}\label{ColDen}
    N_\mathrm{HI}\,[\mathrm{cm^{-2}}]\,= 1.83 \times 10^{18} \int T_\mathrm{b} dv 
\end{equation}

\noindent where $T_\mathrm{b}$ corresponds to the brightness temperature of the emission in Kelvin and $dv$ to the channel width in \kms. Given the proximity of the suveyed volumes, we ignore cosmological corrections to the column density. The brightness temperature $T_\mathrm{b}$ is calculated as follows:

\begin{equation}
    T_\mathrm{b}\,[\mathrm{K}]\,= \frac{605.7}{\Theta_{x}\Theta_{y}} S_{\nu} \bigg( \frac{\nu_{o}}{\nu} \bigg)^2
\end{equation}

\noindent where $\Theta_x$ and $\Theta_y$ are the major and minor axes of the Gaussian beam in arcseconds, $S_{\nu}$ is the flux density in $\mathrm{mJy}\,\mathrm{beam}^{-1}$, and $\nu_{o}$ and $\nu$ are the rest and observed frequency of the \HI\ line emission.

\par In the \HI\ Atlas pages of PP galaxies, the \HI\ column density map is shown in the second and third panel of the top row. In the second panel, the column density map is overlayed on the DECaLS $r-$band image of the galaxy. In this panel the \HI\ center of the galaxy is denoted by a blue cross. The DECaLS $r-$band image of the galaxy is shown in the first panel of the top row, with an orange circle denoting the optical center of the galaxy. In the third panel, the kinematic major axis of the galaxy is illustrated with a purple line and the morphological asymmetry of the galaxy is shown in the top left corner. 

\subsection{HI signal-to-noise maps}

\par In the column density maps created in the above manner a different number of channels is summed along different lines of sight, which results in a non-uniform noise distribution across the map. It is therefore important to measure how the noise, and thus the signal-to-noise, varies across the column density map. The procedure to create a signal-to-noise (S/N) map is as follows. We position the 3-dimensional mask of a galaxy at 8 different locations in the data cube where no \HI\ line emission is detected. Thereafter, we sum the flux along each line of sight in the mask at each of the 8 locations. A noise map is created by calculating the rms in the 8 flux values at each pixel position. A S/N map is obtained by dividing the $N_{\rm{HI}}$ column density map by the noise map.

\par We measure the average column density values along contours of constant S/N ratio for galaxies in the UMa and PP volumes at both the 4.8 kpc and 8 kpc resolution. For each galaxy, the average column density values as a function of S/N ratios are illustrated in Figure \ref{fig:CombinedND}. Since the noise varies among the data cubes of the homogenised sample, a fixed column density value corresponds to different S/N values. For example, at a resolution of 4.8 kpc and 20\,\kms\ , a column density of $5\times10^{19}$\cm\ corresponds to S/N$<$1 in the PP VLA-C data cubes, S/N$<$1.5 in the UMa WSRT data cubes, and S/N$<$4 in the UMa VLA-D data cubes. As described in \cite{Bilimogga2022}, to robustly measure the asymmetries in $N_\mathrm{HI}$ maps of galaxies, we should consider column densities at S/N$ >3$. The median column densities from each panel of Figure \ref{fig:CombinedND} at S/N$=3$ are tabulated in Table \ref{tab:NoiseDiag}. 

\par A signal-to-noise map of the galaxy is illustrated in the fourth panel of the top row of the \HI\ Atlas pages. In this panel, different colours (also shown on the scale to the right of this panel) indicate the signal-to-noise ratios in the homogenised data-cube of PP galaxies. 

\begin{table}
    \centering
    \caption{Median column density $N_\mathrm{HI}^\mathrm{med}(3\sigma)$ levels of the pixels that have a signal-to-noise of 3 in the column density maps for the various homogenised data-sets used in our analysis.}
    \begin{tabular}{lcc}
    \hline
    Data set & Resolution & $N_\mathrm{HI}^\mathrm{med}(3\sigma)$ \\
             & spatial / angular & ($\times10^{19}$ \cm)          \\
    \hline
        \multirow{2}{*}{UMa VLA-D mosaic} & 4.8 kpc / 56$^{\prime\prime}$ &  3.1 \\
                                          & 8.0 kpc / 98$^{\prime\prime}$ &  1.5 \\
                                          &                               &      \\
        \multirow{2}{*}{UMa WSRT cubes}   & 4.8 kpc / 56$^{\prime\prime}$ &  9.7 \\
                                          & 8.0 kpc / 98$^{\prime\prime}$ &  3.2 \\
                                          &                               &      \\
        \multirow{2}{*}{PP VLA-C mosaic}  & 4.8 kpc / 15$^{\prime\prime}$ & 34   \\
                                          & 8.0 kpc / 26$^{\prime\prime}$ & 17   \\
        \hline
        \label{tab:NoiseDiag}
    \end{tabular}
\end{table}

\begin{figure*}[ht!]
    \centering
    \includegraphics[scale=0.9, trim = 2cm 3cm 0cm 4.5cm, clip, width=\textwidth]{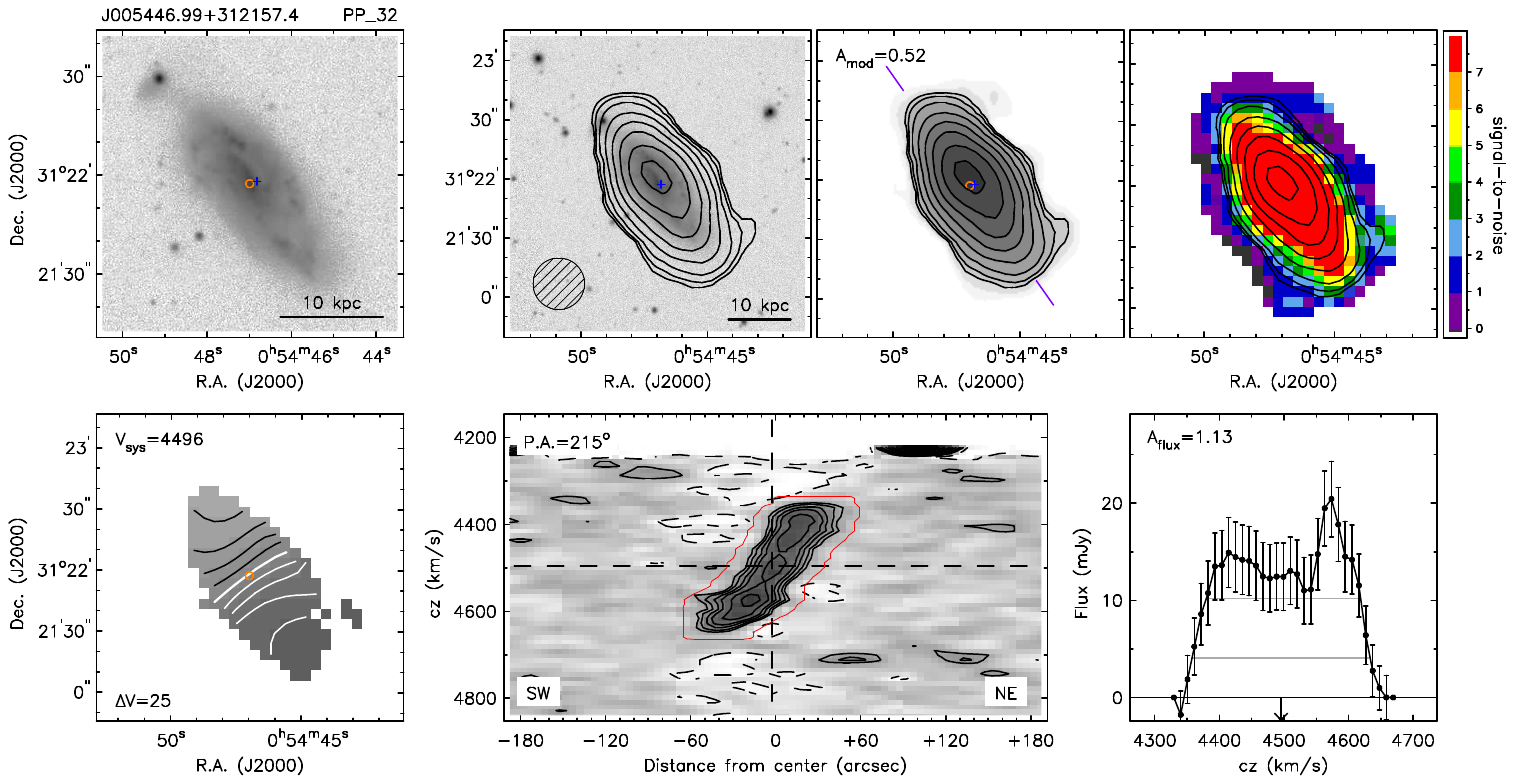}
    \caption{Example of an atlas page for galaxies within the PP volume. A detailed description of the features in the various panels can be found in Appendix \ref{sec:C3AtlasPP} where the atlas pages for all galaxies in the Perseus-Pisces environment are presented.}
    \label{fig:PPHIAtlas}
\end{figure*}

\subsection{Velocity fields}

\par A velocity field for each galaxy in the homogenised sample is produced by fitting a Gaussian to every velocity profile in the data cube within the frequency dependent mask. Initial estimates for the amplitude, central velocity and velocity dispersion of the Gaussian function are required. The initial estimate for the amplitude is the peak flux of the velocity profile, and the initial estimates for the central velocity and dispersion are calculated as the intensity-weighted first and second moments of the profile. Gaussian fits which have an uncertainty of more than 10\,\kms\ in the central velocity or have dispersion values higher than 50\,\kms\ are rejected.

\par In the first panel of the bottom row of the \HI\ Atlas pages, velocity field of the galaxy is illustrated. The systemic velocity of the galaxy is indicated by a thick white line. Several iso-velocity contours are overlaid on the velocity field, with their spacing noted in the lower left corner of the panel. Black contours represent approaching velocities, while white contours represent receding velocities in the illustration. 
\subsection{Position velocity diagrams}

\par A position-velocity diagram is created for each galaxy by extracting a two-dimensional slice from the homogenised data cube. First, the position angle of the major axis of the galaxy was estimated by visually inspecting the receding side of the \HI\ velocity field. In cases where the velocity field did not yield a clear position angle, the optical image of the galaxy was used to determine the position angle. Subsequently, the two-dimensional slice is centered on the \HI center and is made along the major axis determined previously. Position velocity slices are also created for the mask data cube of the galaxy and are overlaid on the diagram from the homogenised data cube in the \HI\ atlas pages. 

\par In the middle panel of the bottom row of the \HI\ Atlas pages, position-velocity diagram of the galaxy is illustrated. The frequency-dependent mask used in the creation of the \HI\ data-products is overlaid as a red line in the panel. The position angle of the major axis is indicated in the top left corner of the panel. 

\subsection{\HI\ atlas pages}

\par A description of the data products acquired from the volume limited observation of the UMa volume and the corresponding atlas pages presenting these products can be found in \cite{Busekool2020}. Similarly, the data products and atlas pages of the targeted observation of the UMa volume can be found in \cite{Verheijen2001}. 

\par In this chapter, we present atlas pages for the \HI\ detections in the VLA-C survey of the PP volume. An example atlas page is presented in Figure \ref{fig:PPHIAtlas} while the atlas pages for all \HI\ detections are presented in Appendix \ref{sec:C3AtlasPP}. A detailed description of the various panels is provided in Appendix \ref{sec:C3AtlasPP}.

\begin{figure*}[ht!]
    \centering
    \includegraphics[width=0.86\textwidth]{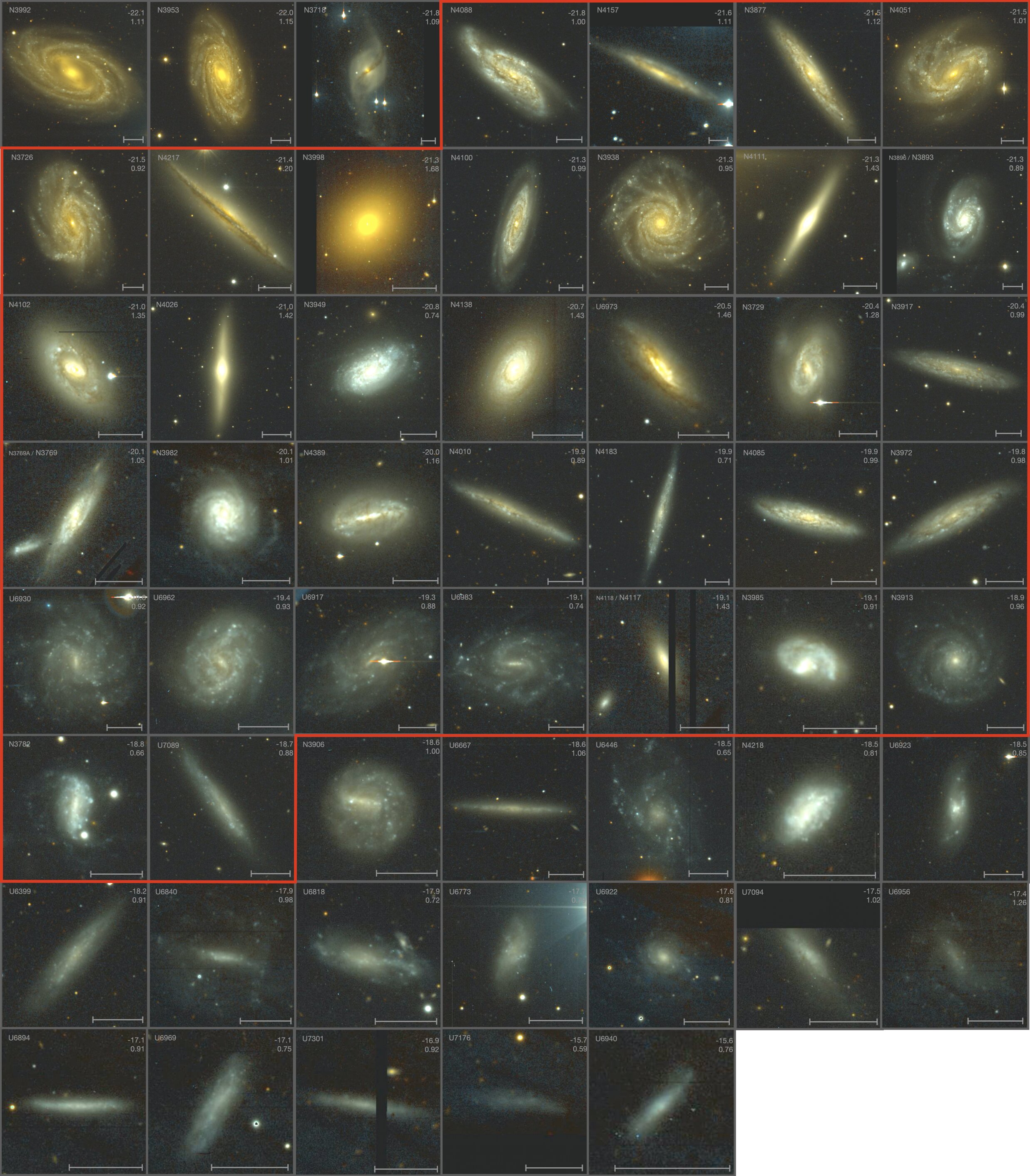}
    \caption{Colour images of galaxies with \logMHI$>8.42$ in the Ursa Major volume, created from photometric imaging data by \cite{Tully1996}, arranged in order of increasing M$^{\rm b,i}_{\rm R}$ magnitude. The red box outlines galaxies with $-21.78 < {\rm M}^{\rm b,i}_{\rm R} < -18.74 $ or $200 > {\rm V}_{\rm max} > 80$ km\;s$^{\rm -1}$ as used in Section \ref{sec:Three-way Comparison}. Within each image, the M$^{\rm b,i}_{\rm R}$ magnitude and M$^{\rm b,i}_{\rm B}-$M$^{\rm b,i}_{\rm R}$ colour are stated in the top right corner. A scale bar indicating 5 kpc is shown in the lower right corner.}
    \label{fig:UMa_magArranged1} 
\end{figure*}

\begin{figure*}[ht!]
    \centering
    \includegraphics[width=0.82\textwidth]{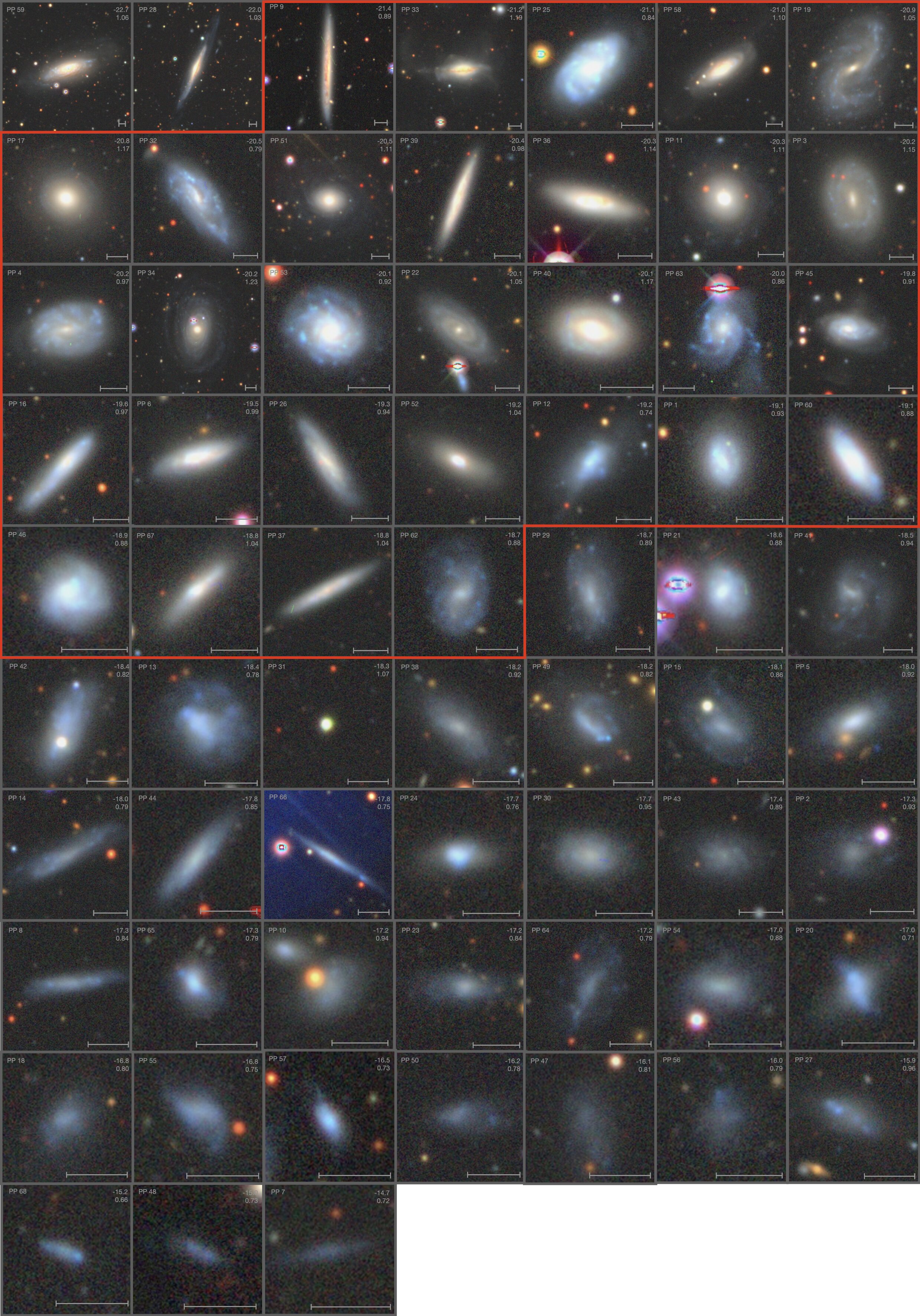}
    \caption{Similar to Figure \ref{fig:UMa_magArranged1} but for galaxies in the Perseus-Pisces volume, using DECaLS images extracted from the SkyViewer tool at http://www.legacysurvey.org.}
    \label{fig:PP_magArranged1}
\end{figure*}

\section{Photometric data}\label{sec:C3Photometry}

\par The UMa and PP volumes surveyed by the \HI\ observations are covered by the SDSS-DR14 photometric imaging survey \citep{SDSSDR14}, which provides total model magnitudes in the $u,g,r,i,z$ bands by fitting either an exponential or a deVaucouleurs profile to a galaxy's light distribution. These sky areas are also covered by the deeper DECaLS-DR9 photometric imaging survey \citep{DECaLSDR9}, which provides total model magnitudes in the $g,r,z$ bands by fitting a more generic Sersic profile. In addition to these two surveys, \cite{Tully1996} provided total Johnson $\mathrm{B_J}$- and Cousin $\mathrm{R_C}$-band magnitudes (B and R hereafter) for 79 UMa galaxies by integrating the observed luminosity profiles, supplemented by carefully fitting and extrapolating the exponential surface brightness profiles of the outer disks of galaxies to infinity.

\par The nearby UMa galaxies can span several arcminutes on the sky and both the SDSS and the DECaLS survey struggle to make proper model fits to these galaxies as illustrated below. Therefore, the B- and R-band magnitudes provided by \cite{Tully1996} are adopted for the UMa galaxies, which will facilitate comparisons with older, pre-SDSS and pre-DECaLS literature studies that used the same B- and R-band filters, as well as in Section \ref{sec:Three-way Comparison} of this paper.

\par For the PP galaxies, only SDSS and DECaLS imaging photometry is available and their $g$- and $r$-band total model magnitudes are converted to B- and R-band magnitudes by using the reworked transformation equations described in Tables 1 and 3 of \cite{Cook2014}, which are stated here:

\begin{equation} \label{Eq:B-Cook}
    \mathrm{B_{\rm Cook}} = g + 0.268(g-r) + 0.218
\end{equation}
\begin{equation} \label{Eq:R-Cook}
    \mathrm{R_{\rm Cook}} = r + 0.214(g-r) - 0.272
\end{equation}

\begin{figure}[ht!]
    \centering
    \includegraphics[width=\columnwidth]{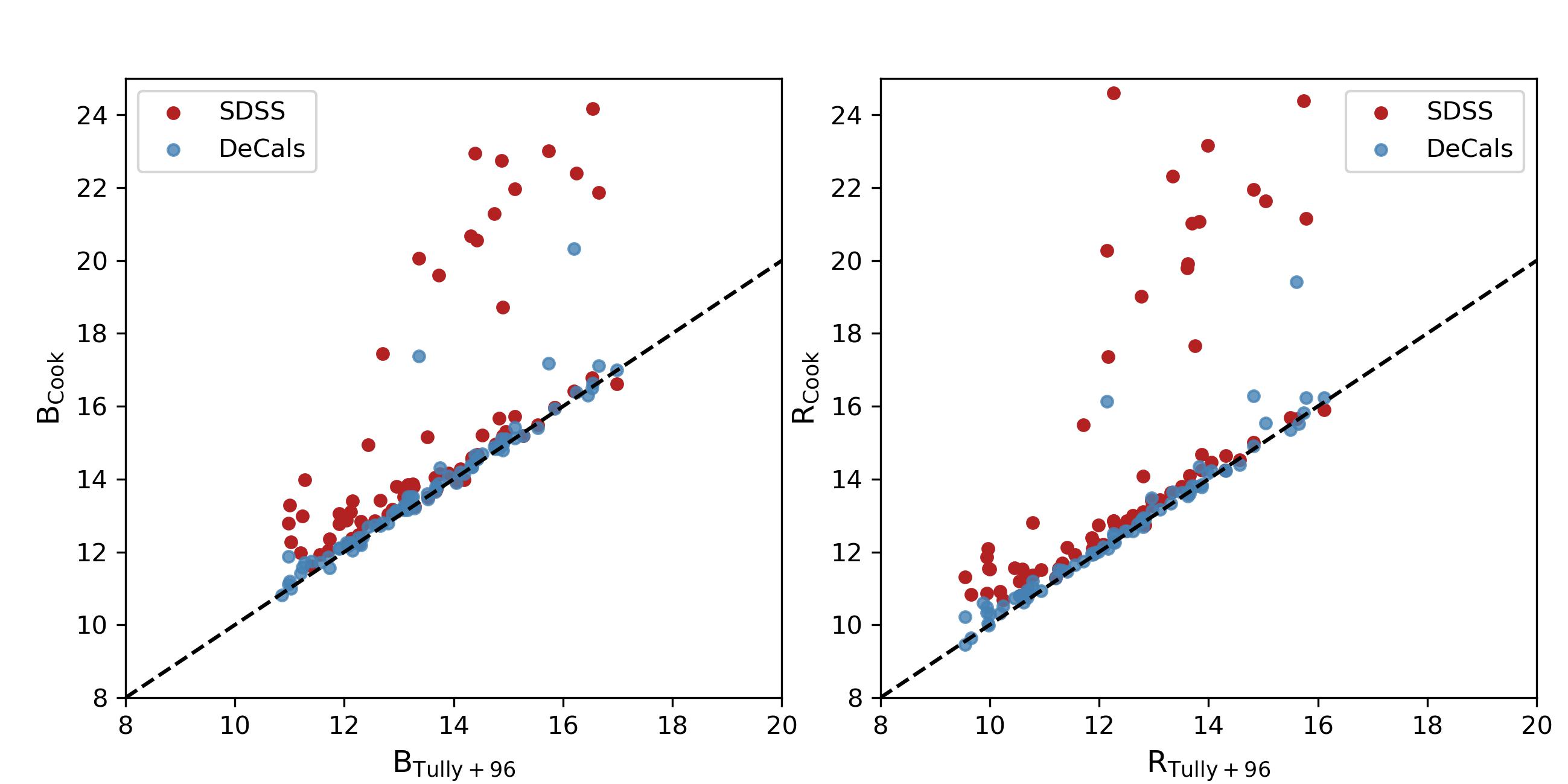}
       \caption{Comparison between the B- and R-band magnitudes for the UMa galaxies from \cite{Tully1996} and magnitudes derived from Equations \ref{Eq:B-Cook} and \ref{Eq:R-Cook}. Blue data points denote magnitudes derived from the DECaLS survey while red points denote magnitudes derived from the SDSS survey.}
    \label{fig:ThesisCookCompare}
\end{figure}

\noindent
The reliability of these transformation equations \ref{Eq:B-Cook} and \ref{Eq:R-Cook} is verified for the UMa galaxies in Figure \ref{fig:ThesisCookCompare}, which compares the B$_{\rm Cook}$ and R$_{\rm Cook}$ magnitudes as derived from the SDSS and DECaLS $g$- and $r$-band model magnitudes against the total B$_{\rm Tully+96}$ and R$_{\rm Tully+96}$ magnitudes provided by \cite{Tully1996}. Figure \ref{fig:ThesisCookCompare} illustrates that the SDSS- and DECaLS-based B$_{\rm Cook}$ and R$_{\rm Cook}$ magnitudes may differ significantly from the total magnitudes measured by \cite{Tully1996}. Many of the big and bright UMa galaxies suffer from fragmentation in the SDSS survey and thus have unreliable B$_{\rm Cook}$ and R$_{\rm Cook}$ magnitudes. At the fainter end, most galaxies are low surface brightness in nature and the limited depth of the SDSS survey results in unreliable magnitudes for these galaxies. On the other hand, magnitudes derived from the deeper DECaLS survey are largely consistent with the magnitudes presented by \cite{Tully1996}. Note that three deviating UMa galaxies (NGC 4010, UGC 7176, UGC 7401) are not fully contained in the DECaLS data bricks and thus their DECaLS $g-$ and $r-$magnitudes could not be determined properly. 

\par From Figure \ref{fig:ThesisCookCompare}, it can be concluded that the \cite{Cook2014} transformations are appropriate to infer B- and R-magnitudes from the $g$- and $r$-band magnitudes for the UMa galaxies, apart from model fitting issues. Furthermore, the derived B- and R-magnitudes from the DECaLS photometry are more reliable than those derived from the SDSS magnitudes. Therefore, DECaLS total $g$- and $r$-band model magnitudes will be used for the PP galaxies to derive their B- and R-magnitudes.

\par The methodology of \cite{Tully2008} is followed to convert the apparent, measured (UMa) or derived (PP), total B- and R- magnitudes to absolute magnitudes. The distance to the UMa sample of galaxies (17.1 Mpc or a distance modulus of 31.16) follows from a cosmic flow field model, largely based on luminosity-linewidth distances with a zero point calibrated using nearby galaxies with Cephe\"id and TRGB distances:

\begin{equation} \label{Eq:TFrb}
    \mathrm{M^{b,i,k}_{\rm B}} = -19.99 - 7.27[{\rm Log}(W^i_R) - 2.5]
\end{equation}
\begin{equation} \label{Eq:TFrR}
    \mathrm{M^{b,i,k}_{\rm R}} = -21.00 - 7.65[{\rm Log}(W^i_R) - 2.5]
\end{equation}

\noindent
where the absolute magnitude M$^{\rm b,i,k}$ is corrected for Galactic extinction ($^{\rm b}$) following \cite{Schlegel1998}, internal extinction ($^{\rm i}$) following \cite{Tully2000}, and cosmological reddening ($^{\rm k}$), while $W^i_R$ is the observed HI line width at 20\% of its peak flux, corrected for turbulent motion ($_R$) following \cite{Tully1985} and inclination ($^i$). The corresponding cosmic flow model prefers a value for the Hubble constant of H$_0$=74 km\,s$^{-1}$Mpc$^{-1}$. Assuming that the PP sample of galaxies partakes in the quiet Hubble flow and adopting an average recession velocity of 4880 km\,s$^{-1}$, places the PP galaxies at a distance of 66 Mpc or a distance modulus of 34.10. Inclinations, corrected line widths and extinctions for the PP and UMa galaxies are calculated following \cite{Tully2008} but the cosmological reddening ($^{\rm k}$) can be neglected. The various observed and derived magnitudes, extinctions and parameters to calculate the extinctions are listed in Tables \ref{tab:PhotometryUMa_ECParams} and \ref{tab:PhotometryUMa} for the UMa sample, and Tables \ref{tab:PhotometryPP_ECParams} and \ref{tab:PhotometryPP} for the PP sample. 

\par Figure \ref{fig:UMa_magArranged1} shows the optical images of the UMa galaxies with \logMHI$>8.42$. Similarly, Figure \ref{fig:PP_magArranged1} shows the optical images for all \HI\ detected PP galaxies. Optical images in these figures are arranged by their M$^{\rm b,i}_{\rm R}$ magnitudes, printed in the upper-right corner of each panel along with their M$^{\rm b,i}_{\rm B}-$M$^{\rm b,i}_{\rm R}$ colours. The scale bar in the bottom-right corner indicates 5 kpc at the distance of the galaxies. Within each of the optical mosaic figures for the UMa and PP samples, the relative brightness and colour scales are the same for all galaxies, but these scales are different for the two samples. Note that NGC 4013 (M$^{\rm b,i}_{\rm R}=-$21.2, \logMHI$=9.46$) is missing from the UMa sample in Figure \ref{fig:UMa_magArranged1}. For NGC 3998 the B- and R-band magnitudes were derived from the DECaLS $g$- and $r$-band model magnitudes due to issues with the sky subtraction in the \citet{Tully1996} photometry. From the HI selected PP sample the optical images of galaxies PP 35 (M$^{\rm b,i}_{\rm R}=-$22.6), and PP 61 (M$^{\rm b,i}_{\rm R}=-$21.57) are missing as they are not fully contained within the VLA passband or the mosaiced sky area.

\par In \cite{Bilimogga2022}, mock galaxies from the \textsc{EAGLE} simulation were selected to have maximum rotational velocities in the range of 200\,\kms$>V_\mathrm{max}>$80\,\kms. To ensure a fair comparison of the UMa, PP and mock galaxy samples, only the 34 UMa galaxies and 31 PP galaxies with rotational velocities in the same range will be considered. As rotation curves are not available for all the galaxies in the UMa and PP volumes, we use equation \ref{Eq:TFrR} with $W^i_R = 2{\rm V}_{\rm max}$ to identify and select galaxies with $V_\mathrm{max}$ within the velocity range mentioned above on the basis of their corresponding $\mathrm{M}^\mathrm{b,i}_\mathrm{R}$ magnitudes ($-21.78 < M_\mathrm{R} < -18.74$). In figures \ref{fig:UMa_magArranged1} and \ref{fig:PP_magArranged1} these galaxies are enclosed by the red boxes.

\begin{figure}[ht!]
    \centering
    \includegraphics[width=\columnwidth]{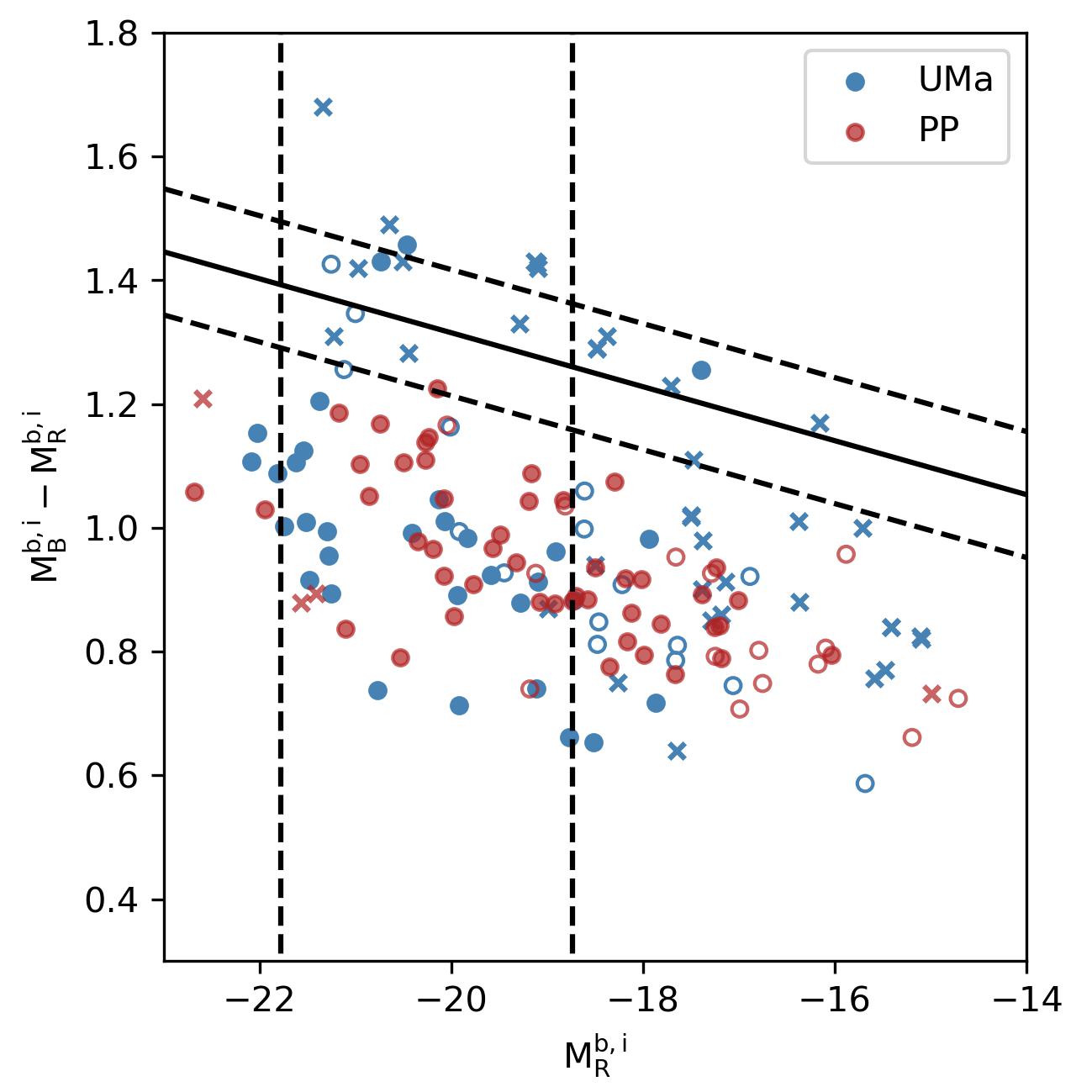}
       \caption{The colour-magnitude diagram of galaxies in the UMa (blue) and PP (red) samples. Blue crosses indicate UMa galaxies with \logMHI $<$8.42 and are often undetected in HI. Red crosses indicate HI-detected PP galaxies that are not completely contained within the VLA passband or the mosaiced sky area. Open circles indicate galaxies with \logMHI$>$8.42 but with less than three beams (3$\times$8 kpc) across the $15\times10^{19}$\cm\, \HI\ column density contour. Filled circles are larger, gas-rich galaxies. Vertical dashed lines enclose the magnitude range corresponding to $200>V_\mathrm{max}>80$\kms\ as applied when selecting mock galaxies from the \textsc{EAGLE} simulation. The solid black line indicates the red sequence of the cluster sample analysed by \cite{ChoqueC2021} with the dotted black lines indicating the $2\sigma$ scatter around the red sequence.}
       \label{fig:CMD}
\end{figure}

\par Figure \ref{fig:CMD} illustrates the Colour$-$Magnitude Diagram (CMD) for the UMa and PP galaxy samples. Recall that the UMa sample is optically selected and complete down to ${\rm m}_{\rm zw} = 14.5$ while the PP sample is HI selected down to $\rm{Log(M}_{\rm{HI}}$)$=8.42$. This means that there could be bright but very gas-poor galaxies in the PP volume, probably on the red sequence, that are not included in the PP sample. The probability that there exist faint (${\rm m}_{\rm zw} > 14.5$) but gas-rich (Log(M$_{\rm HI}$)$>$8.42) galaxies in UMa is small as the UMa volume was surveyed for HI emission by \citet{Wolfinger2013} with 95\% completeness down to Log(M$_{\rm HI}$)$=$8.26 for a distance of 17.1 Mpc.

\par Based on the CMD, there are several subtle but notable differences between the UMa and PP samples. First, at the bright end (M$^{\rm b,i}_{\rm R} < -21$), there are twice as many HI-detected UMa galaxies than gas-bearing PP galaxies (14 versus 7). Second, there are several gas-rich UMa galaxies on the red sequence while the HI-detected PP galaxies are confined to the blue cloud below the red sequence. Third, at the faint end ($-19 < M^{\rm b,i}_{\rm R}$), there are many more PP galaxies with extended HI disks (>24 kpc, solid symbols) than UMa galaxies (22 versus 6). These differences will be further discussed in Section \ref{Discussion}

\section{\HI\ asymmetries in UMa and PP galaxies}\label{sec:Asymmetry}

\subsection{Flux asymmetries in the global profiles}\label{GPAsym}

\par
\begin{figure}[!t]
    \centering
    \includegraphics[width=\columnwidth]{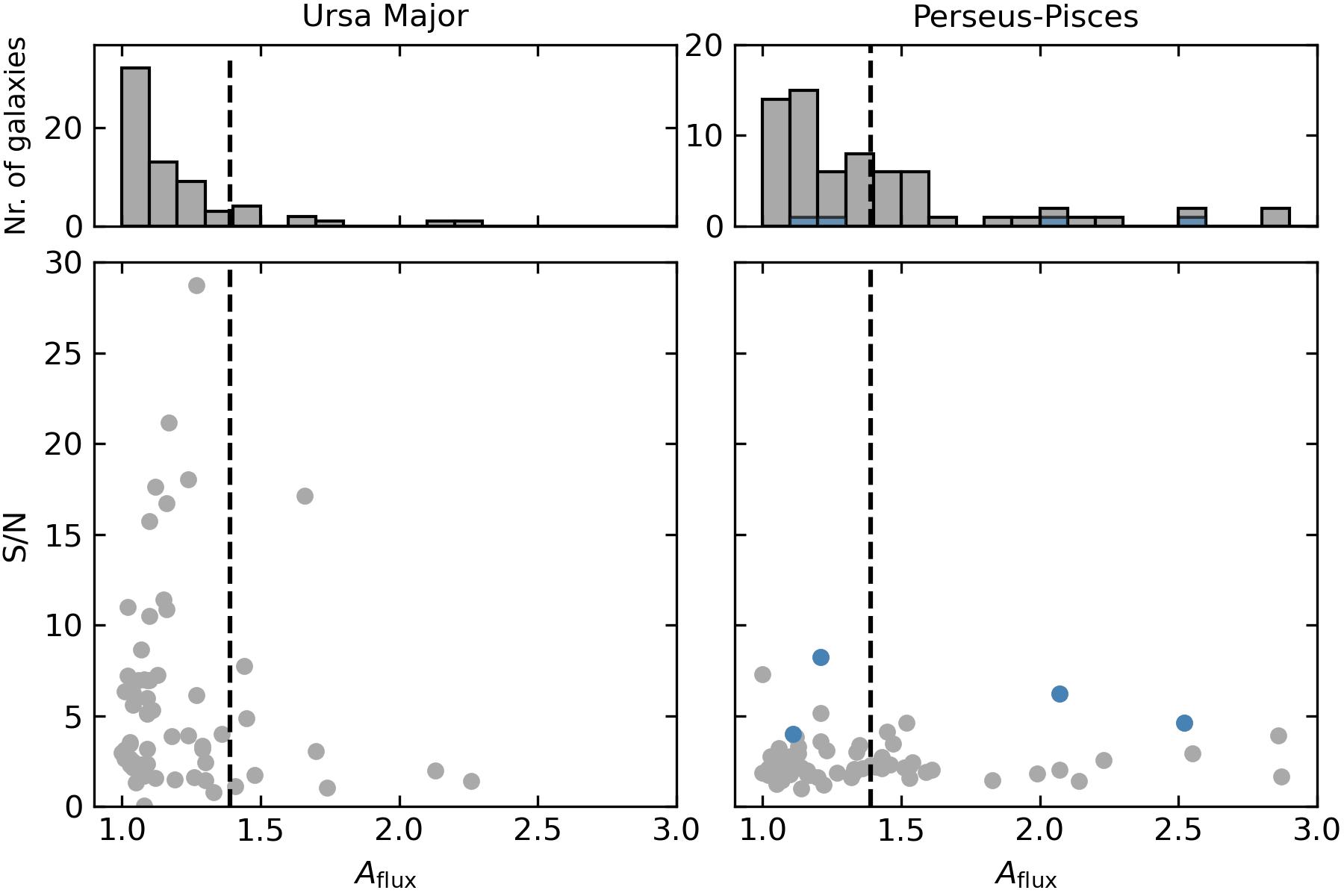}
    \caption{The \Aflux\ and average S/N values of the global profiles of the UMa and PP galaxies, measured at 20\,\kms\ resolution. Galaxies with \Aflux$>$1.39,  indicated with the vertical dashed line, are identified as asymmetric. Galaxies with S/N$\geq$4.0 and $80<V_\mathrm{max}<200$\kms\ are shown in blue. Histograms of the \Aflux\ values are shown in the top panels of the figure.}
    \label{fig:AfluxHist}
\end{figure}

 To measure asymmetries in the global profiles of galaxies we use the integrated flux ratio index (\Aflux) which is defined as follows:
\begin{equation}\label{Aflux}
    A_\mathrm{flux} = \frac{\int_{V_\mathrm{sys}}^{V_\mathrm{high}} S_\nu dv}{\int_{V_\mathrm{low}}^{V_\mathrm{sys}} S_\nu dv}
\end{equation}

\noindent
where $V_\mathrm{low}$ and $V_\mathrm{high}$ are the velocities at the 20 percent of the peak value of the spectrum and $V_\mathrm{sys}$ is the mid-point between the $V_\mathrm{low}$ and $V_\mathrm{high}$ velocities. Note that \Aflux\ always has values larger than 1. In \cite{Bilimogga2022}, we quantified the effects of noise in the \HI\ spectrum on the measured \Aflux\ values. We concluded that a minimum average signal-to-noise ratio of 6 is required to measure \Aflux\ robustly. The average S/N ratio of a global profile is defined as the mean of the S/N values in each channel. We measure \Aflux\ values from global profiles created as described in Section \ref{sec:DataProds} at a velocity resolution of 20\,\kms\,. Figure \ref{fig:AfluxHist} illustrates the measured \Aflux\ values as well as S/N values for the UMa and PP galaxies in the homogenised sample.

\par \cite{Espada2011} used the \Aflux\ index to measure the distribution of intrinsic lopsidedness in global profiles of isolated galaxies from the AMIGA survey. A half-Gaussian can characterise this distribution of asymmetries with a dispersion of $\sigma=0.13$. Only 9 percent and  2 percent of the AMIGA sample have \Aflux$ > 2\sigma$ and $3\sigma$ respectively. In this chapter, we define asymmetric global profiles of UMa and PP galaxies as those with \Aflux$>$1.39 =1+(3$\times$0.13), which is the $3\sigma$ value of the AMIGA sample. Figure \ref{fig:AfluxHist} shows that only 12 percent (8 out of 66) of the UMa galaxies have \Aflux$>1$.39 while 33 percent  (21 out of 63) of the PP galaxies have \Aflux$>$1.39. Considering only the high signal-to-noise global profiles (average S/N $\geq$ 6), we find that 9 percent (4 out of 45) of the UMa galaxies have global profile asymmetries and 33 percent (2 out of 6) of the PP galaxies. The difference in the distribution of the \Aflux\ values of the UMa and PP galaxies is significant even when low S/N global profiles are considered.  

\subsection{Morphological asymmetries in the column density maps}\label{MorphAsym}
We measure the morphological asymmetries in column density maps of galaxies using the modified asymmetry index \Amod\ , defined by \cite{Lelli2014} as:

\begin{equation}\label{AsyMod}
    A_\mathrm{mod} = \frac{1}{N}\sum_{i,j}^{N} \frac{|{I(i,j) - I_{180}(i,j)}|}{|{I(i,j) + I_{180}(i,j)}|}
\end{equation}

\noindent
where, $I(i,j)$ denotes the column density distribution in the original image of the galaxy and $I_{180}(i,j)$ is the column density distribution rotated by 180\degree\ around a chosen center, for which we adopt the optical center. The difference image is normalized with respect to the \textit{local} intensity of the pixels in order to emphasize asymmetries in the low column density outer \HI\ disks. The \Amod\ index is a modification of the asymmetry index that \cite{Conselice2000} measured from optical images of galaxies where the difference image is normalized with respect to the \textit{total} intensity.

\par The motivation for comparing morphological asymmetries of the UMa and PP galaxies is to investigate whether the \Amod\ distributions, which might be indicative of environmental processes acting on galaxies, are different in the two environments. To ensure high-quality measurements of \Amod\ from \HI\ column density maps, ideally the following three observational constraints as derived in \cite{Bilimogga2022} should be met:
1) \HI\ column densities should reach as low as $5\times10^{19}$\cm, \\
2) only include pixels with ${\rm S/N}>3$ in the column density map,\\
3) column density maps should be at least 11 beams across.\\

\par Figure \ref{fig:CombinedND} illustrates the S/N values at different column density levels in the 4.8 and 8 kpc resolution maps of the UMa and PP galaxies. Most of the UMa galaxies in the VLA-D survey meet these observational constraints on column density depth and S/N in both the 4.8 and 8 kpc resolution maps. In the column density maps from the WSRT survey these constraints are only met in the 8 kpc resolution maps. For the PP galaxies, however, pixels with (S/N)$>$3 in the column density maps at 4.8 kpc resolution do not reach below a column density of $25\times10^{19}$\cm\,. At 8 kpc resolution, the pixels with (S/N)$>$3 reach column densities between $10\times10^{19}$\cm\ and $25\times10^{19}$\cm\,. Consequently, a column density threshold of $15\times10^{19}$\cm\ at 8 kpc resolution was chosen for both the UMa and PP galaxies to ensure a fair and consistent comparison of the \Amod\ distributions for both samples. Additionally, a column density threshold of $5\times10^{19}$\cm will be considered for \Amod\ measurements for the UMa galaxies as this column density has $\rm{S/N}>3$ at 8 kpc resolution for these galaxies only. Note that for those UMa galaxies that are detected in both the VLA-D survey and the WSRT observations, the column density maps from the VLA-D survey have higher signal-to-noise and were therefore used in this analysis. One exception is the galaxy NGC 3769, which is only partially detected within the frequency band of the VLA-D survey. Therefore, its column density map from the WSRT observation is used.

\par
\begin{figure*}
    \centering
    \includegraphics[trim={0.8cm 0 0.5cm 0}, clip, width=0.9\textwidth]{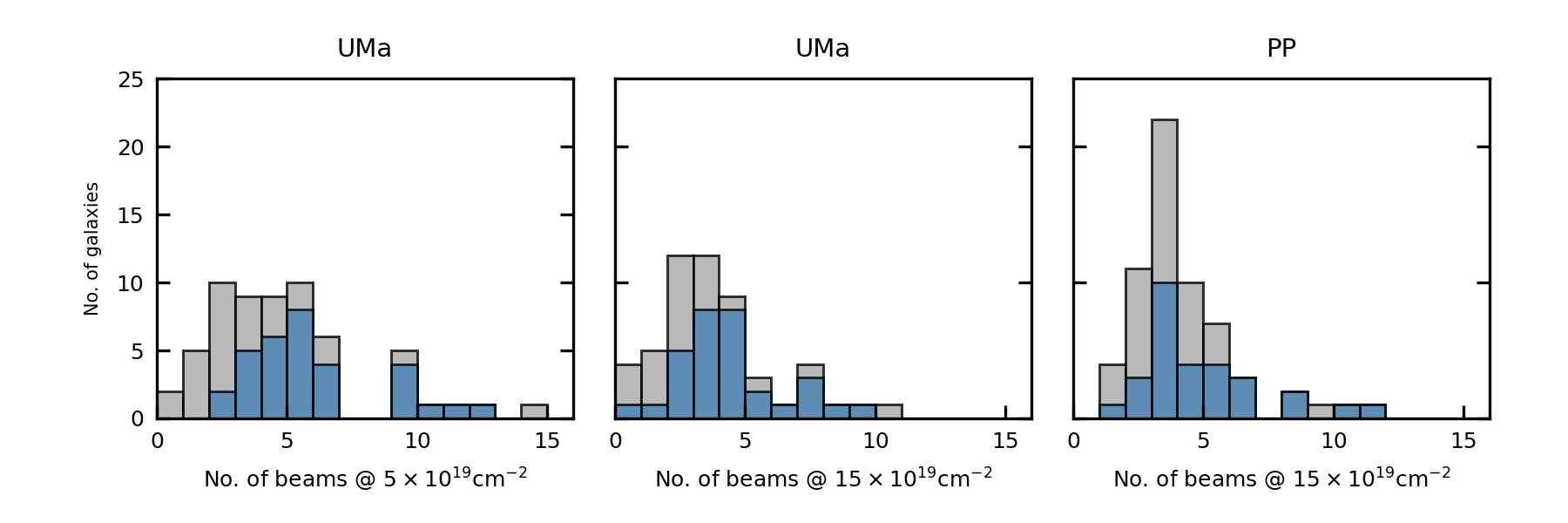}
    \caption{This figure illustrates the distribution of the number of beams across the UMa and PP galaxies. The beams are measured from column density maps where pixels below a certain column density value (mentioned in the panels) are clipped. Blue histograms illustrate the distribution of the number of beams of UMa and PP galaxies that have maximum rotational velocities in the range of 80\,\kms$<V_\mathrm{max}<$200\,\kms.}
    \label{fig:BeamDistribution}
\end{figure*}

 Figure \ref{fig:BeamDistribution} shows the distribution of the number of 8 kpc beams across the column density maps of the UMa galaxies at $5\times10^{19}$\cm\ and at $15\times10^{19}$\cm, and for PP galaxies at $15\times10^{19}$\cm. At the column density threshold of $15\times10^{19}$\cm\ none of the UMa galaxies and only one PP galaxy satisfies constraint 3 mentioned above. Similarly, at $5\times10^{19}$\cm, only three UMa galaxies have more than 11 beams across their column density maps. To circumvent the issue of small number statistics, \Amod\ values are therefore measured with a lower constraint of just 3 beams across the column density maps. 

\par To further match the UMa and PP samples in terms of observational constraints and galaxy properties, we will only consider UMa galaxies with a \logMHI$>$8.42, which is the \HI\ mass detection limit at the distance of the Perseus-Pisces volume. Henceforth, we will consider the \Amod\ distributions of the UMa and PP galaxies with $80<V_\mathrm{max}<200$\kms\ and \logMHI$>8.42$ based on pixels with $N_{\rm{HI}}>15\times10^{19}$\cm\ in their \HI\ column density maps at 8 kpc resolution, provided that these maps are at least 24 kpc (3 beams) across. These asymmetry values will be denoted as \AmodH\ hereafter. For the UMa galaxies only we will also consider the \Amod\ distribution considering pixels down to 5$\times$10$^{19}$\cm. These asymmetry values will be denoted as \AmodL\ hereafter.

\begin{figure*}
    \centering
    \includegraphics[width=0.9\textwidth]{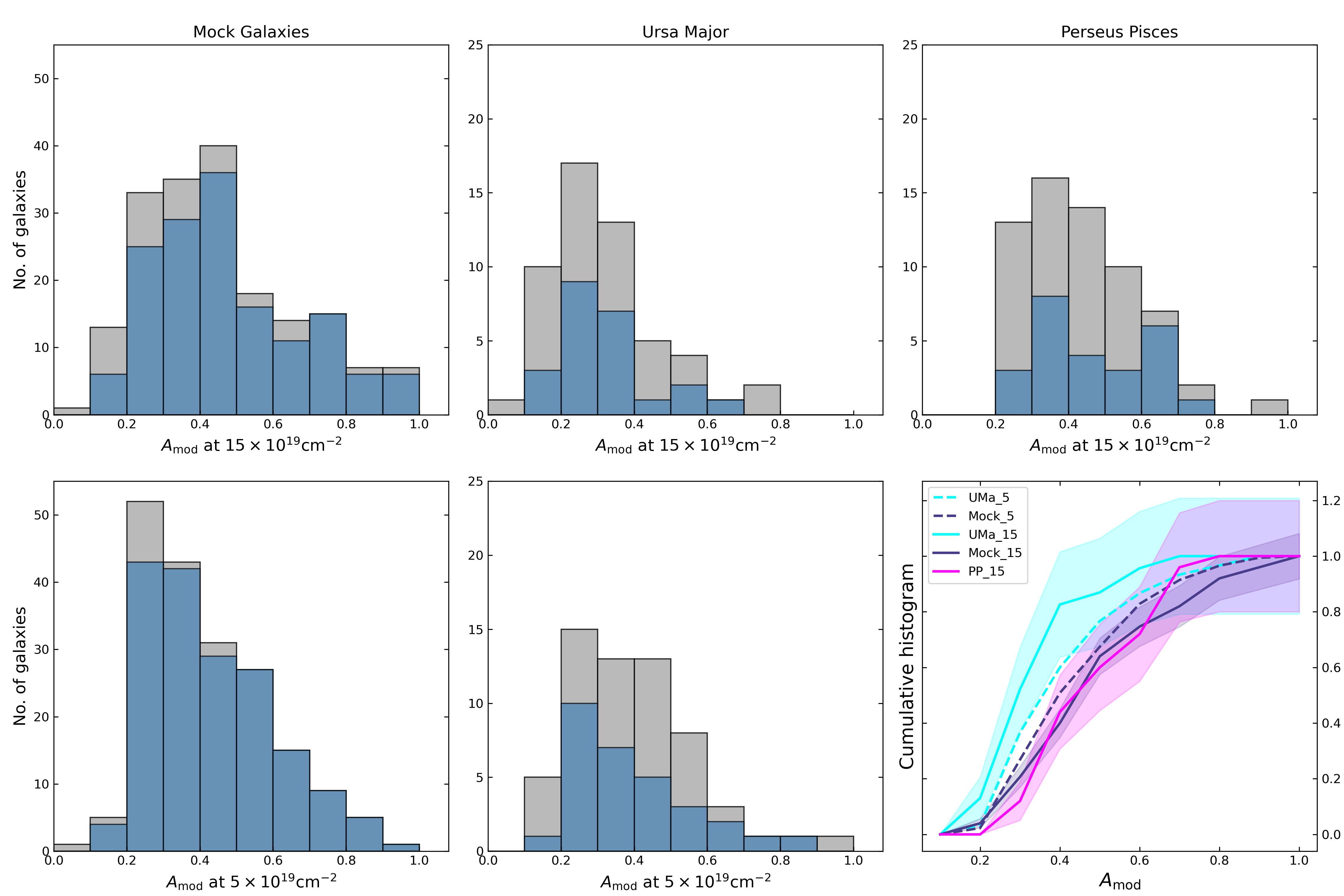}
    \caption{Comparisons of the \AmodH\ and \AmodL\ distributions of the mock galaxies from the \textsc{EAGLE} simulations, the UMa galaxies and the PP galaxies at 8 kpc resolution. The top row illustrates the \AmodH\ distributions while the bottom row shows the \AmodL\ distributions for the mock and UMa galaxies only. In both rows, the \Amod\ distributions of the full samples of galaxies are shown in grey while in blue the \Amod\ distributions of sub-samples of galaxies with $80<V_{\rm{max}}<200$\kms, \logMHI$>$8.42 and >3 beams across their major axis are shown. The bottom right panel shows the cumulative distribution functions of the \AmodH\ (solid lines) and \AmodL\ (dashed lines) values of the mock, UMa and PP sub-samples. The shaded regions indicate 1$\sigma$ errors on the cumulative distribution functions of the \AmodH\ values for the mock, UMa, and PP sub-samples.} 
    \label{fig:ComapreAmodMockUMaPP}
\end{figure*}

\subsection{Comparison of morphological asymmetries in UMa, PP and mock galaxies} \label{sec:Three-way Comparison}

\par The \AmodH\ and \AmodL\ distributions for the UMa and PP galaxies are illustrated in Figure \ref{fig:ComapreAmodMockUMaPP}. In addition, these distributions are also calculated and shown for a sample of noise-free mock galaxies from the EAGLE simulations, taken from \cite{Bilimogga2022} and processed such that they meet the same observational and mass constraints as the homogenised observed galaxy samples. We use 2-sample KS tests to determine if the various distributions shown in Figure \ref{fig:ComapreAmodMockUMaPP} are substantially different. These comparisons will demonstrate whether the sample of mock galaxies used in this study is representative of the observed galaxies in the UMa or the PP environments. We will first compare the \AmodH\ distributions for the UMa, PP and mock galaxies, followed by a comparison of the \AmodL\ distributions for the UMa and mock galaxies.

\par The 2-sample KS test contrasting the \AmodH\ distributions for the full UMa and PP samples (gray histograms) yields a p-value of 0.0023, implying that the two asymmetry distributions are not identical. This is also illustrated by the diverging cumulative distribution functions (CDFs) of the two samples shown in the bottom right panel of Figure \ref{fig:ComapreAmodMockUMaPP} (the banded blue and pink curves). The \AmodH\ distribution of the mock galaxies also appears to be different from that of the UMa galaxies, but seems to be similar to that of the PP galaxies. This is confirmed by two 2-sample KS tests. The p-value for the mock and UMa \AmodH\ distributions is 0.0002, indicating that the two distributions are indeed different. The p-value for the mock and PP \AmodH\ distributions is 0.55, implying that the two distributions are similar. The CDF for the \AmodH\ distribution of the mock galaxies is shown by the solid black line with its narrow shaded band in the bottom right panel of Figure \ref{fig:ComapreAmodMockUMaPP}. Evidently, it closely follows the CDF of the PP galaxies (pink solid line) but it is clearly distinct from the CDF of the UMa galaxies (solid blue line). At this point, we like to remark already that a higher column density threshold of $N_{\rm{HI}}>15\times10^{19}$\cm\ corresponds to the inner region of an \HI\ disk, which is not easily affected by external environmental processes, making the differences in the \AmodH\ distributions even more striking. This will be discussed further in a wider context in Section \ref{Discussion}. 

\par As mentioned before, the column density maps of the UMa galaxies at 8 kpc resolution reach column densities as low as $5\times10^{19}$\cm\ at S/N values well above 3. Therefore, \AmodL\ values of galaxies in the UMa environment can be safely compared to those of the noise-free mock galaxies at the same spatial resolution. Again, we compare the \AmodL\ distributions for UMa and mock galaxies with $80<V_{\rm{max}}<200$\kms\ and illustrate this in the bottom row of Figure \ref{fig:ComapreAmodMockUMaPP}. The \AmodL\ distributions of both samples appear similar and a 2-sample KS test yields a p-value of 0.47, implying that the two \AmodL\ distributions are likely to results from the same parent distribution. Similarly, the CDFs of the two samples follow each other closely (the dashed curves in Figure \ref{fig:ComapreAmodMockUMaPP}). Thus, we conclude that the mock and UMa galaxies have similar \AmodL\ distributions.

\par To summarize this subsection, we first note that the \AmodH\ distributions of the UMa and PP galaxies are distinct from each other, while the \AmodH\ distribution of the mock galaxies resembles that of the PP galaxies. At the same time we note that the \AmodL\ distribution of the mock galaxies resembles that of  the UMa galaxies. Since the CDF of the UMa galaxies rises more slowly for the \AmodL\ values than for the \AmodH\ values, we conclude that the outer \HI\ disks of the UMa galaxies tend to be slightly more asymmetric than the inner HI disks. We shall discuss possible astrophysical reasons for these differences in Section \ref{Discussion}.

\section{Discussion}\label{Discussion}

 In this section, we will first examine the relation between the two asymmetry indices \Aflux\ and \Amod. Next, we will discuss the differences between the \AmodH\ and \AmodL\ values for the UMa galaxies, and how these compare to those of the mock galaxies. Subsequently, we discuss possible astrophysical reasons for the different \AmodH\ and \AmodL\ distributions of the UMa and PP galaxies. Finally, we will investigate if galaxies with morphological \HI\ asymmetries also show peculiarities in their optical images.
 
\subsection{Comparing \Aflux\ and \Amod\ values}

\par For distant and spatially unresolved galaxies, we can only measure the asymmetries in their global profiles. It is therefore important to investigate the relation between asymmetries in the global profile (\Aflux) and the morphological asymmetries (\Amod) measured from spatially resolved \HI\ maps. In \cite{Bilimogga2022}, we have shown that for the mock galaxies from the \textsc{EAGLE} simulations, the \Aflux\ and the \Amod\ values are unrelated. The same conclusion was reached by \cite{Hank2025} on the basis of simulated SIMBA galaxies. Observationally, \cite{Reynolds2020} have shown that only a weak correlation exists between global profile and morphological asymmetries. Since we have measured both the global profile asymmetries and the morphological asymmetries at 8 kpc resolution for both the UMa and PP galaxies, we can now investigate for our samples how these indices relate to each other. 

\par
\begin{figure}
    \centering
    \includegraphics[width=\columnwidth]{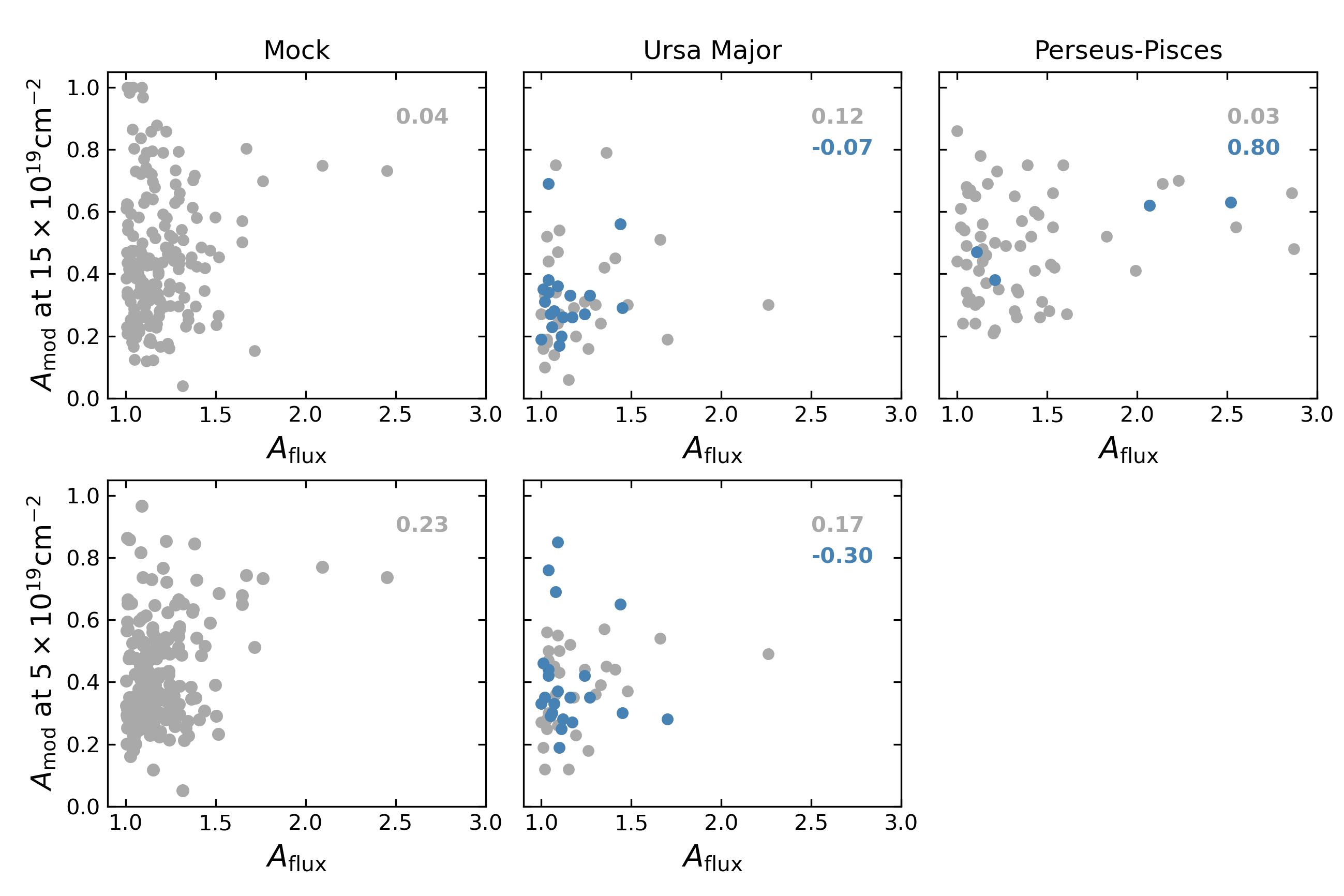}
    \caption{This figure compares the \Aflux\ and the \Amod\ values of the mock, UMa and PP samples. The \Aflux\ values are measured with a velocity resolution of 20\,\kms. Similarly, the \Amod\ values are measured at 8 kpc resolution at high and low thresholds of $15\times10^{19}$\cm\ and $5\times10^{19}$\cm\ respectively. Points shown in blue indicate galaxies that satisfy the constraints on both observational parameters as well as on the $V_{\text{max}}$ range. The Spearman correlation coefficient between the \Aflux\ and the \Amod\ values are shown in the top right corner of each panel.}
    \label{fig:AmodAflux_combined}
\end{figure}

 In the top row of Figure \ref{fig:AmodAflux_combined} we compare the \Aflux\ and the \AmodH\ values of the UMa and PP galaxies measured at 8 kpc and 20\,\kms\ resolution as described in Section \ref{sec:Asymmetry}. The comparison of the \Aflux\ and the \Amod\ values of the noise-free mock galaxies from \cite{Bilimogga2022} at the same resolution is also illustrated in this figure. The \textbf{19} UMa and the \textbf{4} PP galaxies that satisfy the observational constraints for \textit{both} the \Aflux\ and \AmodH\ indices, as well as the $V_{\text{max}}$ range, are shown with blue symbols. The \textbf{34} UMa galaxies and the \textbf{64} PP galaxies that do not satisfy all these constraints are shown with grey symbols. Considering only the blue points in the UMa and PP panels of the top row of Figure \ref{fig:AmodAflux_combined}, due to the low number of points, we do not find a statistically significant correlation between the \Aflux\ and the \AmodH\ values of these galaxies. Even when considering all galaxies (blue and grey points) in the UMa and PP panels, there is no correlation between the two asymmetry indices as indicated by the Spearman correlation coefficients, which are $0.12$ and $0.026$ for the UMa and the PP galaxies respectively.
 
\par We also compare the \Aflux\ values of the UMa galaxies to the \AmodL\ values at the lower column density threshold. This is illustrated in the bottom row of Figure \ref{fig:AmodAflux_combined}, which also includes a panel comparing the \Aflux\ and \AmodL\ values of the noise-free mock galaxies. Again, we do not see a correlation between the two asymmetry indices for both the UMa and mock samples. The Spearman correlation coefficient for the UMa galaxies is $0.174$ for the full sample and -0.227 for the sample of galaxies satisfying observational constraints on both the \Aflux\ and the \AmodL\ indices. We remind the reader that the global profiles are constructed from 3-dimensional masks applied to the observed data cubes. The boundaries of these masks correspond to typical column densities comparable to or less than $5\times10^{19}$\cm\ for both the UMa and PP galaxies. 
 
\par The shape of a global profile is influenced by both the distribution and the kinematics of the \HI\ gas disk of a galaxy (\cite{Watts2020oct}, figure 2). Therefore, to gain some understanding as to why the \Aflux\ and \Amod\ values are unrelated, we visually inspect the position-velocity diagrams and 2-dimensional velocity fields of a few galaxies selected from Figure \ref{fig:AmodAflux_combined}. This kinematic information is presented in the atlas pages of \cite{Verheijen2001} for the UMa galaxies and in Appendix \ref{sec:C3AtlasPP} for the PP galaxies. We define four categories of asymmetry as described below. We use \Aflux$=1.39$ \citep{Espada2011} to distinguish galaxies with low or high \Aflux\ values. Additionally, we adopt \Amod$ = 0.5$ to distinguish galaxies with low or high \Amod\ values.

\par
\begin{itemize}
    \item Low \Aflux\ and low \Amod: Galaxies in our sample that have a low \Aflux\ and a low \Amod\ value display a symmetric velocity field and a symmetric position-velocity diagram. It should be noted that a velocity field does not convey information about the column density distribution, while the position-velocity diagram only displays the flux distribution and kinematics along a slice through the cube. Some examples of such systems in UMa are UGC 6446, UGC 6922, UGC 6930, and UGC 7094. PP 28 and PP 11 are examples of such systems in the PP volume.
    \item High \Aflux\ and low \Amod: Galaxies that have low \Amod\ but a high \Aflux\ value must have a significant kinematic asymmetry. The asymmetry in kinematics can also be observed in the unequal flux distribution in the position-velocity diagram across the major axis of the galaxy as well as in the isovelocity contours of the velocity field. NGC 3729 and NGC 4085 as well as PP 43 and PP 60 are examples of such systems. 
    \item High \Aflux\ and high \Amod: Galaxies that have high \Aflux\ and high \Amod\ values could have either symmetric or asymmetric kinematics. After all, the global profile is a convolution of the 2-dimensional kinematic and spatial distribution of the gas. One such example is PP 17 where the galaxy shows a fairly symmetric velocity field but the gas disk is extended to the north (along the receding side of the kinematic major axis). On the other hand, NGC 4088 has an asymmetric velocity field and also an asymmetric morphology. 
    \item Low \Aflux\ and high \Amod: A galaxy with asymmetric morphology can nevertheless have a symmetric global profile owing to conspiring asymmetry in its kinematics. In our homogeneous sample, we come across galaxies such as NGC 3893, which is an interacting system with a warped disk and a tidal tail. Similarly, PP 57 also has a nearly symmetrical global profile yet its morphology and velocity field show asymmetries. 
\end{itemize}

\par The above examples illustrate that the asymmetries in morphology and kinematics of a galaxy can conspire to result in different levels of asymmetry in the global profile. This explains the lack of correlation when comparing the \Aflux\ and \Amod\ indices for the homogeneous sample, which is also observed in previous studies \citep{Reynolds2020}. This presents a clear challenge for asymmetry studies of high redshift and spatially unresolved galaxies for which global profile asymmetries can not be used unambiguously to infer morphological and/or kinematic asymmetries.

\subsection{\Amod\ values of the mock and UMa galaxies}\label{sec:AsyMockUMa}

\begin{figure}
    \centering
    \includegraphics[width=\columnwidth]{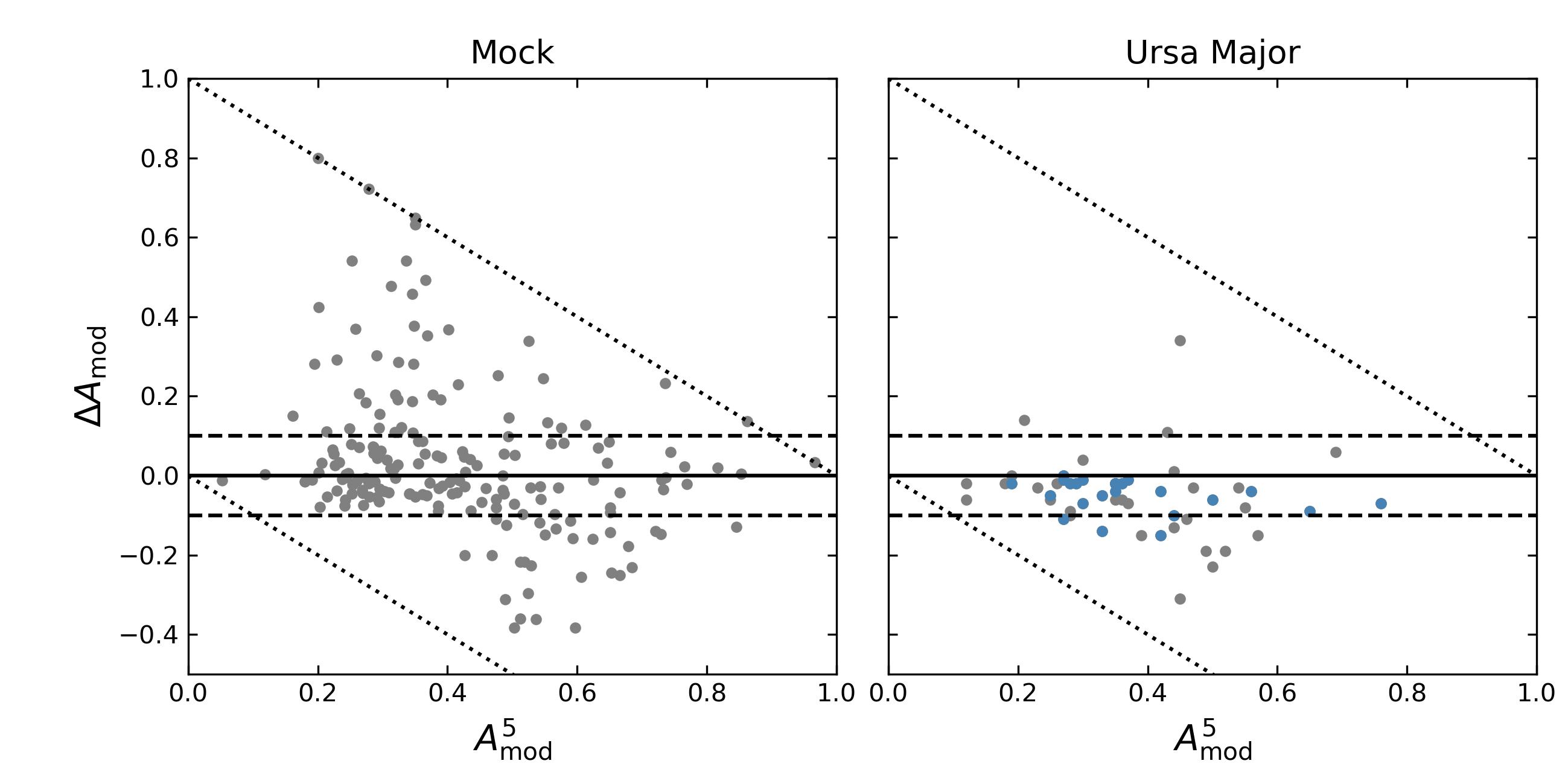}
    \caption{In this figure we show the change in \Amod\ values at $15\times10^{19}$\cm\ with respect to those at $5\times10^{19}$\cm\ in the mock and the UMa galaxies. In the UMa sample, galaxies that have maximum rotational velocities in the range of 80\,\kms$<V_\mathrm{max}<$200\,\kms, with \logMHI$>8.42$ and that have more than 3 beams across their major axis are shown in blue. The region between the dotted lines indicates the area where valid $\Delta A_{\text{mod}} = A^{15}_{\text{mod}}-A^{5}_{\text{mod}}$ values can be found.}
    \label{fig:CompareAmodMockUMa}
\end{figure}

\par In Section \ref{sec:Three-way Comparison}, we found that the \AmodL\ distributions are similar for the UMa and the mock galaxies. On the contrary, the \AmodH\ distributions are different. This is also illustrated by the CDFs shown in the bottom right panel of Figure \ref{fig:ComapreAmodMockUMaPP}. This observation begs the question: What causes the \AmodH\ distributions to be different at the higher column density threshold? Is the \Amod\ distribution of the UMa or the mock sample changing with an increasing threshold, or both? The CDFs in Figure \ref{fig:ComapreAmodMockUMaPP} reveal that the \Amod\ distribution of the UMa sample changes significantly such that the UMa galaxies become less asymmetric at higher column density. However, the \Amod\ distribution of the mock sample only changes marginally such that the mock galaxies become slightly more asymmetric at lower column densities. We will now look into this in more detail.

\par To gain a better understanding as to why the \Amod\ distributions are changing, we measure the difference $\Delta$\Amod\ = \AmodH$-$\AmodL\ for each individual galaxy in the UMa and the mock sample. We illustrate these differences in Figure \ref{fig:CompareAmodMockUMa} and find some surprising results. There is a striking difference in how the \Amod\ values are changing for the UMa and the mock galaxies. Where nearly all UMa galaxies show $\Delta$\Amod$<$0, the \Amod\ values vary erratically for the mock galaxies. We find that mock galaxies with \AmodL$\leq$0.4 tend to become more asymmetric at the higher column density threshold, which is not observed at all in the UMa sample. On the other hand, mock galaxies with \AmodL$>$0.4 tend to be less asymmetric at the higher column density threshold. The latter is also observed in the UMa sample but the $\Delta$\Amod\ values are, however, not as extreme as in the mock sample. Although the global correlation seen in the mock galaxy sample can be partly explained by the exclusion zones in Figure \ref{fig:CompareAmodMockUMa}, these zones do not limit the scattering of the UMa galaxies. To quantify these differences, we consider the number of galaxies with $\Delta$\Amod$>$0.1 (Category +1), |$\Delta$\Amod|$\leq$0.1 (Category 0), and $\Delta$\Amod$<$$-$0.1 (Category -1). For the mock galaxies, these ratios are 51:110:28 while for the UMa galaxies, these ratios are 3:39:11 for the full sample shown in grey in Figure \ref{fig:CompareAmodMockUMa} and 0:20:3 for the subsample shown in blue. 

\par We select a handful of UMa galaxies from the three categories defined above and visually inspect their \HI\ column density maps. In Category 0, most of the UMa galaxies have \AmodL$<$0.5, indicating that these UMa galaxies are only mildly disturbed. Also, within this Category 0 are four 'blue' galaxies with \AmodL$>$0.5 (NGC 3769, NGC 3893, UGC 6973 and UGC 6962) that are undergoing strong tidal interactions. It is surprising that their \Amod\ values do not change much and that we observe tidal features at both column density thresholds. All four 'blue' UMa galaxies belonging to Category $-1$ have \AmodL$<$<0.5, indicating that they are also only mildly disturbed. Nevertheless, the {\it relative} change in \Amod\ is significant, which suggests that the gas above column densities of $15\times10^{19}$\cm\ is distributed quite symmetrically around the optical center. Within Category $-1$, three of the UMa galaxies (NGC 3718, NGC 3729, and NGC 4088) have a neighboring galaxy that may have perturbed the outer low column density gas but did not disturb the gas at higher column densities, contrary to the strongly interacting galaxies in Category 0 mentioned above.

\par A small but notable subsample of UMa galaxies is comprised of NGC 3998, NGC 4026, and NGC 4111. These lenticular galaxies present peculiar cases as they do not fall in the three categories we described above yet they are spectacular examples of the environmental influence on the \HI\ disk of a galaxy. At the resolution of 8 kpc, the column density of the \HI\ gas in their tails does not exceed $15\times10^{19}$\cm\ while there is very little gas in the disk above $15\times10^{19}$\cm. Therefore, their \AmodH\ values cannot be measured and used to categorise these galaxies. These galaxies are lenticular systems with long tails of \HI\ gas that extend beyond the stellar disk. It is important to note that such striking lopsided features in \HI\ are not detected in any other UMa galaxy.

\par
\begin{figure*}
    \centering
    \includegraphics[width=\textwidth]{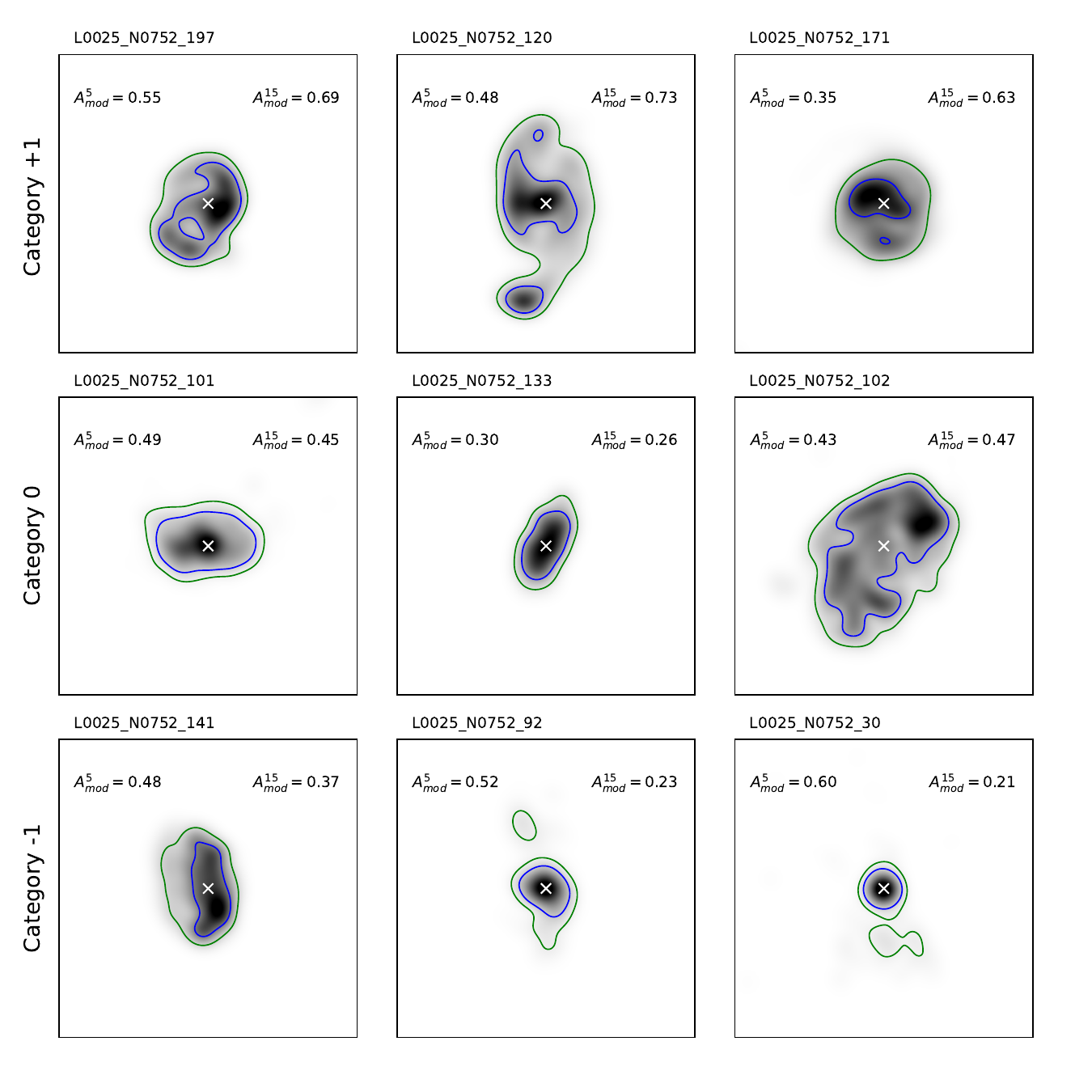}
    \caption{Column density maps of mock galaxies that fall into the three categories defined based on Figure \ref{fig:CompareAmodMockUMa}. The top row shows galaxies in Category +1, the middle row shows those in Category 0 and the bottom row shows those in Category -1. In each row, the galaxies are ordered by increasing differences in their \Amod\ values.}
    \label{fig:MockMosaic}
\end{figure*}

 Turning our attention to the sample of mock galaxies we note that many fall into Category +1, contrary to the UMa sample of galaxies. We visually inspect the column density maps of mock galaxies from the three categories and illustrate a handful of these in Figure \ref{fig:MockMosaic}. Mock galaxies in Category +1 have high \AmodH\ values due to the presence of clumps of high column density gas that are offset from the center as illustrated in the top row of Figure \ref{fig:MockMosaic}. It is not entirely clear what leads to the creation of mock galaxies such as those in Category +1. \cite{Bahe2016} find that galaxies in the \textsc{EAGLE} simulations have uncharacteristically large holes with low \HI\ surface density in their gas disk, which originate due to heating events included in the star-formation feedback mechanism. Even though we do not observe obvious holes in the gas disks of mock galaxies in our sample, it is possible that the feedback mechanisms may be responsible for the clumpiness that we observe.

\par For galaxies in Category 0, \Amod\ values do not change significantly, indicating that the high and low column density gas is equally affected (or unaffected) by the environment. This is shown in the middle row of Figure \ref{fig:MockMosaic}. For mock galaxies in Category -1 we observe that the low column density gas disk shows signatures of disturbances whereas the disk at the high column density is more symmetric with respect to the center. This is illustrated in the bottom row of Figure \ref{fig:MockMosaic}. We observe features in the mock galaxies in Category 0 and Category -1 that are similar to those of the UMa galaxies in these two respective categories. 

\subsection{Asymmetries in the mock, UMa and PP galaxies}\label{AsymmHypothesis}

\par In Section \ref{sec:Three-way Comparison}, we found that the \AmodL\ distribution of the mock galaxies is similar to the \AmodL\ distribution of the UMa galaxies, while the \AmodH\ distribution of the mock galaxies is similar to the \AmodH\ distribution of the PP galaxies as illustrated in Figure \ref{fig:ComapreAmodMockUMaPP}. In the previous section, we identified the strong star-formation feedback mechanism in the \textsc{EAGLE} simulation (described by \cite{Bahe2016}) as a likely cause for the higher \AmodH\ values in the mock galaxies. On the other hand, the \AmodH\ distribution of the UMa galaxies shifts to lower \Amod\ values compared to its \AmodL\ distribution. We found that for most UMa galaxies, their \AmodH\ values are close to or lower than their \AmodL\ values, i.e. UMa galaxies are in Category 0 or -1. Intuitively, this is the expected behavior of galaxies where the tidal interactions have not yet affected the gas at high column densities, which is typically located deeper in a galaxy's potential.

\par Due to the lower sensitivity of the VLA-C survey of the PP volume, we cannot robustly measure the asymmetry values below $15\times10^{19}$\cm\,. The PP \AmodH\ distribution is distinct from the UMa distribution and is shifted to higher \Amod\ values. Note that these results are aligned with the results of \cite{Angiras2007} who concluded, although with a different methodology, that galaxies in UMa are less asymmetric compared to galaxies in the Eridanus group.

The \AmodH\ distribution of the mock galaxies is similar to that of the PP galaxies. Based on our conclusions from Section \ref{sec:AsyMockUMa}, we put forth two hypotheses to explain the shift to higher \AmodH\ values for the PP sample compared to the \Amod\ distributions of the UMa sample:
\begin{itemize}
    \item Hypothesis 1: PP galaxies are similar to the mock galaxies, where strong star-formation feedback causes clumpy gas disks with large, low surface density holes and thus high \AmodH\ values. Therefore, we would expect a higher star formation rate in the PP galaxies than in the UMa galaxies. This would also mean that the underlying cause of asymmetries is different for the PP sample than for the UMa sample where in the latter, tidal interactions seem to be the leading cause for asymmetric \HI\ disks. This hypothesis of enhanced star formation in PP galaxies will be investigated further in a forthcoming study.
    \item Hypothesis 2: Just like the asymmetric UMa galaxies in Category 0, many more PP galaxies may also be undergoing strong tidal interactions which are affecting the high column density gas. These strong interactions would not only cause disturbances in the lower column density gas disk as well, which would show up in more sensitive \HI\ imaging observations, but also in the stellar disk. In the following subsection, we will investigate the presence of optical disturbances in the UMa and PP galaxies in relation to the asymmetries of their \HI\ disk. In a forthcoming study, we will investigate the role of nearby neighbors and the global environment to dive deeper into the root cause of the increased asymmetries in the PP sample compared to the UMa sample.
\end{itemize}

\subsection{Asymmetries in the optical images}

To investigate the presence of optical asymmetries in relation to \HI\ asymmetries, we assess the optical images of galaxies shown within the red boxes in Figures \ref{fig:UMa_magArranged1} and \ref{fig:PP_magArranged1}. These galaxies have inferred rotational velocities in the range of $80<V_{\rm{max}}<200$\kms\ and have \logMHI$>$8.42. In the UMa sub-sample, \cite{Tully1996} identify 28 spiral galaxies, 4 early-type galaxies, and 2 dwarf irregular (NGC 3782 and UGC 7089) galaxies. In the PP sub-sample however, most of the faintest galaxies do not have an optical morphological classification in the literature. Therefore, we visually classify the PP galaxies ourselves and identify 18 spiral galaxies, 8 early-type galaxies, and 4 dwarf irregular (PP 12, PP 1, PP 46, and PP 62) galaxies. 

\par The optical images of the 'red box' sub-samples, hereafter referred to as the `normal-contrast' images, are re-arranged in Figures \ref{fig:UMaOptMosaic1} and \ref{fig:PPOptMosaic1} by their \AmodH\ values to identify disturbances in the high-surface brightness parts of their stellar disks. We also present `high-contrast' DECaLS images of these galaxies in Figures \ref{fig:UMaOptMosaic15_HC1} and \ref{fig:PPOptMosaic15_HC1} to investigate if subtle tidal interactions result in disturbances at lower optical surface brightness levels. From hypothesis 2 postulated in Section \ref{AsymmHypothesis}, we would expect to find optical disturbances if galaxies are undergoing strong tidal interactions that would disrupt their \HI\ disk as well as their stellar disk. From Figures \ref{fig:UMaOptMosaic1}, \ref{fig:PPOptMosaic1}, \ref{fig:UMaOptMosaic15_HC1}, and \ref{fig:PPOptMosaic15_HC1} we visually identify optical disturbances as those features which may not be present on the opposite side of the disk, i.e. are not axisymmetric, or, if present, have a notably different surface brightness. This assessment is admittedly subjective but the reader is invited to assess our assessments.

\par First, we discuss the UMa sub-sample in the following paragraphs. Based on the above guidelines, we indicate galaxies with visually disturbed 'normal-contract' images, with an orange exclamation mark in each panel of Figure \ref{fig:UMaOptMosaic1}. We find that 10 out of the 28 spiral galaxies in the UMa sub-sample have asymmetries in their high-surface brightness stellar disk. Similarly, the two dwarf irregular galaxies have optical disturbances (basically by definition), while none of the early-type galaxies are optically disturbed. Next, we divide the UMa sub-sample shown in Figure \ref{fig:UMaOptMosaic1} into two subsets: those with \AmodH$\leq$0.33 (18 galaxies), and those with \AmodH$>$0.33 (16 galaxies). Each of these two subsets has 6 optically disturbed galaxies with a range of \AmodH\ values. Notably, the three UMa galaxies with the highest \AmodH\ or \AmodL\ values are early-type galaxies (NGC 3998, NGC 4111, NGC 4026) that host long tails of \HI\ gas, while their optical images show no disturbed features. We conclude that the \AmodH\ values of the UMa galaxies have no relation to the presence of optical disturbances. Such a lack of correlation is also found when considering the \AmodL\ values. 

\par In the `high-contrast' optical images of UMa galaxies displayed in Figure \ref{fig:UMaOptMosaic15_HC1}, disturbances are indicated by orange arrows. The panels in Figure \ref{fig:UMaOptMosaic15_HC1} are also ordered by \AmodH\ but zoomed out with respect to the images in Figure \ref{fig:UMaOptMosaic1}. We find that none of the UMa galaxies in the first subset (\AmodH$\leq$0.33) have disturbances at lower surface brightness levels. However, in the `normal-contrast' images, 6 galaxies from this subset showed optical disturbances. This implies that there are no UMa galaxies with \AmodH$\leq$0.33 that have optical disturbances in both the low and high surface brightness parts of their stellar disk. In the above-defined second subset (\AmodH$>$0.33), 7 of the 16 galaxies show low surface brightness disturbances. Of these, only 4 UMa galaxies exhibit deviations in both the low and high surface brightness parts of their stellar disk, while the remaining three galaxies have disturbances only in their outer, low surface brightness disk. We tabulate these numbers in Table \ref{tab:OpticalDisturbanceClasses} and discuss the possible implications of these disturbances later in this subsection.

\begin{table*}
\centering
\caption{Classification of UMa and PP galaxies from the two \HI\ asymmetry subsets into different classes of optical disturbance. Here HSB and LSB refers to the high surface brightness and low surface brightness parts of the stellar disk of galaxies.}
\label{tab:OpticalDisturbanceClasses}
\begin{tabular}{|l|c|c|c|c|}
\hline
    & \multicolumn{2}{c}{UMa} & \multicolumn{2}{|c|}{PP} \\\cline{2-5}
    & \AmodH\ $\leq$0.33 & \AmodH\ $>$0.33 & \AmodH\ $\leq$0.52 & \AmodH\ $>$0.52 \\[0.8mm]
    \hline
 Total & 18 & 16 & 16 & 14 \\
 Undisturbed  & 12 &  7 &  3 &  7 \\
 HSB only     &  6 &  2 &  4 &  3 \\
 HSB and LSB    &  0 &  4 &  5 &  2 \\
 LSB only     &  0 &  3 &  4 &  2 \\ 
\hline
\end{tabular}
\end{table*}

\par Now we will discuss the PP sub-sample in the subsequent paragraphs. We indicate galaxies with disturbances in their `normal-contrast' images with an exclamation mark in each panel of Figure \ref{fig:PPOptMosaic1}. We find that 9 out of 20 spiral galaxies and all four dwarf irregular galaxies exhibit optical disturbances at high surface brightness levels. Similar to the UMa sub-sample, we find that none of the 6 early-type galaxies exhibit an optical disturbance. We also divide the PP sub-sample into two subsets: those with \AmodH$\leq$0.52 (16 galaxies) and those with \AmodH$>$0.52 (14 galaxies). This \AmodH\ value that divides the PP sub-sample into almost equal halves lies at a higher \AmodH\ value compared to the UMa sub-sample. This is a natural consequence of the \AmodH\ distribution of the PP sub-sample, shown in Figure \ref{fig:ComapreAmodMockUMaPP}, which is shifted to higher \AmodH\ values compared to the UMa sub-sample. In the first subset (\AmodH$\leq$0.52), we find that 9 out of the 16 PP galaxies are optically disturbed while 5 out of the 14 PP galaxies are optically disturbed in the second subset(\AmodH$>$0.52). Similar to the UMa sub-sample, PP galaxies with high surface brightness disturbances have a range of \AmodH\ values and there is no correlation between the presence of optical disturbances and the \HI\ asymmetry. 

\par We display the `high-contrast' images of the PP galaxies in Figure \ref{fig:PPOptMosaic15_HC1} where the panels are also ordered by \AmodH\ values and optical disturbances identified at lower surface brightness are again indicated by orange arrows. In the low-\AmodH\ subset, we find that 9 of the 16 PP galaxies show low surface brightness optical disturbances, contrary to the low-\AmodH\ UMa galaxies. Five out of these 9 galaxies also exhibit high surface brightness disturbances. The remaining 4 PP galaxies in the first subset show disturbances only in their outer, low surface brightness stellar disk. In the high-\AmodH\ subset, 4 out of 14 PP galaxies show low surface brightness disturbances, while two of those galaxies have disturbances in both their low and high surface brightness disks. We tabulate the number of galaxies in different classes of optical disturbances in Table \ref{tab:OpticalDisturbanceClasses}. From this table, it is apparent that a larger fraction of galaxies in the low \HI\ asymmetry subset of the PP sub-sample displays optical disturbances compared to the low \HI\ asymmetry subset of the UMa sub-sample. The high-\AmodH\ subsets of both the UMa and the PP sub-samples display comparable fractions of optical disturbances.  

\par In comparing the UMa and PP sub-samples, we note that 15 out of 34 UMa galaxies (44\%) show optical disturbances while this occurs in 20 out of 30 PP galaxies (67\%). Apart from this difference, we also note that the nature of the stellar disturbances is different in the UMa and PP sub-samples. Contrary to the UMa sub-sample, we find many stellar streams in the PP sub-sample that indicate tidal interactions or ongoing minor mergers (PP 63, PP 58, PP 22, PP 45, PP 32, PP 12 and PP 33). Optical disturbances in the UMa sub-sample mainly seem to be due to disk instabilities (e.g. NGC 3949, NGC 3983, NGC 4051 and NGC 3985.) or mild tidal interactions (e.g. NGC 3769, NGC 3893 and UGC 6962). In conclusion, we do not find a direct relation between stellar and \HI\ asymmetries but we postulate that the higher fraction of \HI\ asymmetries in PP galaxies is due to more effective interactions and a higher merger rate in the PP environment. 

\par
\begin{figure}
    \centering
    \includegraphics[width=\columnwidth]{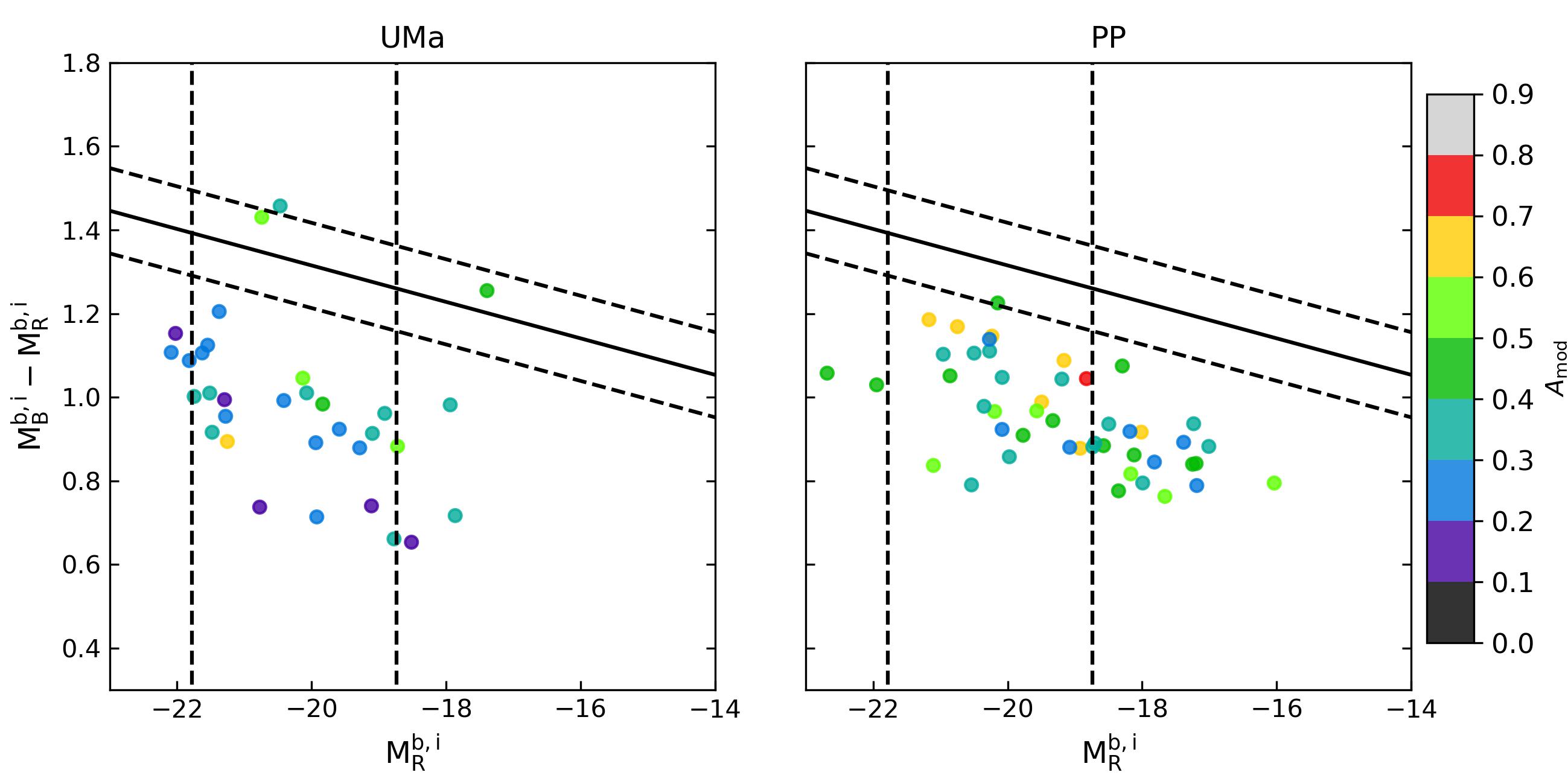}
    \caption{In this figure, the colour-magnitude diagrams of the UMa and PP galaxies are shown with respect to the red sequence of nearby clusters described by \cite{ChoqueC2021}. The colours indicate the \Amod\ values of the galaxies measured at a threshold of $15\times10^\mathrm{19}$\cm\ and having more than 3 beams across their major axis. }
    \label{fig:CMD_Amod}
\end{figure}

 We recreate Figure \ref{fig:CMD} with only the galaxies belonging to the UMa and PP sub-samples and present this in Figure \ref{fig:CMD_Amod}, where colours indicate the \AmodH\ values of galaxies. We find that galaxies in the UMa sub-sample either belong to the blue-cloud region or the red sequence region of the colour-magnitude diagram. We also find several bright, massive galaxies with low \AmodH\ values (blue and purple symbols) in the UMa sub-sample. On the contrary, none of the galaxies in the PP sub-sample are on the red sequence but are exclusively found in the blue cloud or green valley regions of the colour-magnitude diagram. We also find that PP galaxies in the green valley tend to have higher \AmodH\ values (yellow and red symbols). Based on these findings, we extend our hypothesis that the UMa galaxies appear to have transitioned beyond the stage where tidal interactions are effective and mergers are dominant in their local environment, suggesting that the most massive, bright UMa galaxies are no longer growing through merging or \HI\ gas accretion. On the other hand, the PP galaxies are actively undergoing tidal interactions and merging as can be inferred from their higher \AmodH\ values and can be seen in their high-contrast images. This hypothesis of Ursa Major and Perseus-Pisces being different dynamical environments is also supported by the findings of \cite{Hank2025} on the basis of the SIMBA simulations. In a forthcoming paper, we will delve deeper into the underlying reasons for the inherent differences in the two environments studied here and how they become evident through the distinct properties of their constituent galaxies.

\section{Summary and conclusion}\label{Conclusion}

\par In this paper, we presented the results from two volume-limited, high-resolution \HI\ imaging surveys of the Ursa Major (UMa) and Perseus-Pisces (PP) environments. Combining targeted observations of galaxies in the Ursa Major environment with the WSRT with \HI\ detected galaxies from the two VLA \HI\ imaging surveys, we construct a homogeneous sample of galaxies with $80<V_{\rm{max}}<200$\kms\ and \logMHI$>$8.42 for which their \HI\ column density maps and global profiles have been brought to the same spatial and velocity resolution. With the same observational constraints on signal-to-noise, column density levels and spatial resolutions, we measured spectral \Aflux\ asymmetries in the global profiles and morphological \Amod\ asymmetries above two column density threshold levels in the \HI\ maps of galaxies in both environments. We also considered these \HI\ asymmetries in a sample of mock galaxies from the EAGLE simulations. Finally, we related these asymmetries in the \HI\ distributions to asymmetries in the stellar disks at high and low surface brightness levels. Our main results are as follows:

\begin{itemize}
    \item We find a higher fraction of galaxies with asymmetric global profiles in the Perseus-Pisces environment than in the Ursa Major environment. Using the $3\sigma$ value of \Aflux\ of the AMIGA sample to identify asymmetric global profiles, we find that only 12 percent of the UMa galaxies have asymmetric global profiles compared to 33 percent of the PP galaxies. However, the S/N values of the global profiles in both samples is rather low. Considering only the high S/N profiles, we still find that 9 percent of UMa galaxies and 33 percent of the PP galaxies have asymmetric global profiles.
    \item Due to the low sensitivity of the PP VLA-C observations, we are compelled to choose a high column density threshold of $15\times10^{19}$\cm\ for the measurement of morphological asymmetry using the \Amod\ index. Similarly, very few UMa and PP galaxies are resolved beyond 11 beams and thus we chose to include galaxies with at least 3 beams across. With this threshold and relaxed constraint on the number of beams, we find that the \AmodH\ distributions of UMa and PP galaxies are different in the sense that the PP galaxies are more asymmetric. This is further confirmed by the 2-sample KS test yielding a p-value of 0.0023. 
    \item For the UMa galaxies it was possible to robustly measure morphological asymmetry at a lower column density threshold of $5\times10^{19}$\cm. We compare the \AmodL\ distribution of UMa galaxies to the noise-free \AmodL\ distribution of mock galaxies from the \textsc{EAGLE} simulations at the same threshold. We find that at the \AmodL\ distributions of UMa and mock galaxies are similar with a p-value of 0.47 for the 2-sample KS test.
    \item Comparing the global profile and morphological asymmetry values, we find that no correlation exists between the two, similar to the lack of correlation observed for the mock sample in \cite{Bilimogga2022} as well as in previous studies \citep{Reynolds2020}. This lack of correlation arises due to the dependence of the shape of the global profile on both the distribution and kinematics of the \HI\ disk. 
    \item From a comparison of the morphological asymmetries of the mock galaxies to those of galaxies in the UMa and PP environment we find that there is a category of mock galaxies which are not representative of real galaxies. These mock galaxies have higher asymmetries at a threshold of $15\times10^{19}$\cm\ arising from clumpy gas disk with pockets of low density gas. This may be a result of unrealistic feedback in the models as noted by \cite{Bahe2016}. 
    \item We visually identify optical asymmetries for galaxies in UMa and PP environments and find that there is no correlation between the presence of asymmetry in the optical and asymmetry in the \HI\ gas disk of any particular galaxy. However, it is to be noted that a higher fraction (48 percent) of galaxies in the PP environment have an asymmetric \HI\ or stellar distribution compared to galaxies in the UMa environment.
\end{itemize}

\par We conclude that the differences in \HI\ asymmetry fractions and distributions of the galaxy populations in the Ursa Major and Perseus-Pisces environments are significant and may result from processes that are more or less effective in these two environments. In a forthcoming paper we will report our ongoing investigation of the correlation between local environmental processes and the asymmetries of galaxies in these two environments.

\begin{acknowledgements}

\par PB and MV acknowledge support by the Netherlands Foundation for Scientific Research (NWO) through VICI grant 016.130.338. PB acknowledges the Leids-Kerhoven Bosscha Fund for financial travel support. JMvdH acknowledges support from the European Research Council under the European Union's Seventh Framework Programme (FP/2007-2013) / ERC Grant Agreement nr. 291531 (HIStoryNU).

\end{acknowledgements}

%
%

\begin{appendix}
\onecolumn

\section{Details of the WSRT observations}\label{sec:WSRTObs}

\begin{longtable}{c c c c c c c} 
\caption{Observational details of the WSRT sample}\\
\hline
\textbf{Name} & \textbf{Length of observation} & \textbf{Frequency} & \textbf{Bandwidth} & \textbf{Channel width} & \textbf{Beam size} & \textbf{RMS noise}\\
& (hrs) & (MHz) & (MHz) & (\kms) & (arcsec) & ($\mathrm{mJy}\,\mathrm{beam}^{-1}$) \\
\hline
\endfirsthead

\caption{continued}\\
\hline
\textbf{Name} & \textbf{Length of observation} & \textbf{Frequency} & \textbf{Bandwidth} & \textbf{Channel width} & \textbf{Beam size} & \textbf{RMS noise}\\
\hline
\endhead
\hline
\endfoot

\label{tab:obsWSRT_details}
    NGC 3718 & $1\times12$ & 1415.3 & 5 & 16.61 & $12.9\times17.2$ & 0.20\\
    NGC 3729 & $1\times12$ & 1415.3 & 5 & 16.61 & $12.9\times17.2$ & 0.20\\
    NGC 3726 & $1\times12$ & 1416.4 & 2.5 & 4.13 & $12.2\times16.5$ & 0.60\\
    NGC 3769 & $1\times12$ & 1416.9 & 2.5 & 4.14 & $11.9\times15.9$ & 0.40\\
    NGC 3877 & $2\times12$ & 1416.1 & 5 & 16.59 & $12.2\times17.6$ & 0.14\\
    NGC 3893 & $1\times12$ & 1415.9 & 2.5 & 4.15 & $12.2\times16.1$ & 0.58\\
    NGC 3896 & $1\times12$ & 1415.9 & 2.5 & 4.15 & $12.2\times16.1$ & 0.58\\
    NGC 3917 & $1\times12$ & 1415.8 & 2.5 & 4.15 & $11.7\times15$ & 0.36\\
    UGC 6840 & $1\times12$ & 1415.8 & 2.5 & 4.15 & $11.7\times15$ & 0.36\\
    NGC 3949 & $1\times12$ & 1416.6 & 2.5 & 4.14 & $11.9\times16$ & 0.36\\
    NGC 3953 & $1\times12$ & 1415.8 & 5 & 16.55 & $13.8\times16.4$ & 0.20\\
    NGC 3972 & $1\times12$ & 1416.4 & 2.5 & 4.15 & $11.7\times15.2$ & 0.38\\
    NGC 3992 & $1\times12$ & 1415.5 & 5 & 8.31 & $12.1\times15.1$ & 0.58\\
    UGC 6923 & $1\times12$ & 1415.5 & 5 & 8.31 & $12.1\times15.1$ & 0.58\\
    UGC 6940 & $1\times12$ & 1415.5 & 5 & 8.31 & $12.1\times15.1$ & 0.58\\
    UGC 6969 & $1\times12$ & 1415.5 & 5 & 8.31 & $12.1\times15.1$ & 0.58\\
    NGC 4010 & $1\times12$ & 1416.1 & 2.5 & 4.15 & $12.5\times17$ & 0.38\\
    NGC 4051 & $1\times12$ & 1417.1 & 2.5 & 4.15 & $10.5\times15.1$ & 0.68\\
    NGC 4088 & $1\times12$ & 1416.7 & 5 & 16.5 & $12\times15.7$ & 0.29\\
    NGC 4085 & $1\times12$ & 1416.7 & 5 & 16.5 & $12\times15.7$ & 0.29\\
    NGC 4100 & $1\times12$ & 1415.2 & 5 & 16.61 & $12.1\times15.9$ & 0.30\\
    NGC 4102 & $4\times12$ & 1416.4 & 5 & 4.15 & $12.1\times15.1$ & 0.19\\
    NGC 4138 & $1\times12$ & 1416.3 & 5 & 16.59 & $12.1\times17.3$ & 0.52\\
    NGC 4157 & $1\times12$ & 1416.8 & 5 & 16.57 & $11.8\times16.7$ & 0.32\\
    NGC 4183 & $1\times12$ & 1415.9 & 2.5 & 4.15 & $11.8\times17.2$ & 0.39\\
    NGC 4217 & $1\times12$ & 1415.5 & 5 & 16.6 & $13.2\times18.6$ & 0.19\\
    NGC 4218 & $1\times12$ & 1416.8 & 2.5 & 4.15 & $11.9\times16.4$ & 0.36\\
    NGC 4220 & $1\times12$ & 1416.8 & 5 & 16.57 & $12.3\times16.5$ & 0.19\\
    NGC 4389 & $2\times12$ & 1416.9 & 2.5 & 4.14 & $12\times16.9$ & 0.26\\
    UGC 6962 & $5\times12$ & 1416.8 & 2.5 & 4.15 & $12\times18.8$ & 0.19\\
    UGC 6973 & $5\times12$ & 1416.8 & 2.5 & 4.15 & $12\times18.8$ & 0.19\\
    UGC 6917 & $1\times12$ & 1416.0 & 2.5 & 4.15 & $11.6\times17.2$ & 0.33\\
    UGC 6930 & $1\times12$ & 1416.7 & 2.5 & 4.14 & $12.3\times16.8$ & 0.38\\
    UGC 6399 & $1\times12$ & 1416.6 & 2.5 & 4.15 & $12.9\times16.6$ & 0.38\\
    UGC 6446 & $1\times12$ & 1417.4 & 2.5 & 4.14 & $12.2\times15$ & 0.60\\
    UGC 6667 & $1\times12$ & 1415.8 & 2.5 & 4.15 & $12\times15.4$ & 0.63\\
    UGC 6773 & $1\times12$ & 1416.0 & 2.5 & 4.15 & $12.6\times16.6$ & 0.36\\
    UGC 6818 & $1\times12$ & 1416.5 & 2.5 & 4.15 & $12.6\times17.6$ & 0.36\\
    UGC 6894 & $1\times12$ & 1416.7 & 2.5 & 4.15 & $12.5\times14.3$ & 0.37\\
    UGC 6983 & $2\times12$ & 1415.3 & 2.5 & 4.16 & $12.2\times15.3$ & 0.44\\
	\hline
\end{longtable}

\newpage


# Appendix B: H I atlas pages of PP galaxies

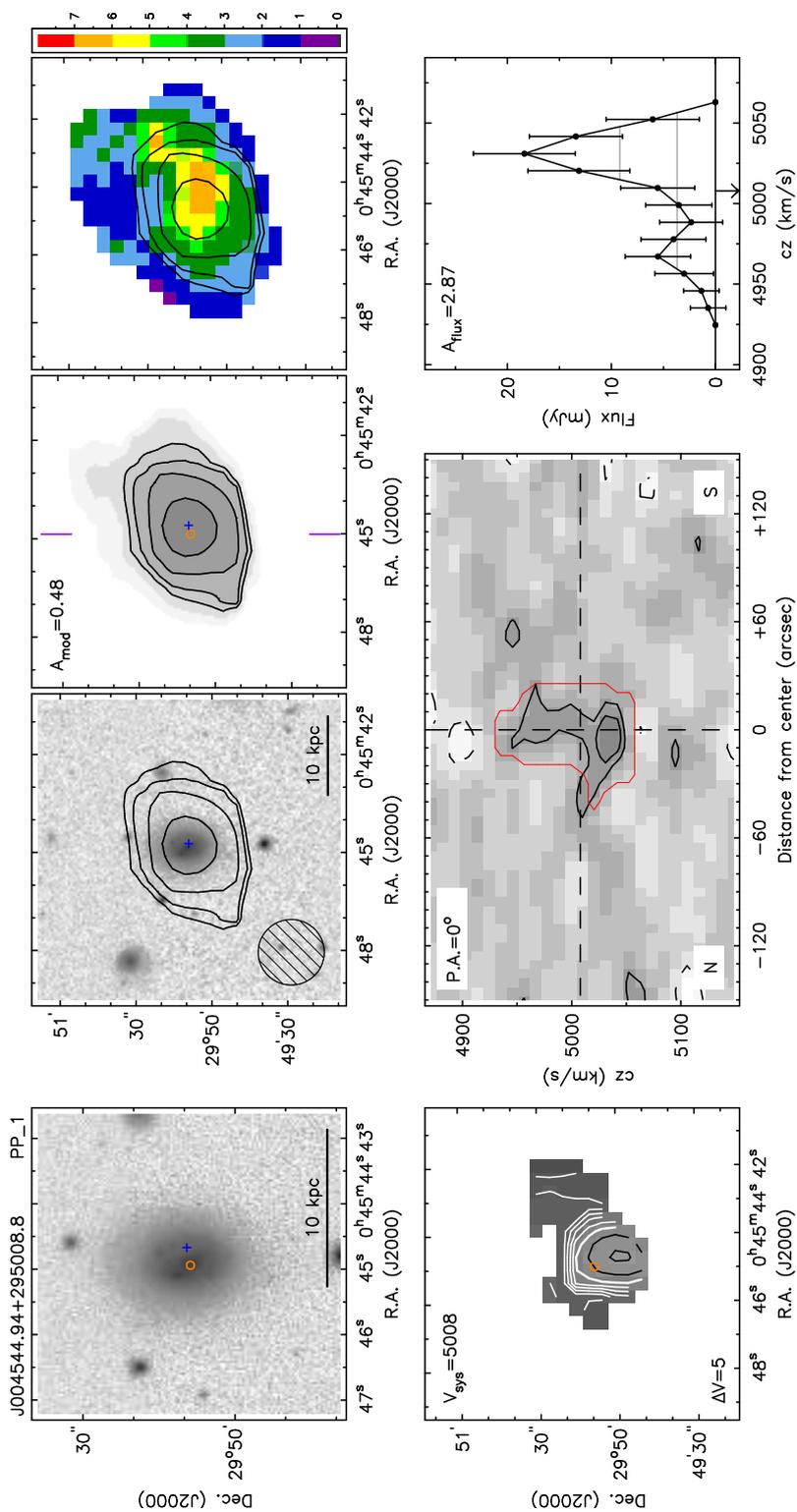

The Atlas Pages of the PP galaxies (see also Section 3 for detailed descriptions of the various panels) show the following: Top row from left to right: an r-band image taken from DECaLS, H I column density contours (1.5, 2, 3, 5, 8, 12, 17, 23, 30, 38 × $10^{20}$ cm$^{-2}$) superposed on the r-band image, the H I image and a signal-to-noise image (S/N scale is given at the right hand side). The orange circle indicates the optical center, and the blue plus sign the centroid of the H I distribution. Bottom row from left to right: the velocity field (contour interval is 1 mJy/beam) and a global profile with an indication of the systemic velocity (down arrow) indicated in the second panel from the top right (contour interval is 1 mJy/beam) and a global profile with an indication of the systemic velocity (down arrow) and the FWHM at 50% (W50) and 20% (W20) of the peak (grey horizontal lines). Values for the systemic velocity, $A_{mod}$ and $A_{flux}$ are indicated in the velocity field, the H I image and the global profile plot respectively. The red contour in the lower middle panel indicates the extent of the frequency-dependent mask within which the flux was collected to construct the column density map and the global profile.

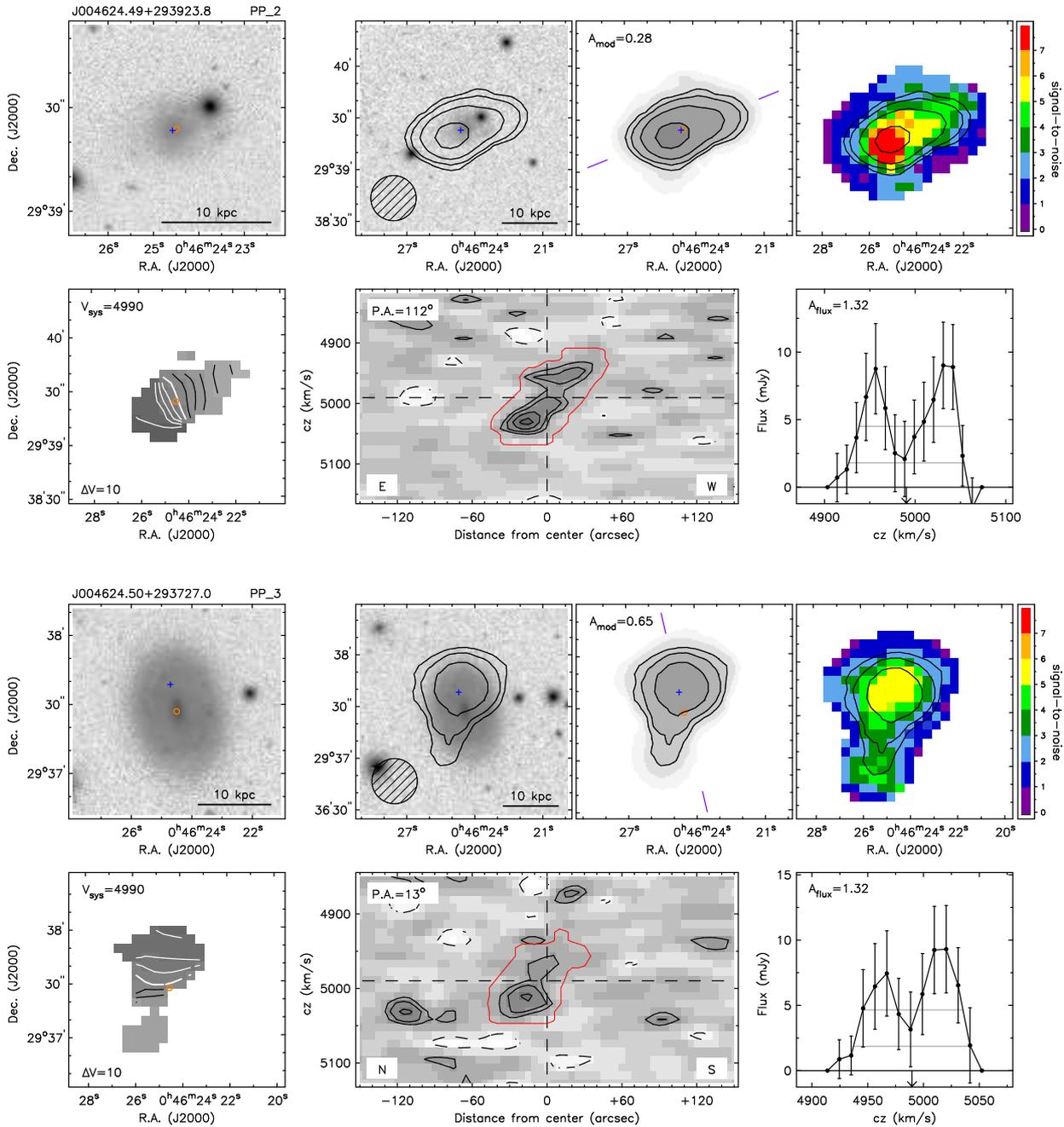

Atlas page 2 and 3

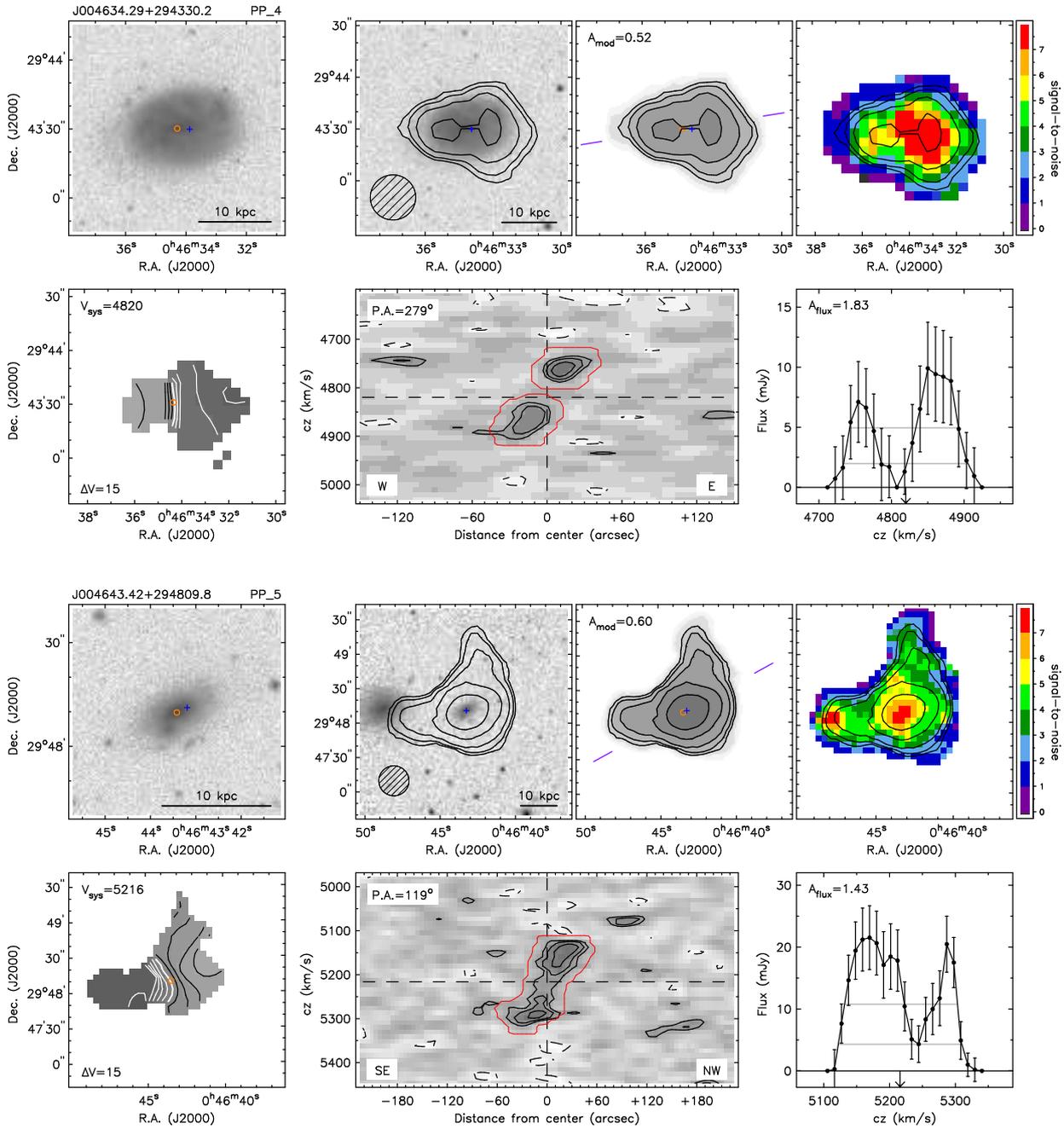

Atlas page 4 and 5

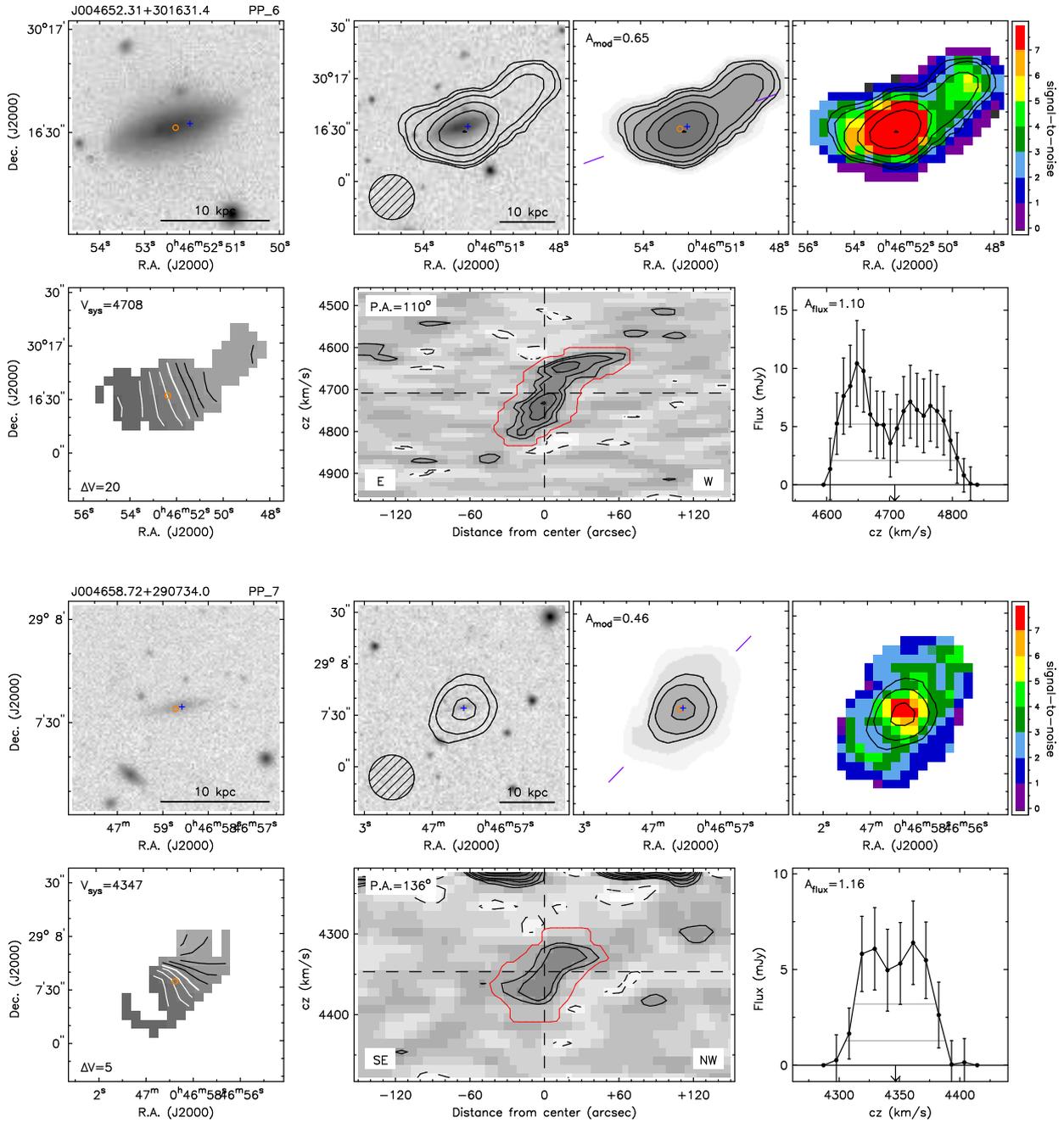

Atlas page 6 and 7

4

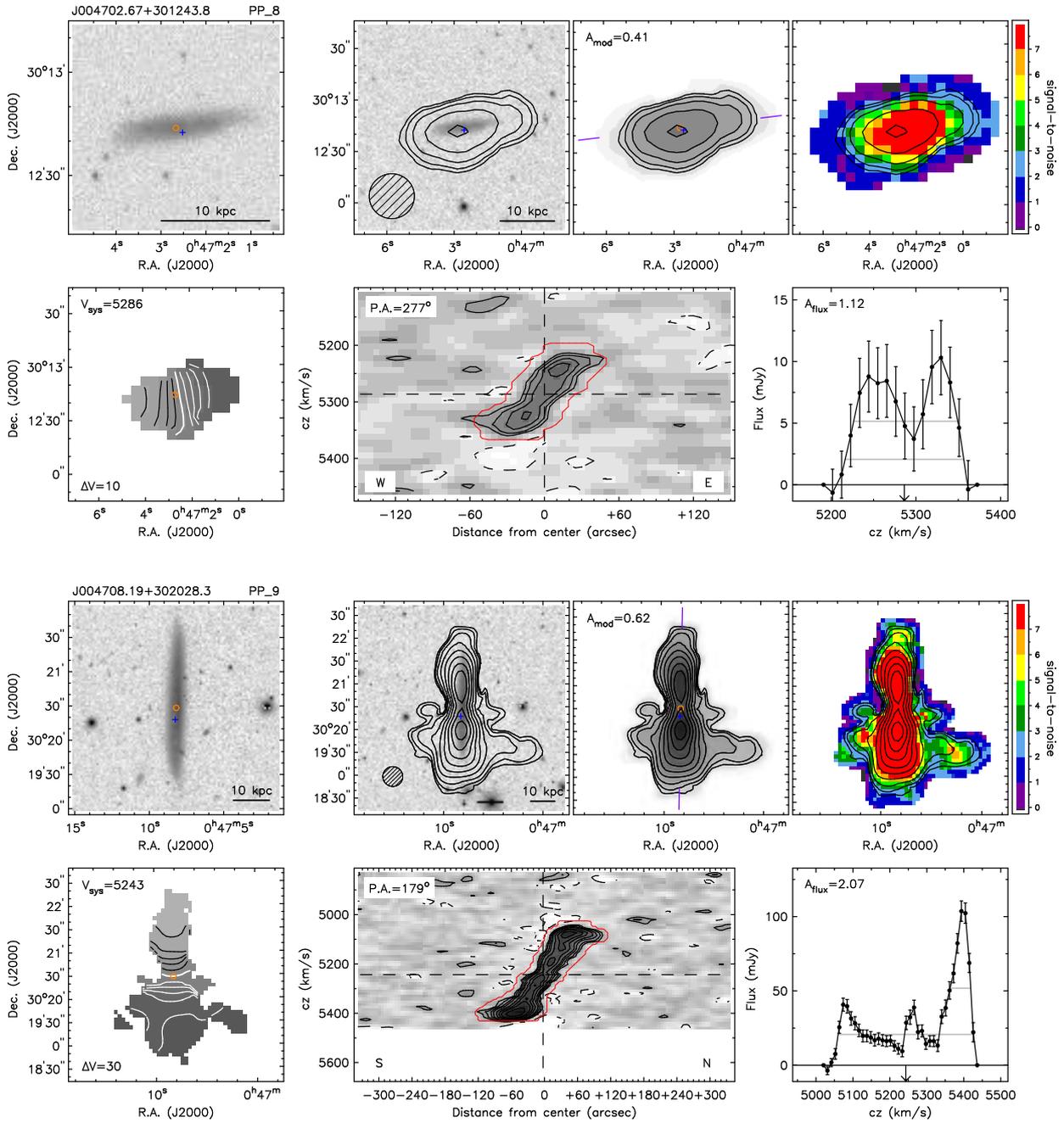

Atlas page 8 and 9

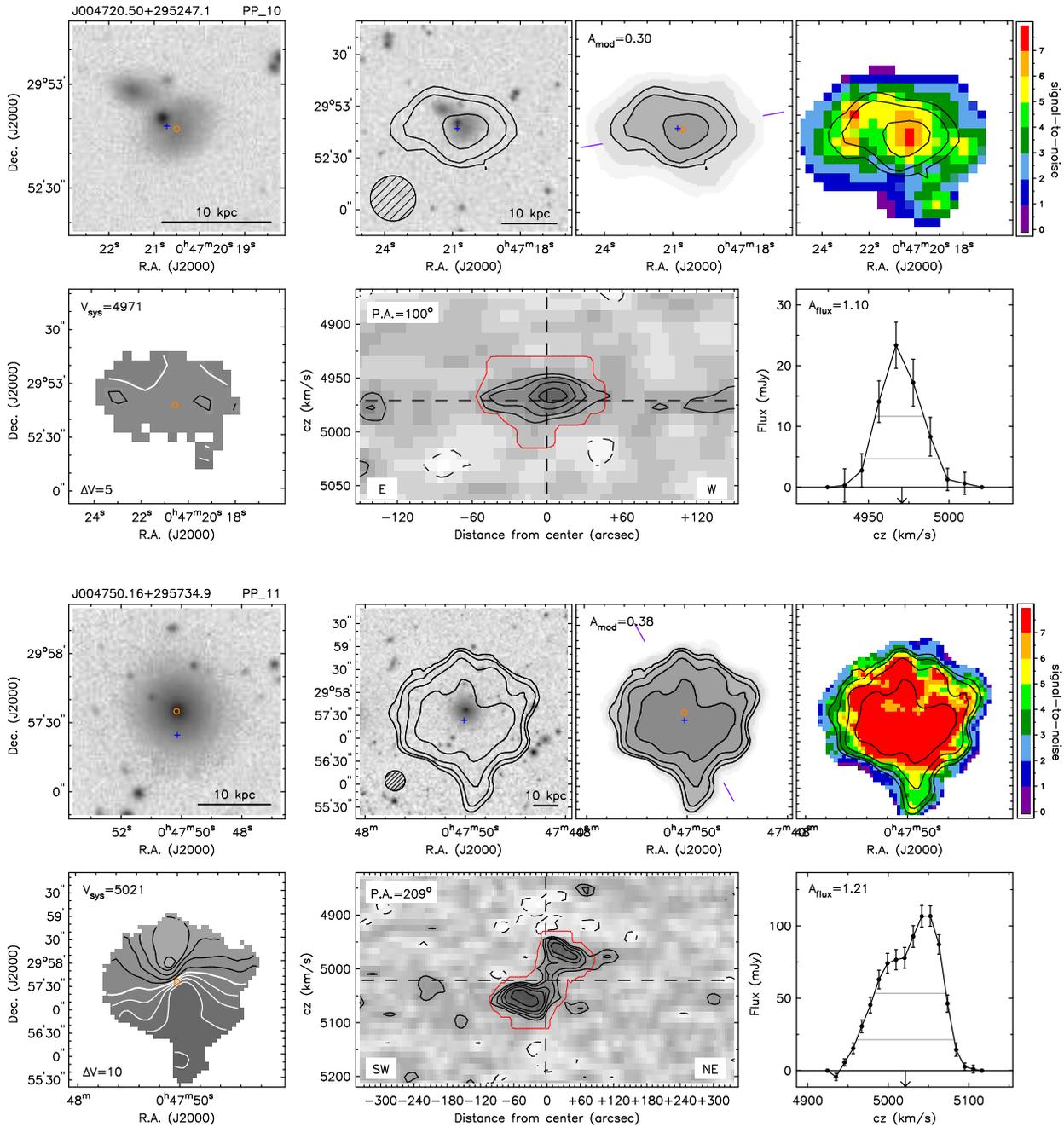

Atlas page 10 and 11

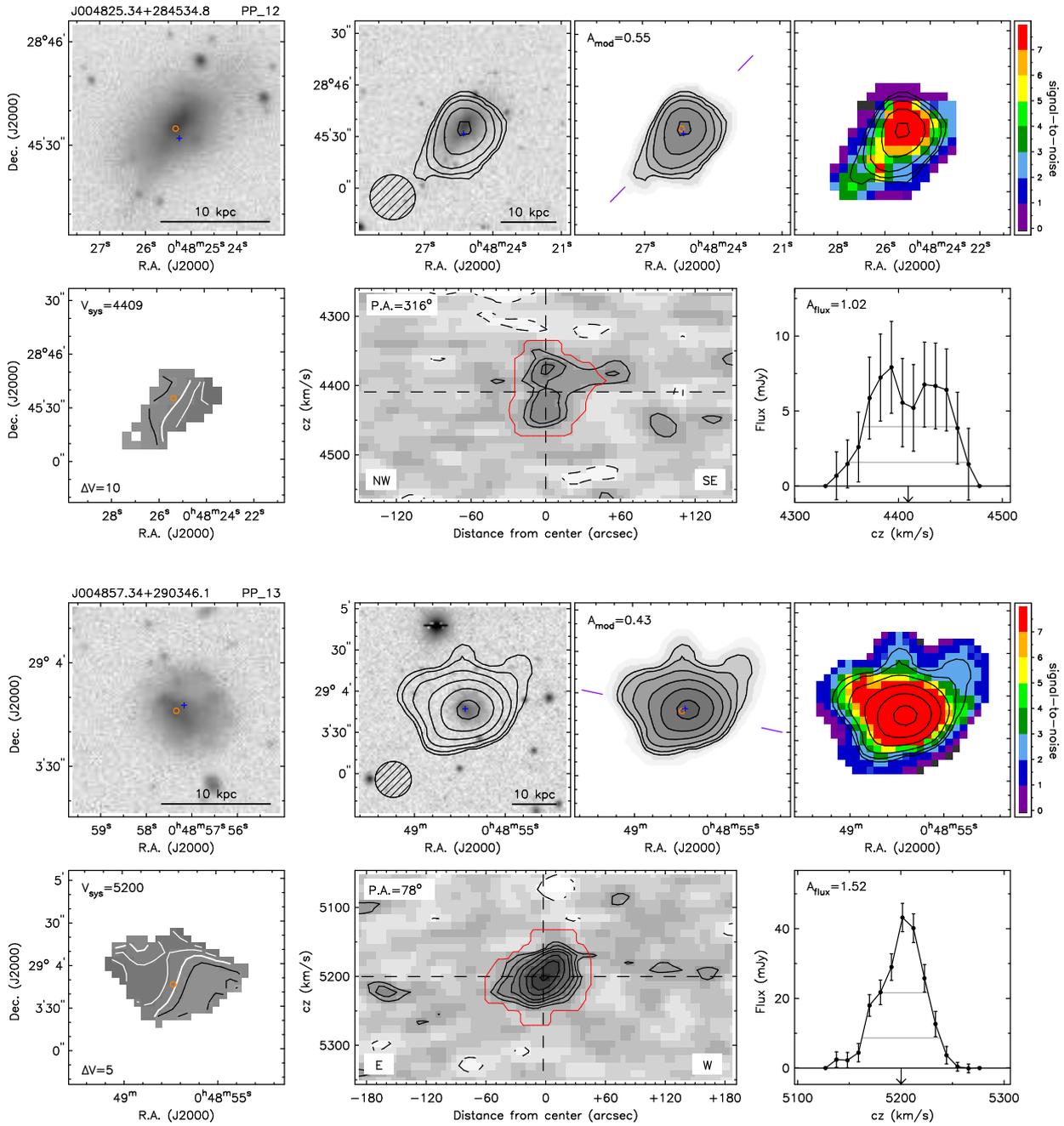

Atlas page 12 and 13

7

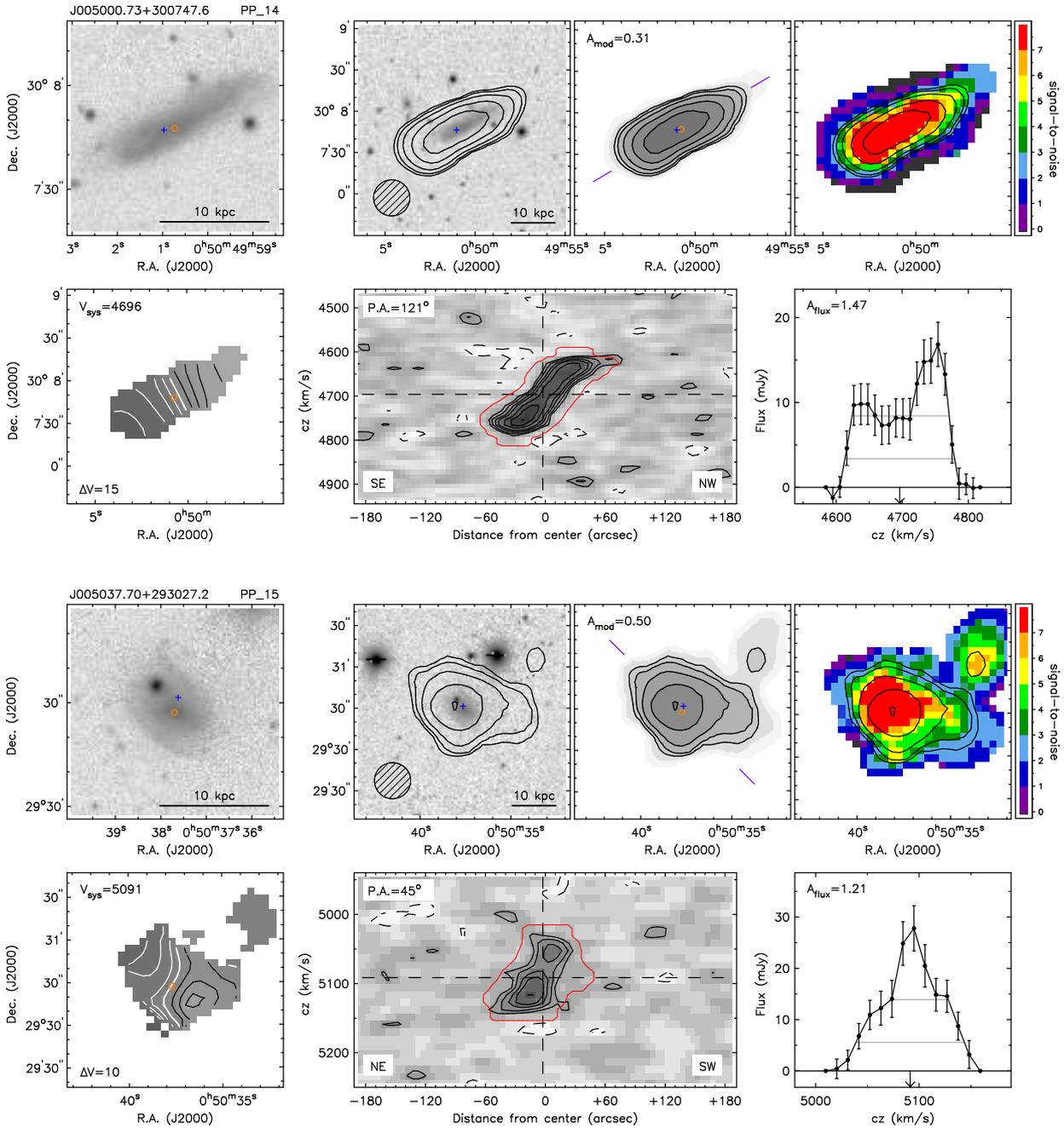

Atlas page 14 and 15

8

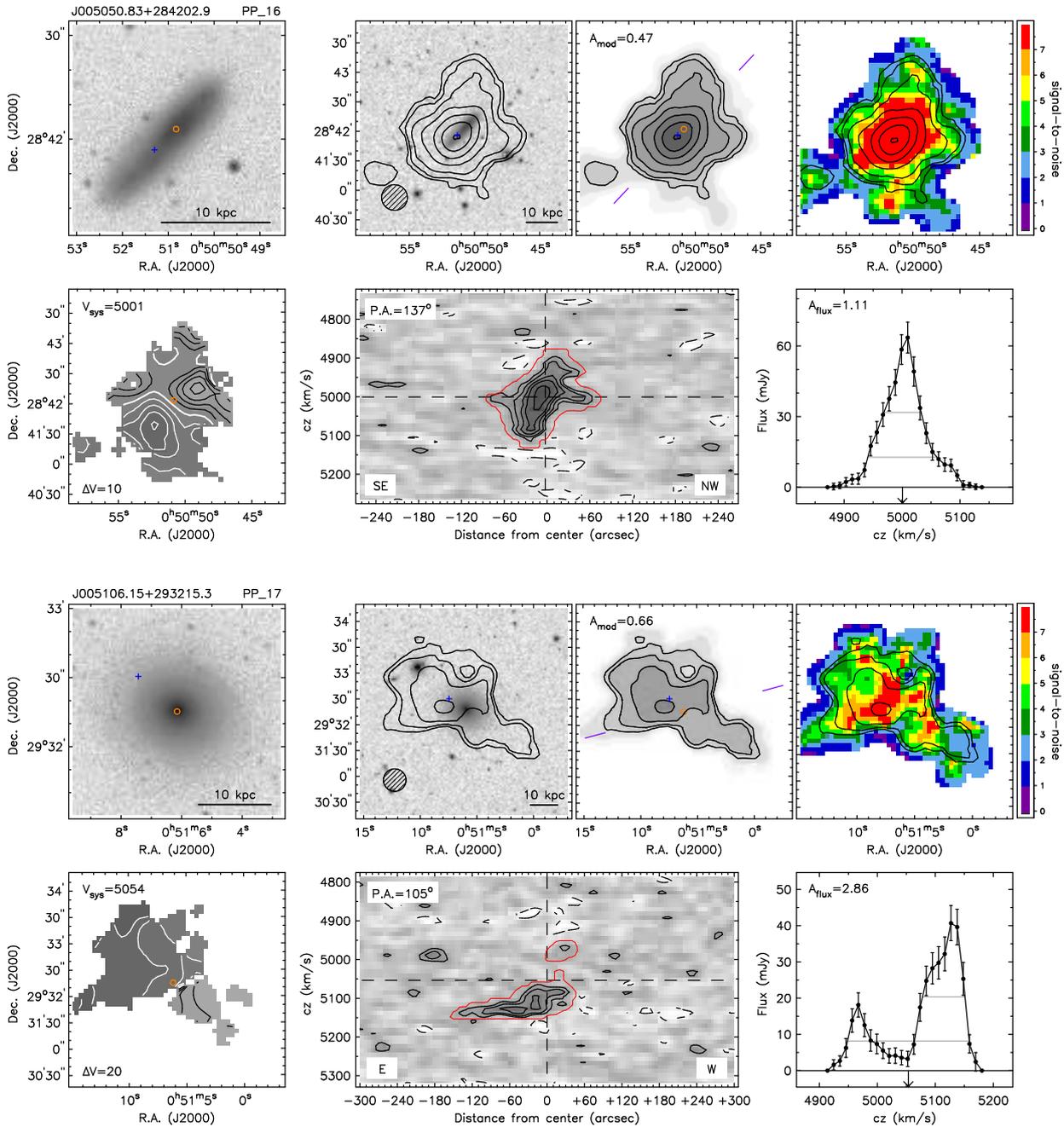

Atlas page 16 and 17

9

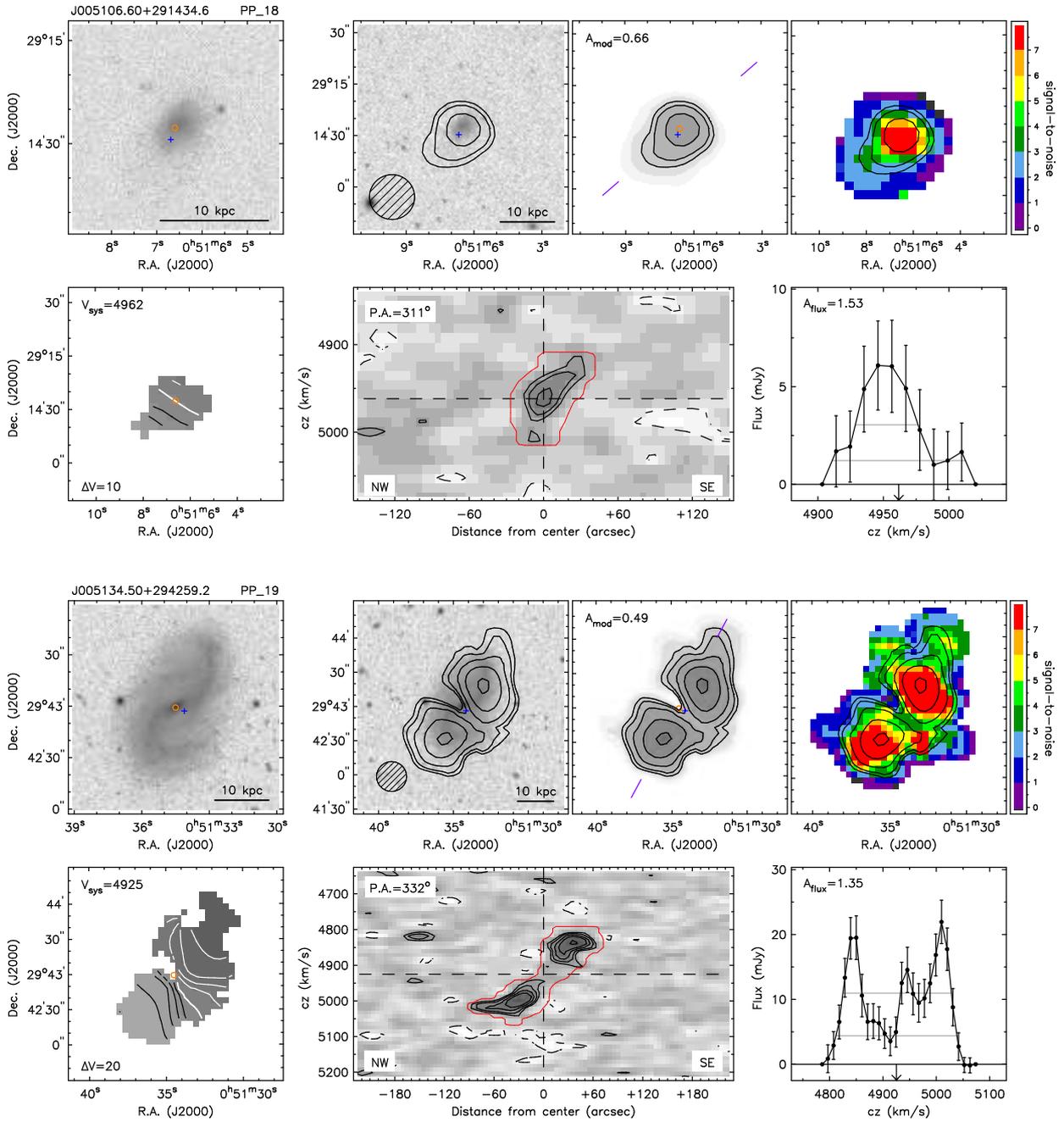

Atlas page 18 and 19

10

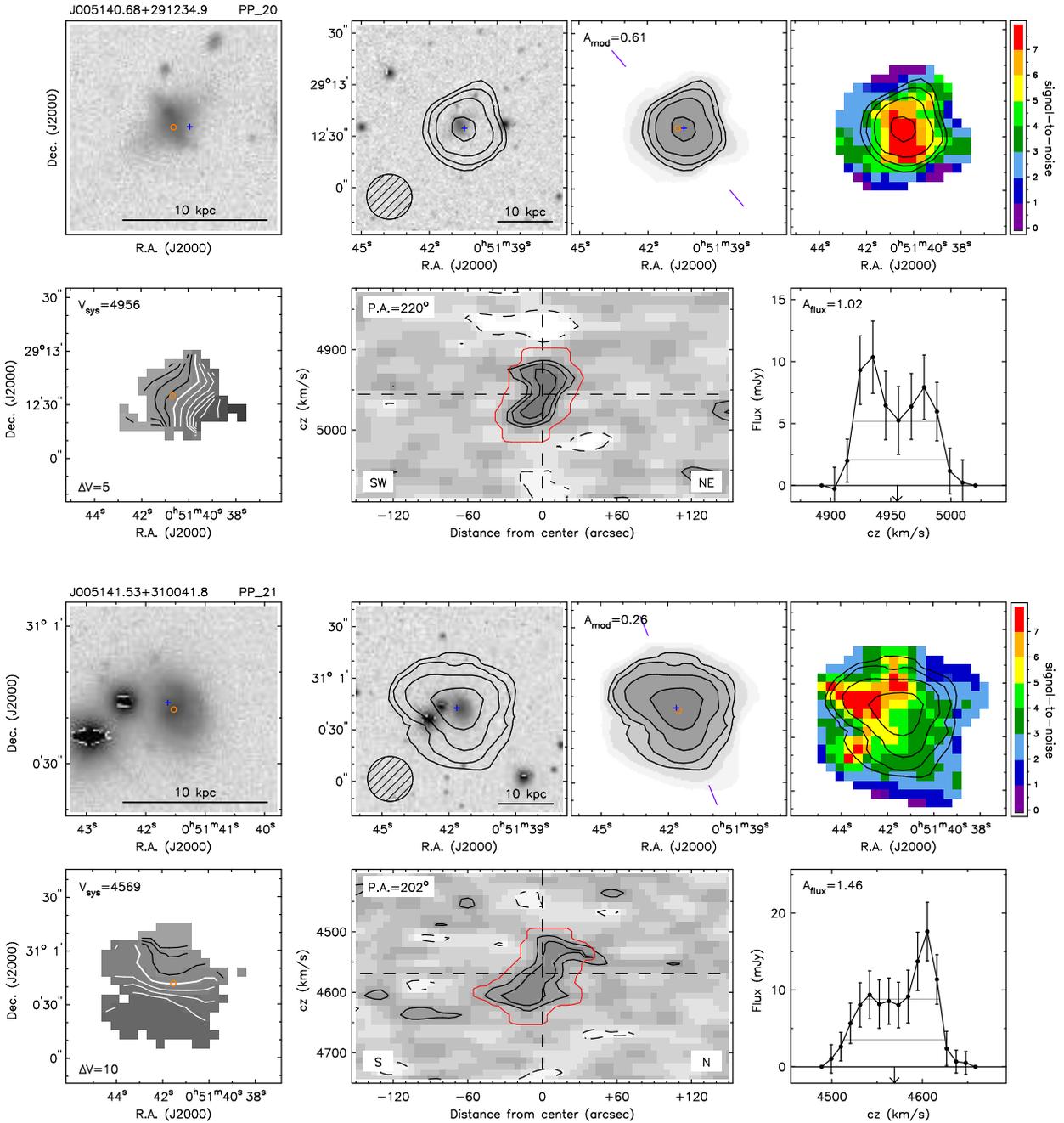

Atlas page 20 and 21

11

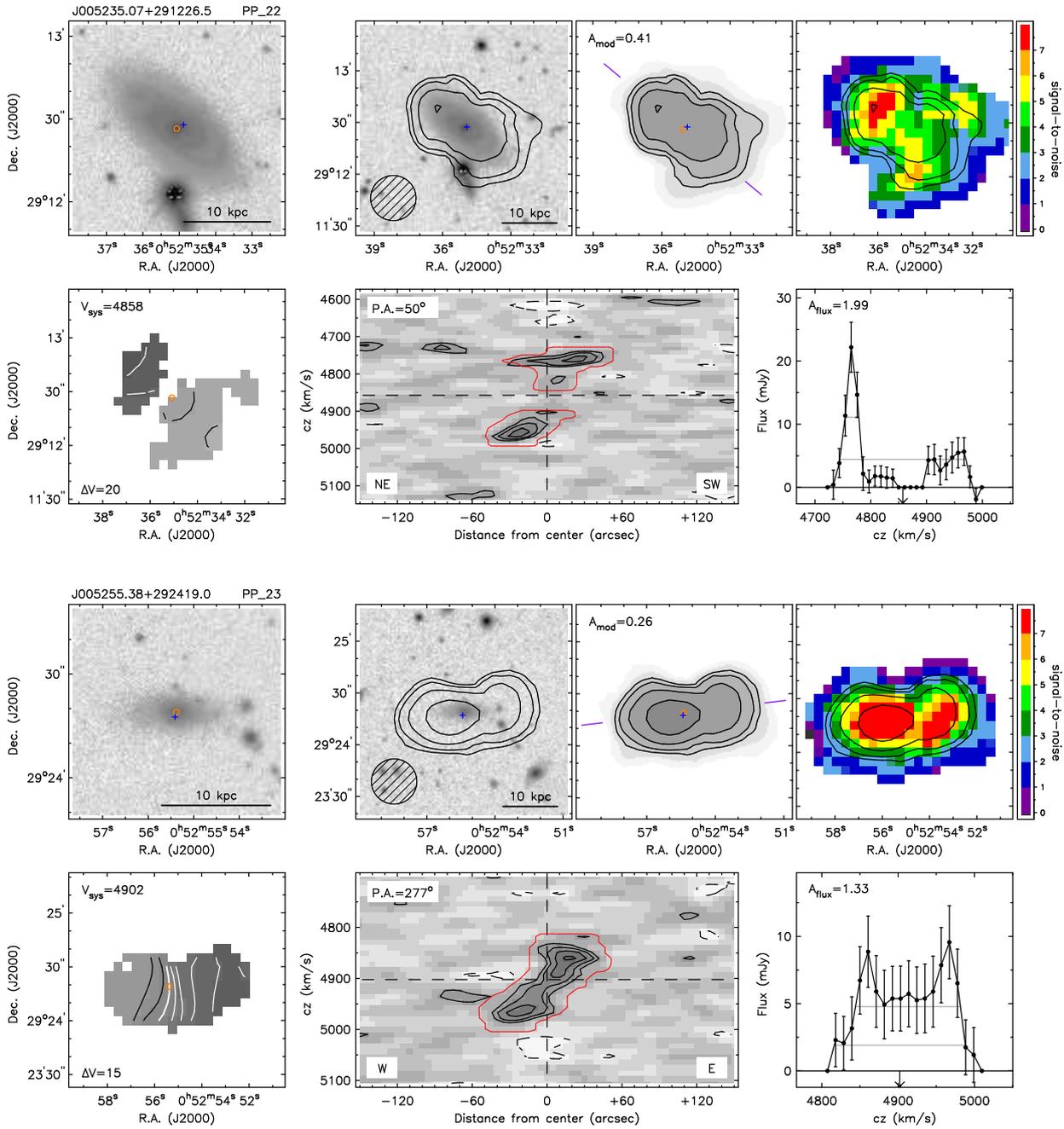

Atlas page 22 and 23

12

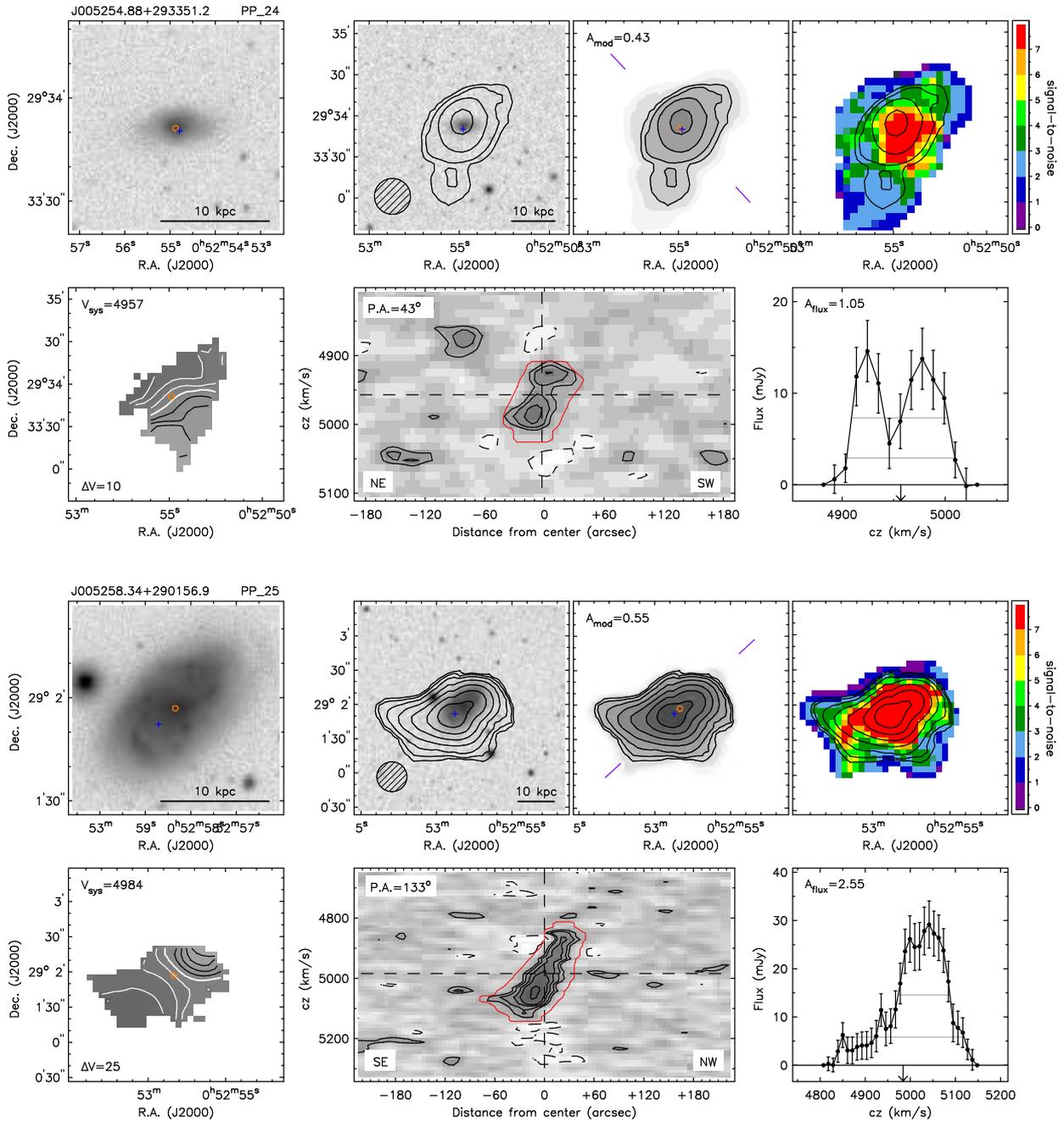

Atlas page 24 and 25

13

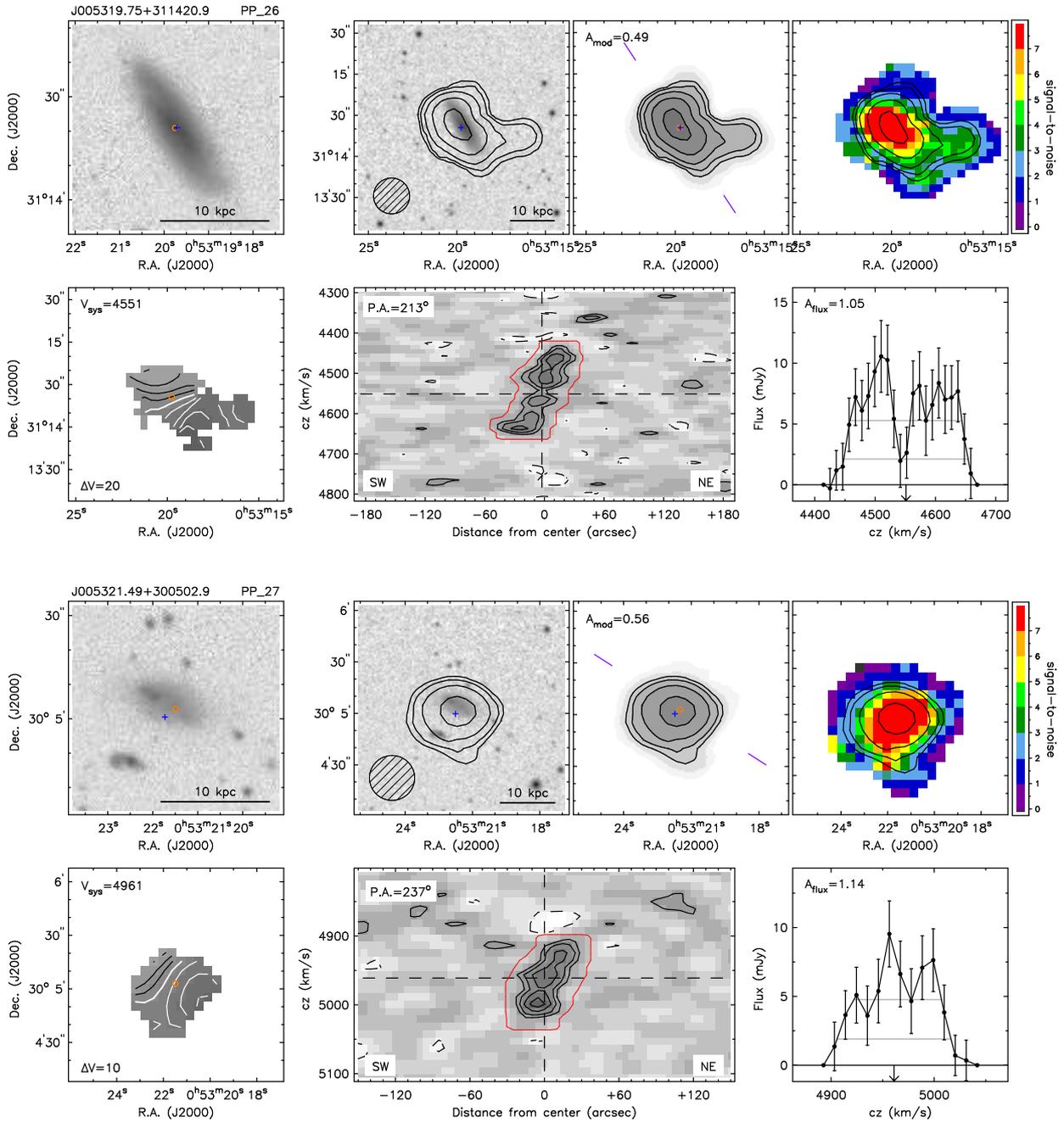

Atlas page 26 and 27

14

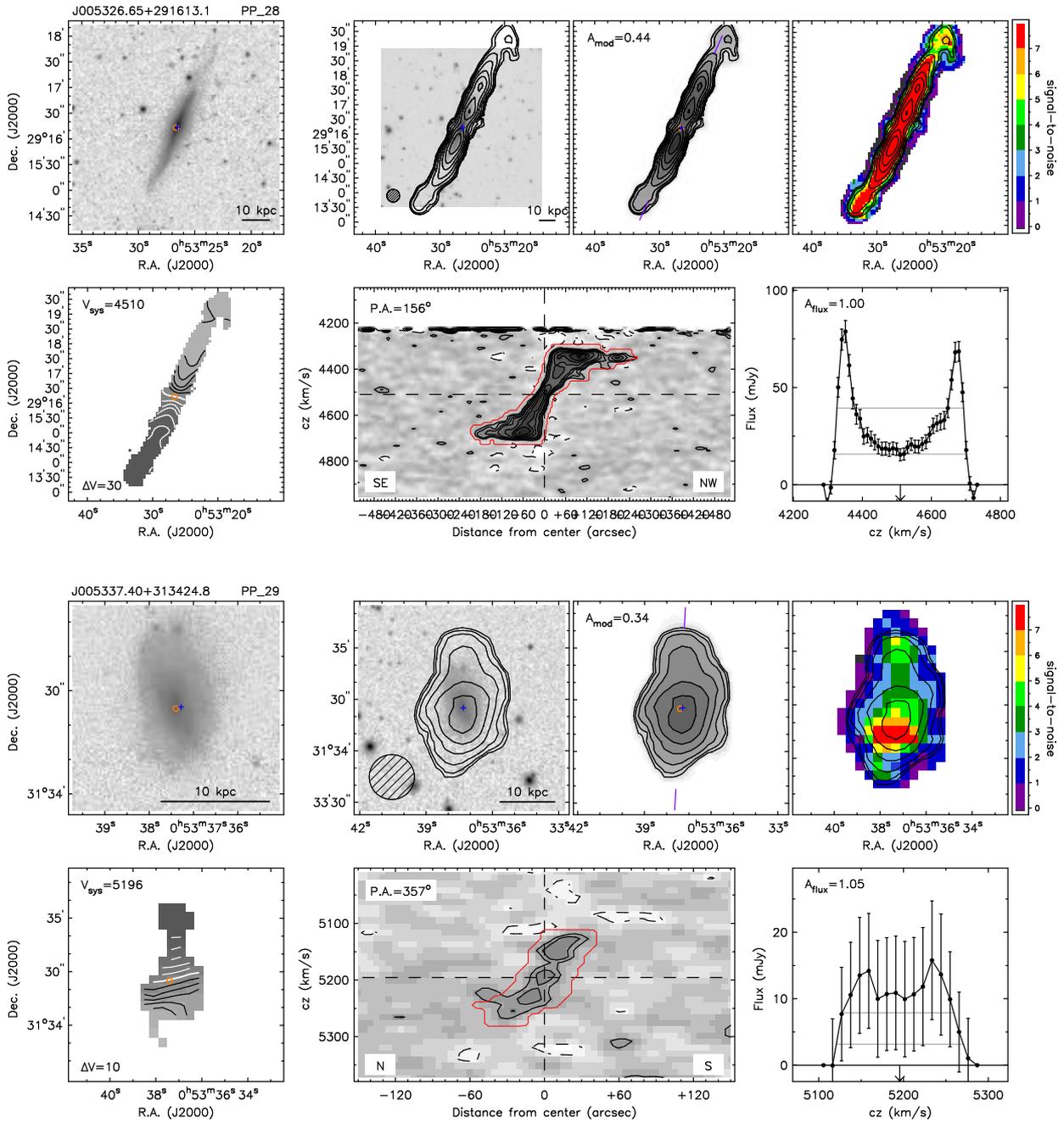

Atlas page 28 and 29

15

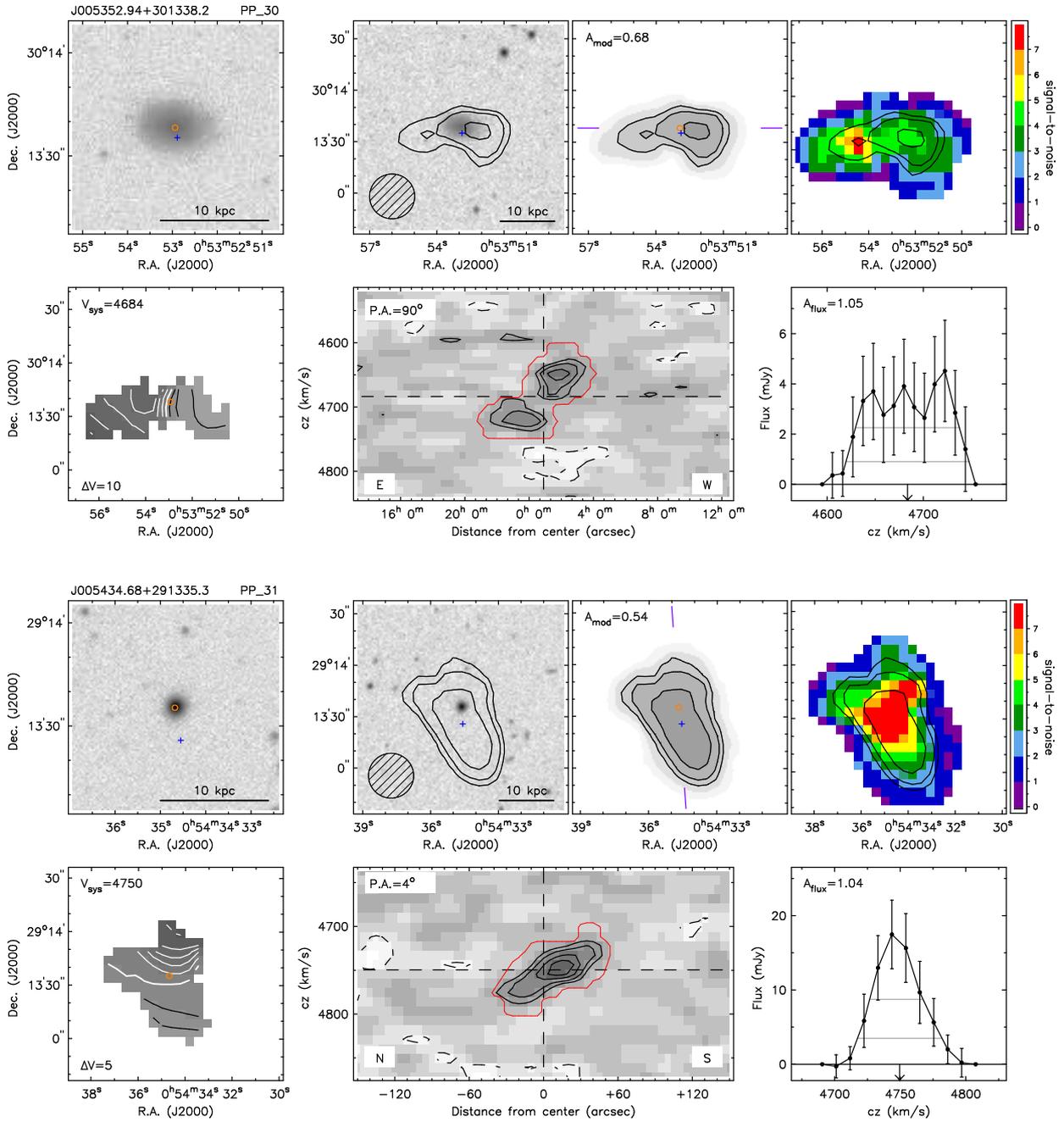

Atlas page 30 and 31

16

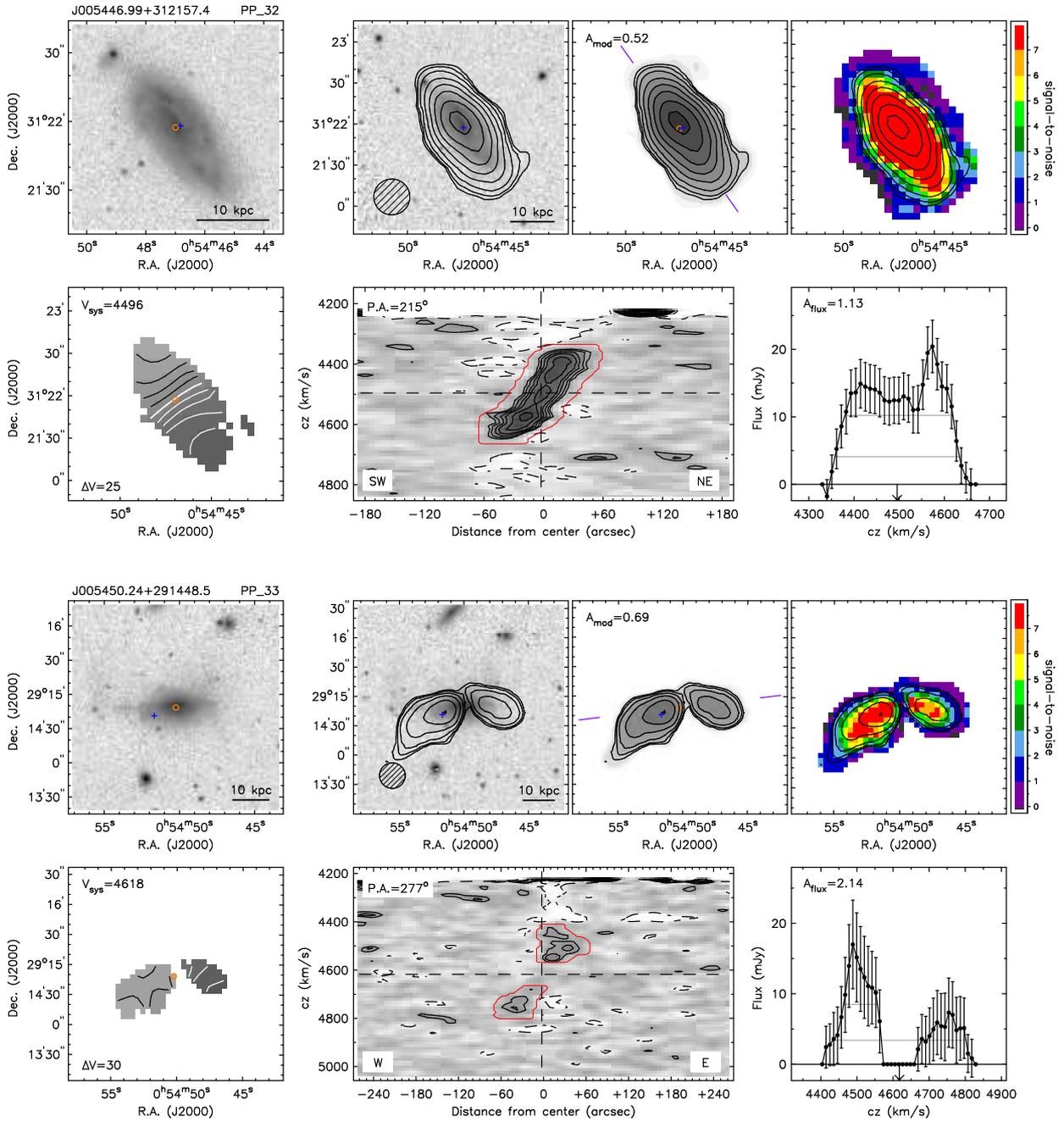

Atlas page 32 and 33

17

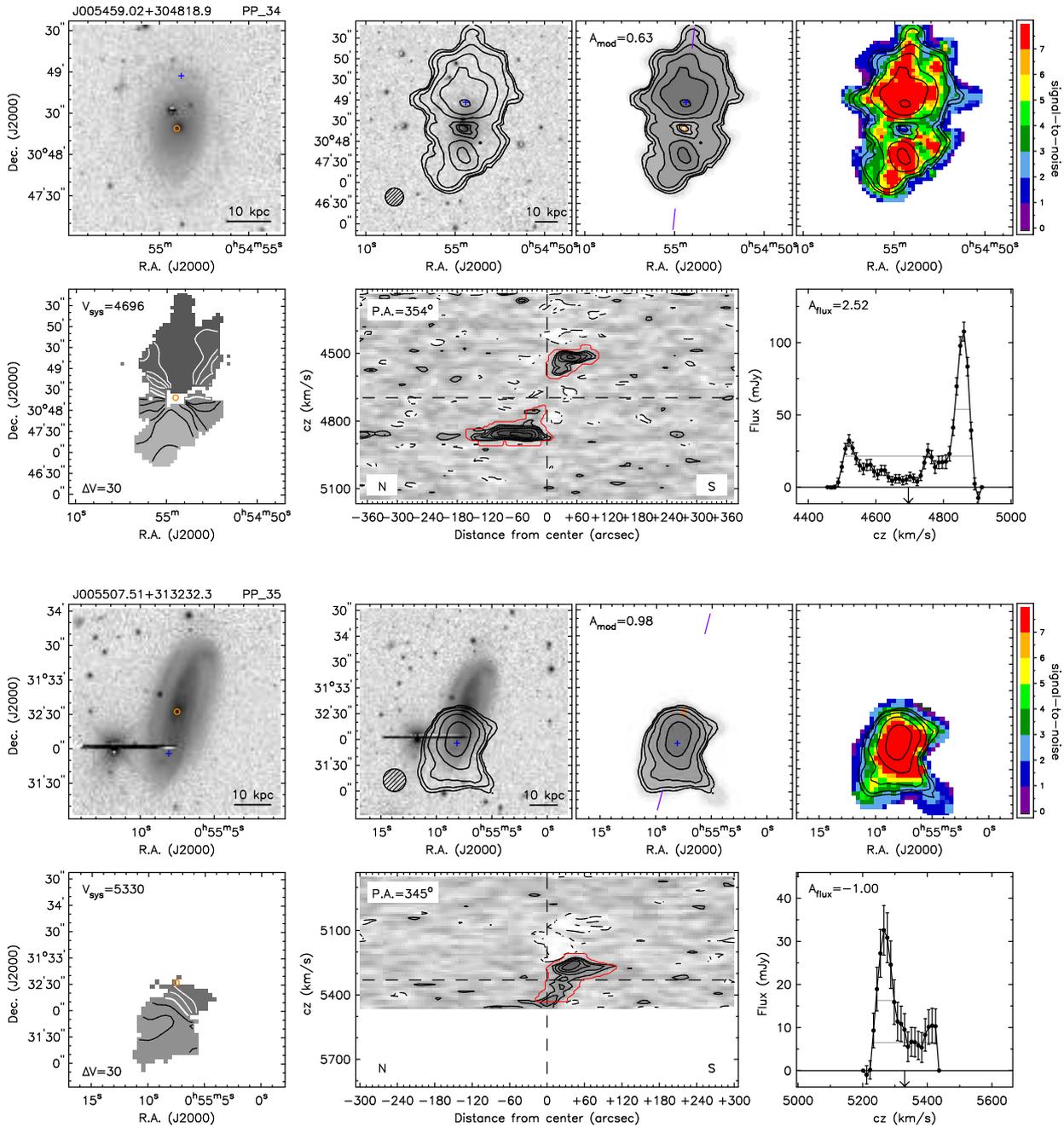

Atlas page 34 and 35

18

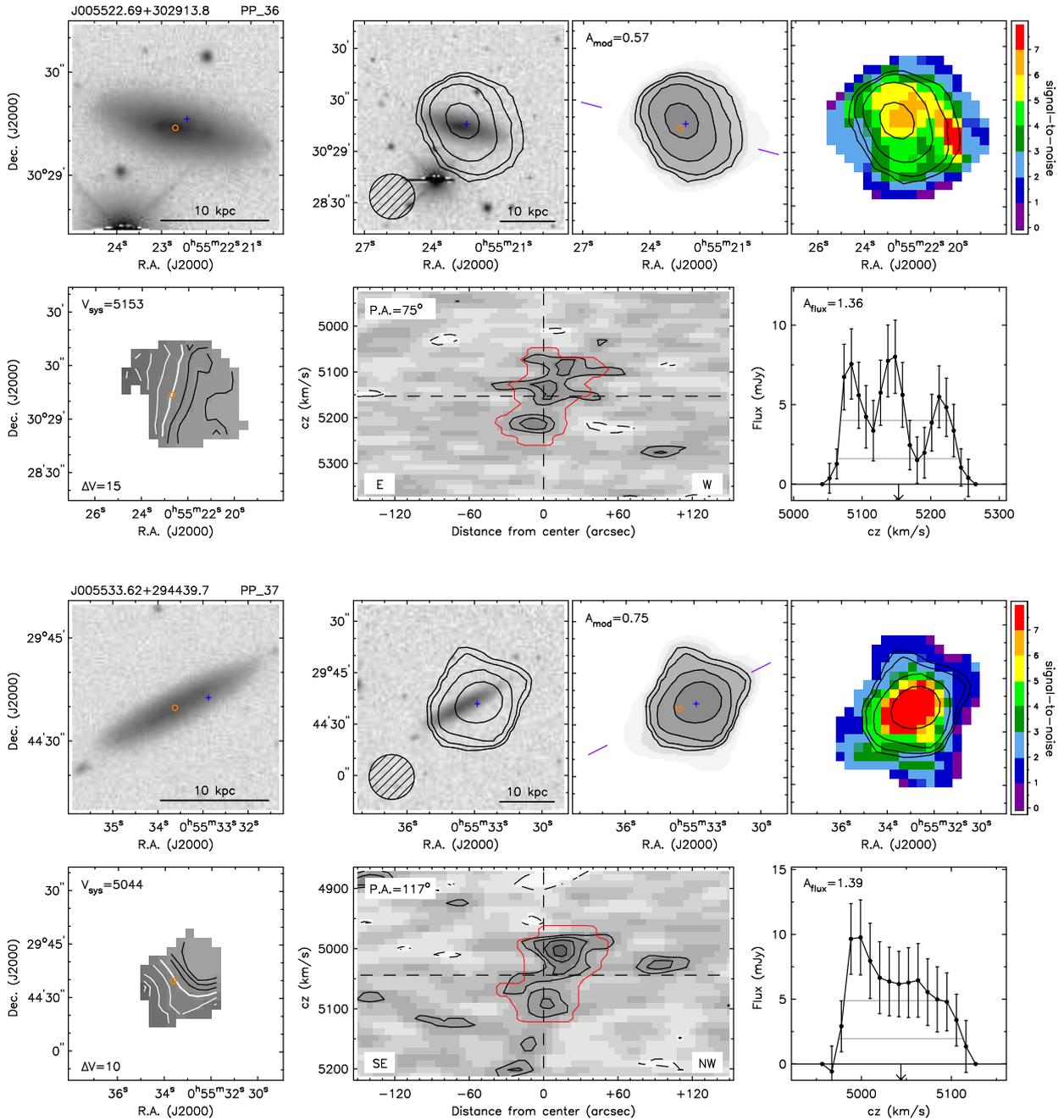

Atlas page 36 and 37

19

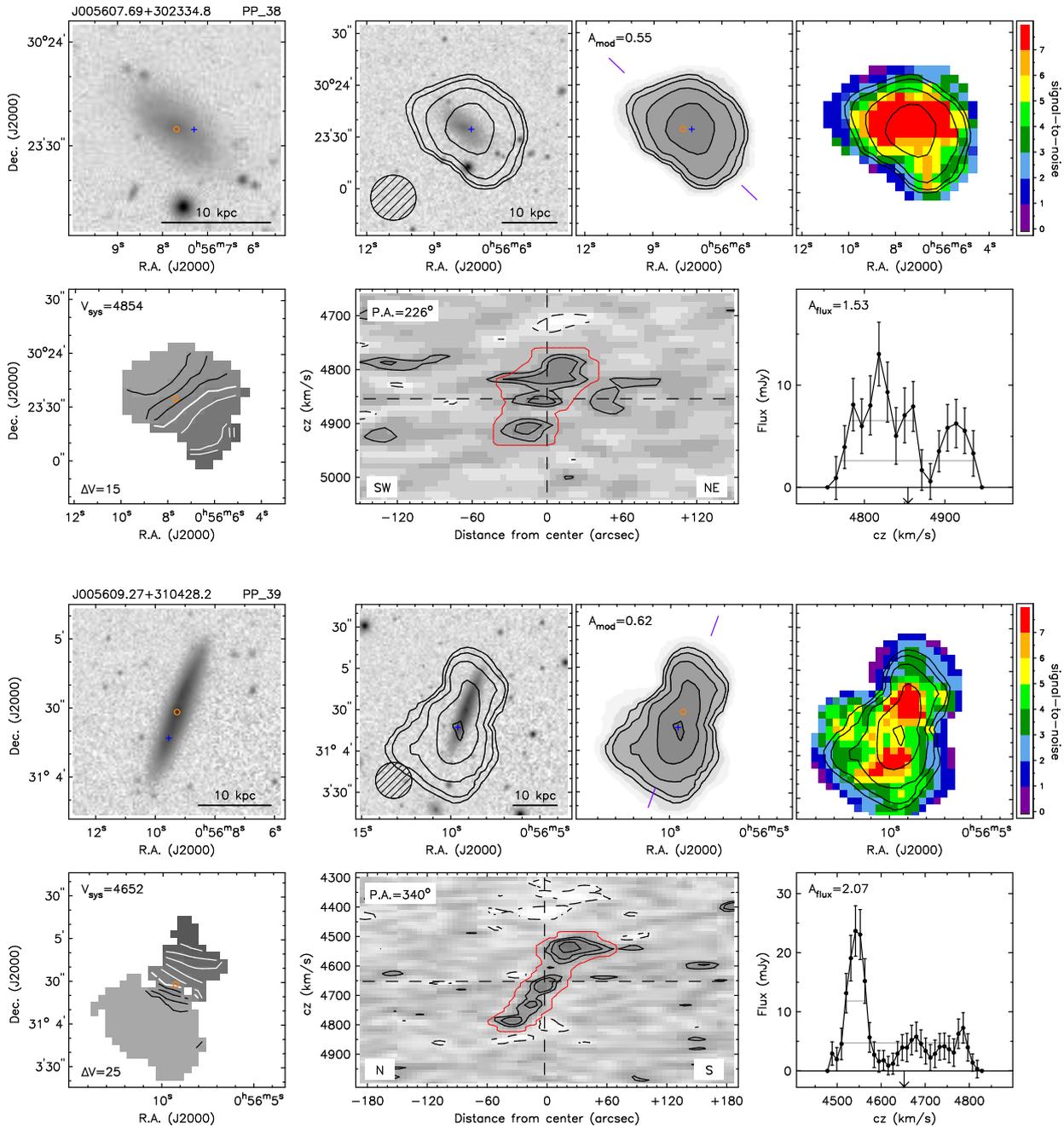

Atlas page 38 and 39

20

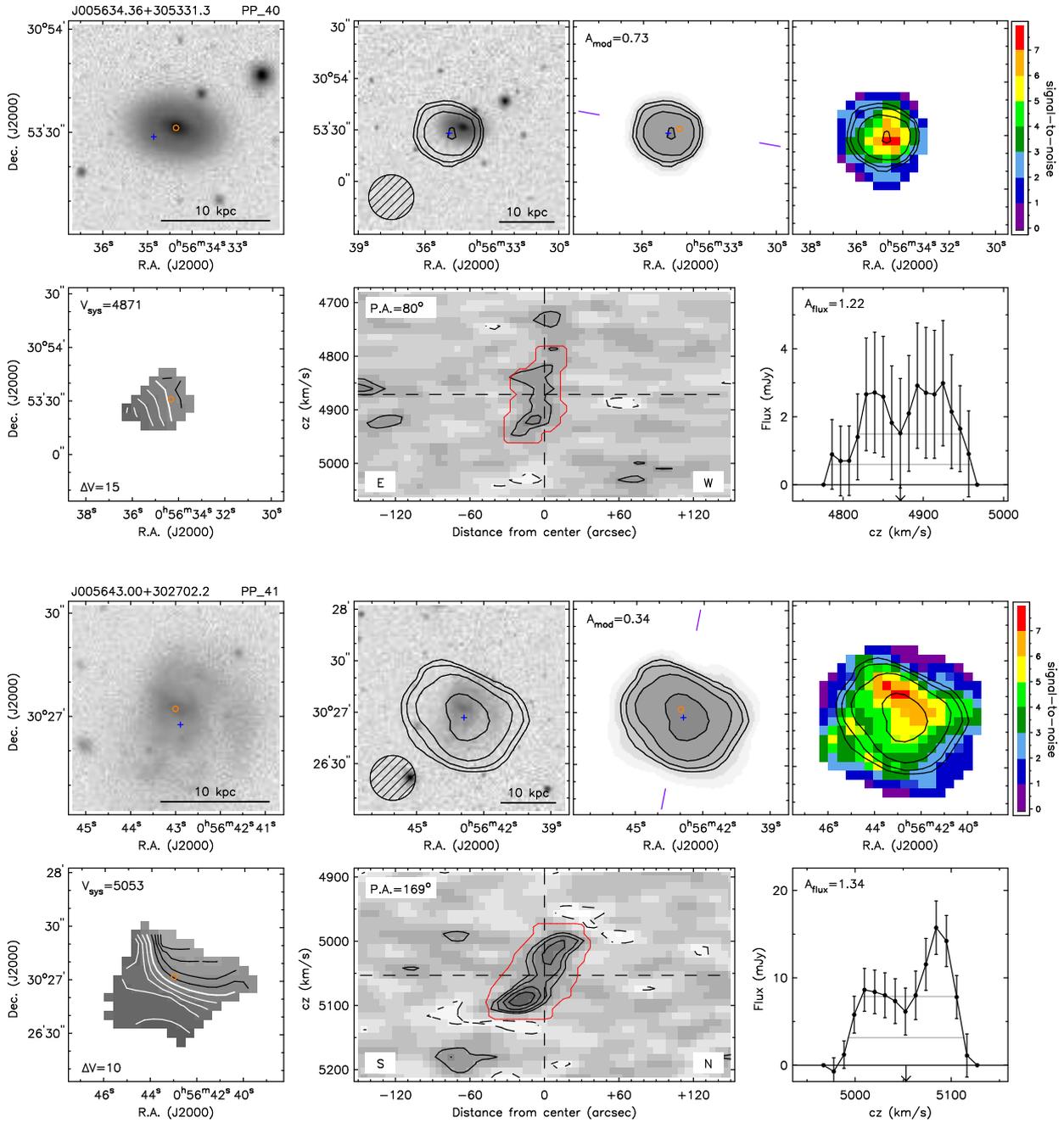

Atlas page 40 and 41

21

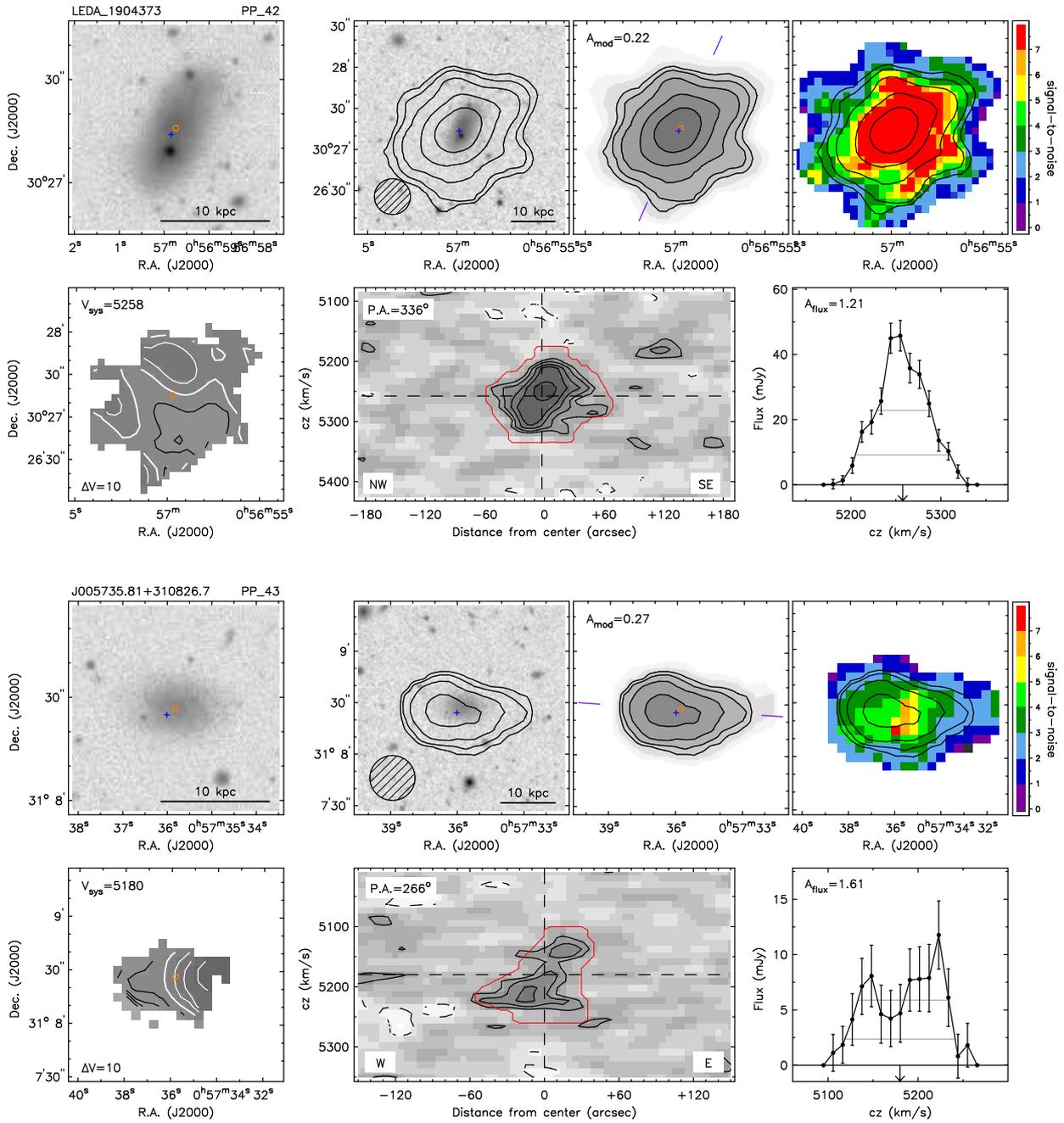

Atlas page 42 and 43

22

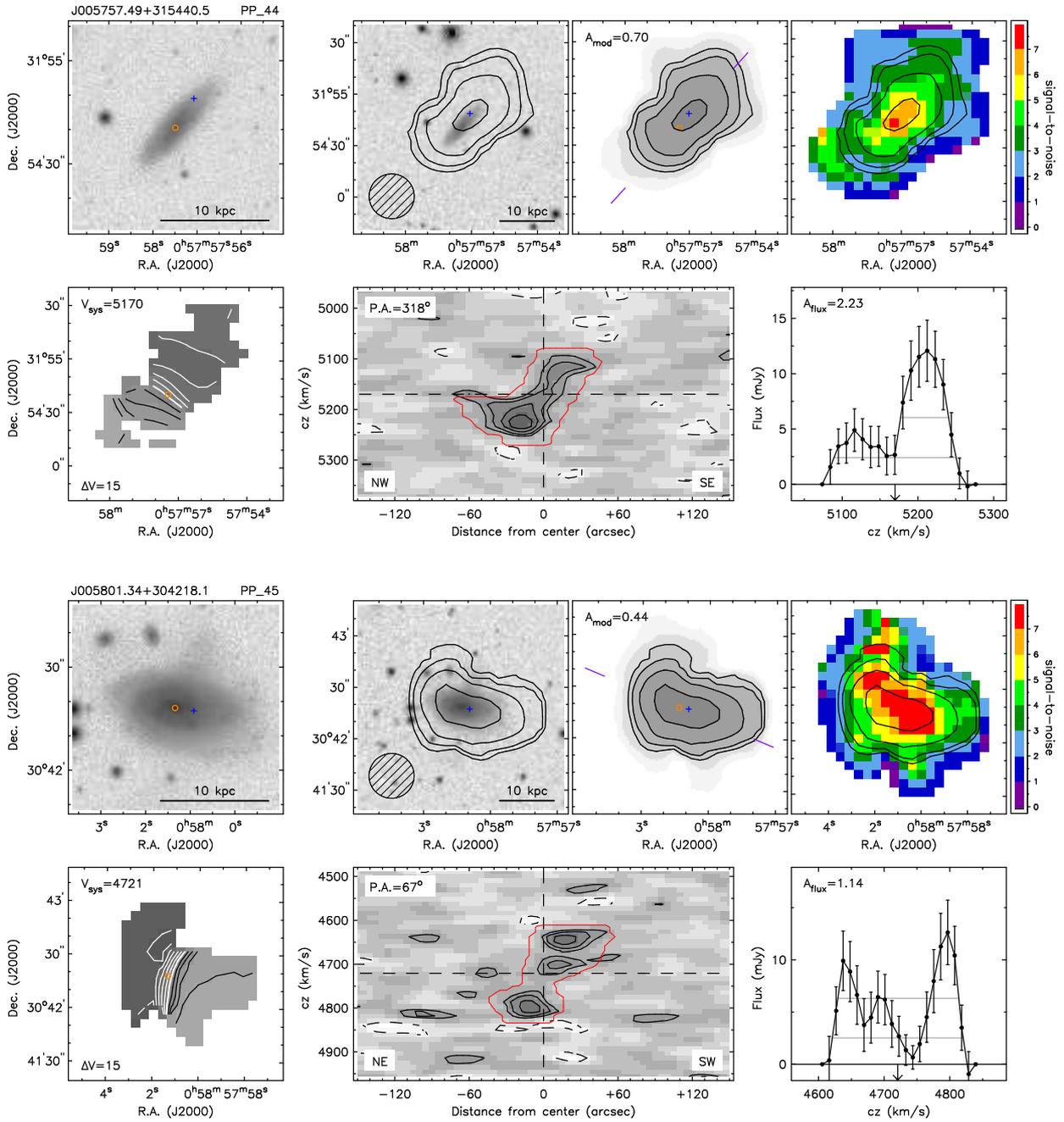

Atlas page 44 and 45

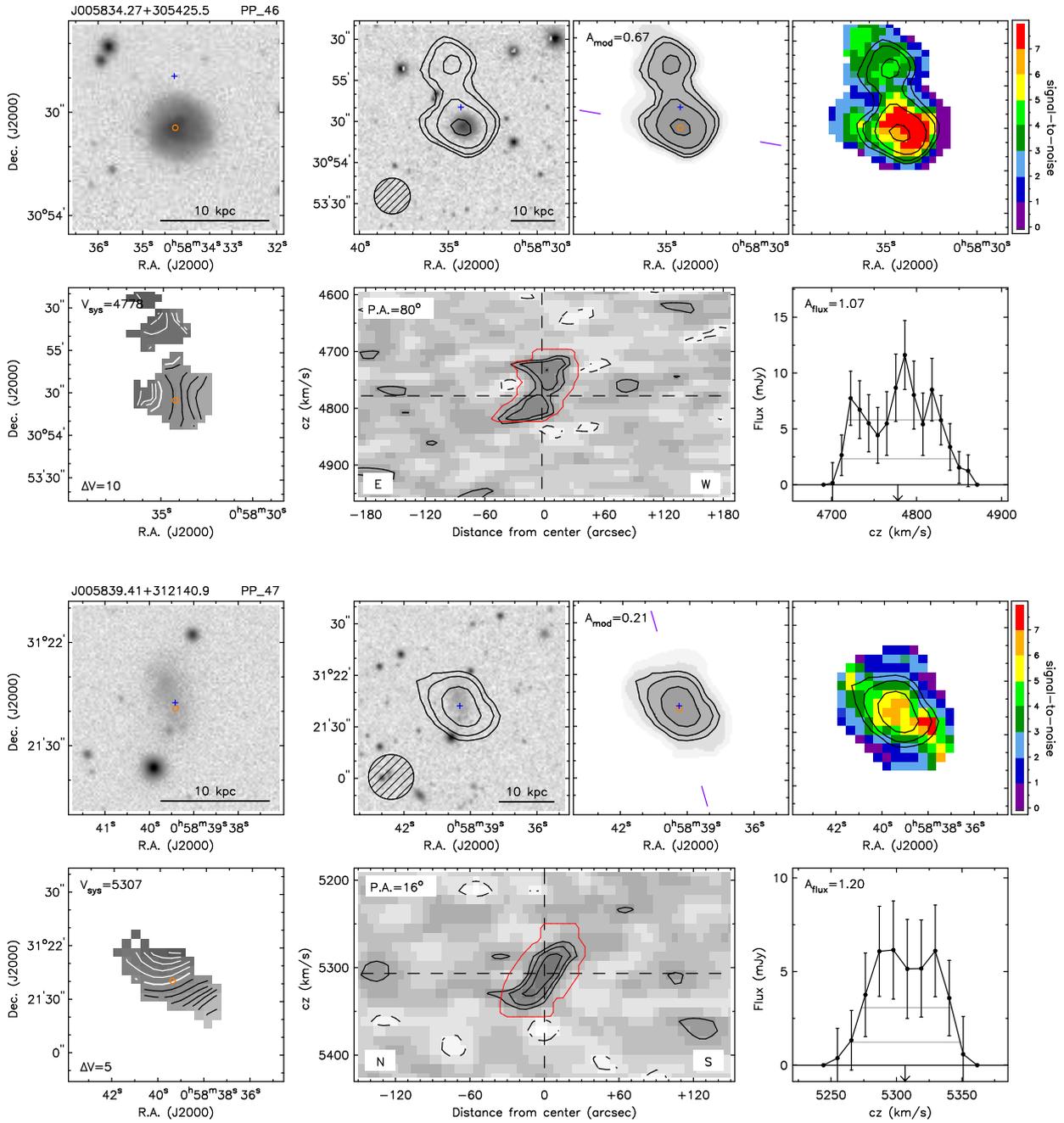

Atlas page 46 and 47

24

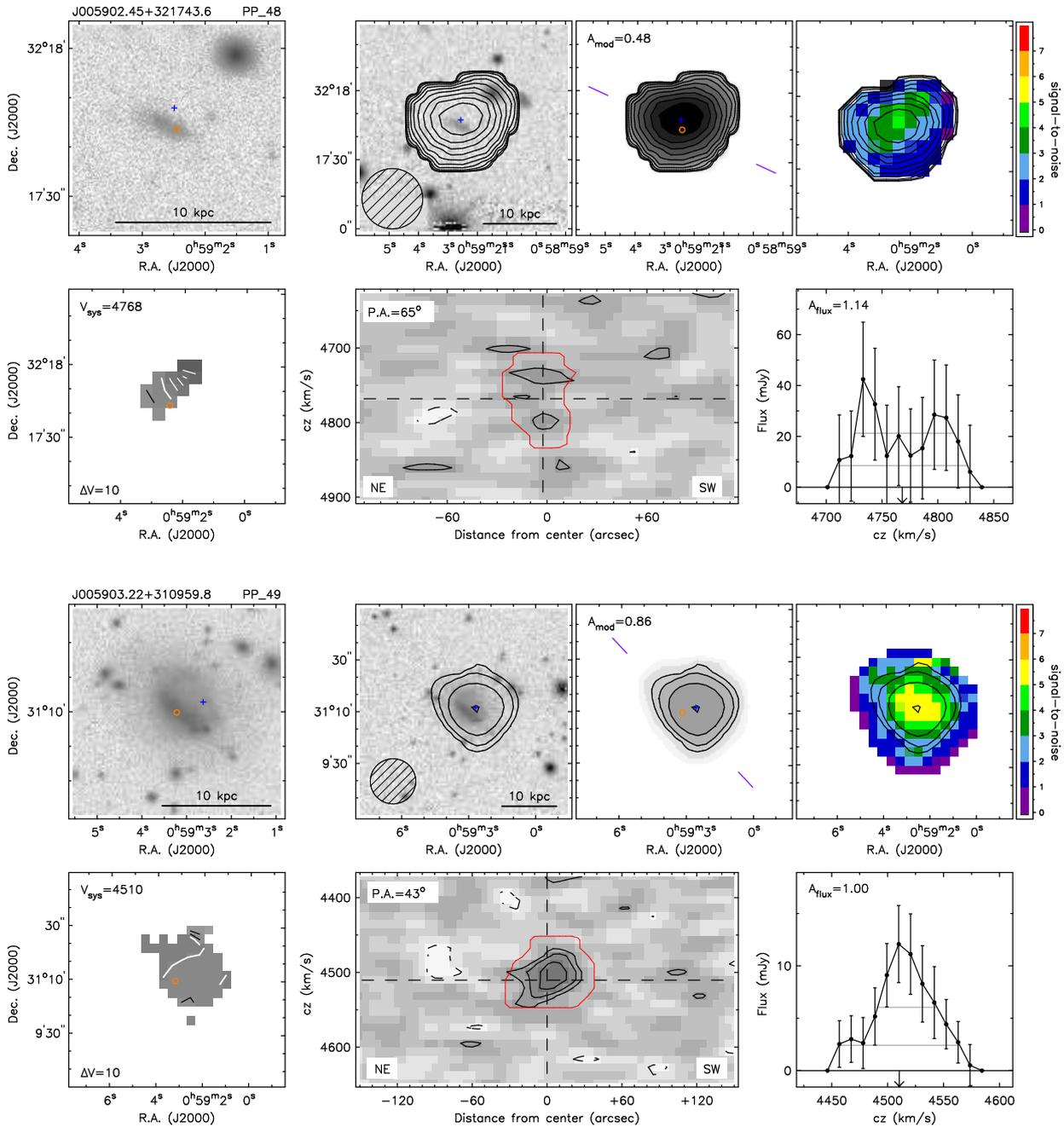

Atlas page 48 and 49

25

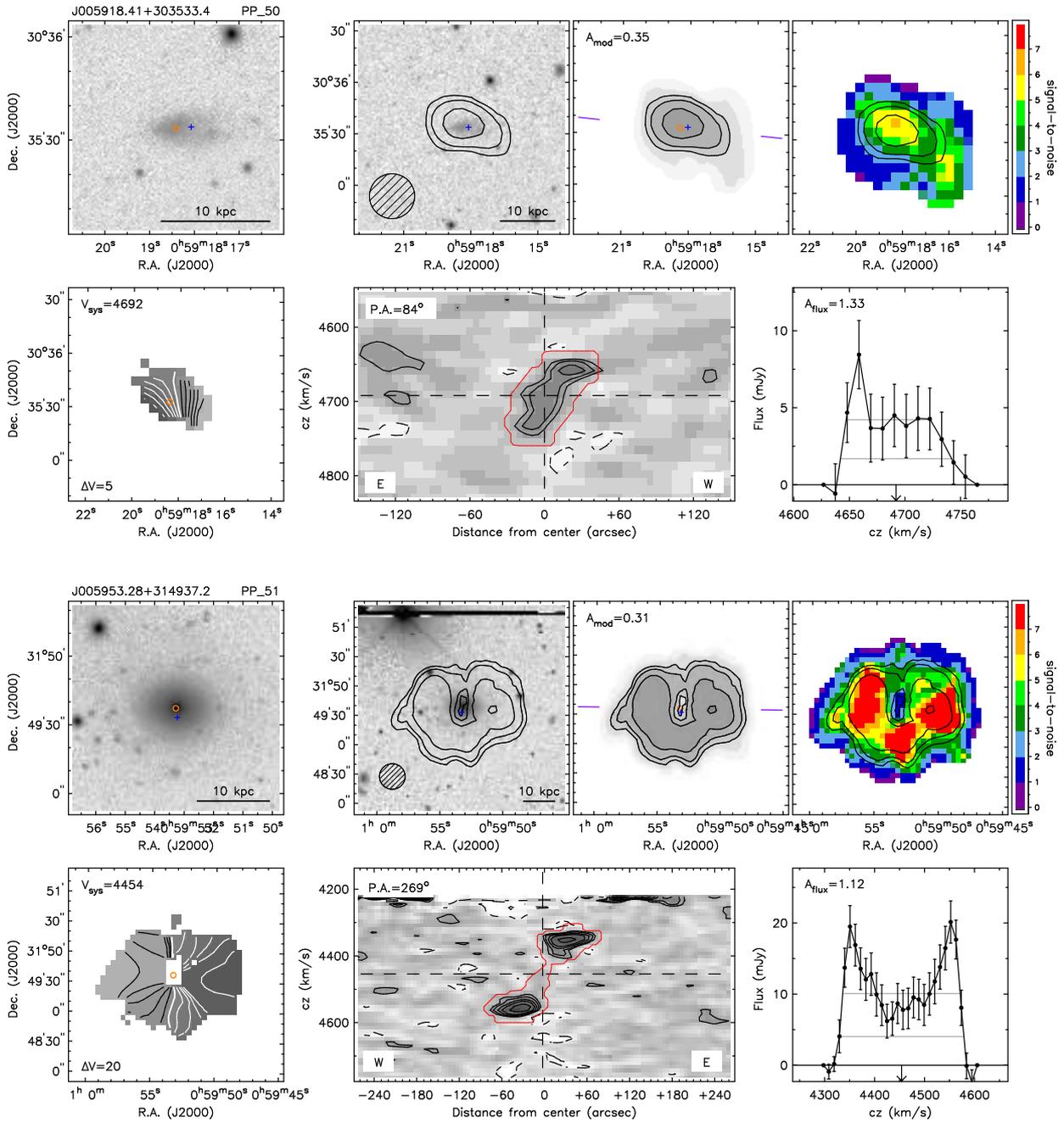

Atlas page 50 and 51

26

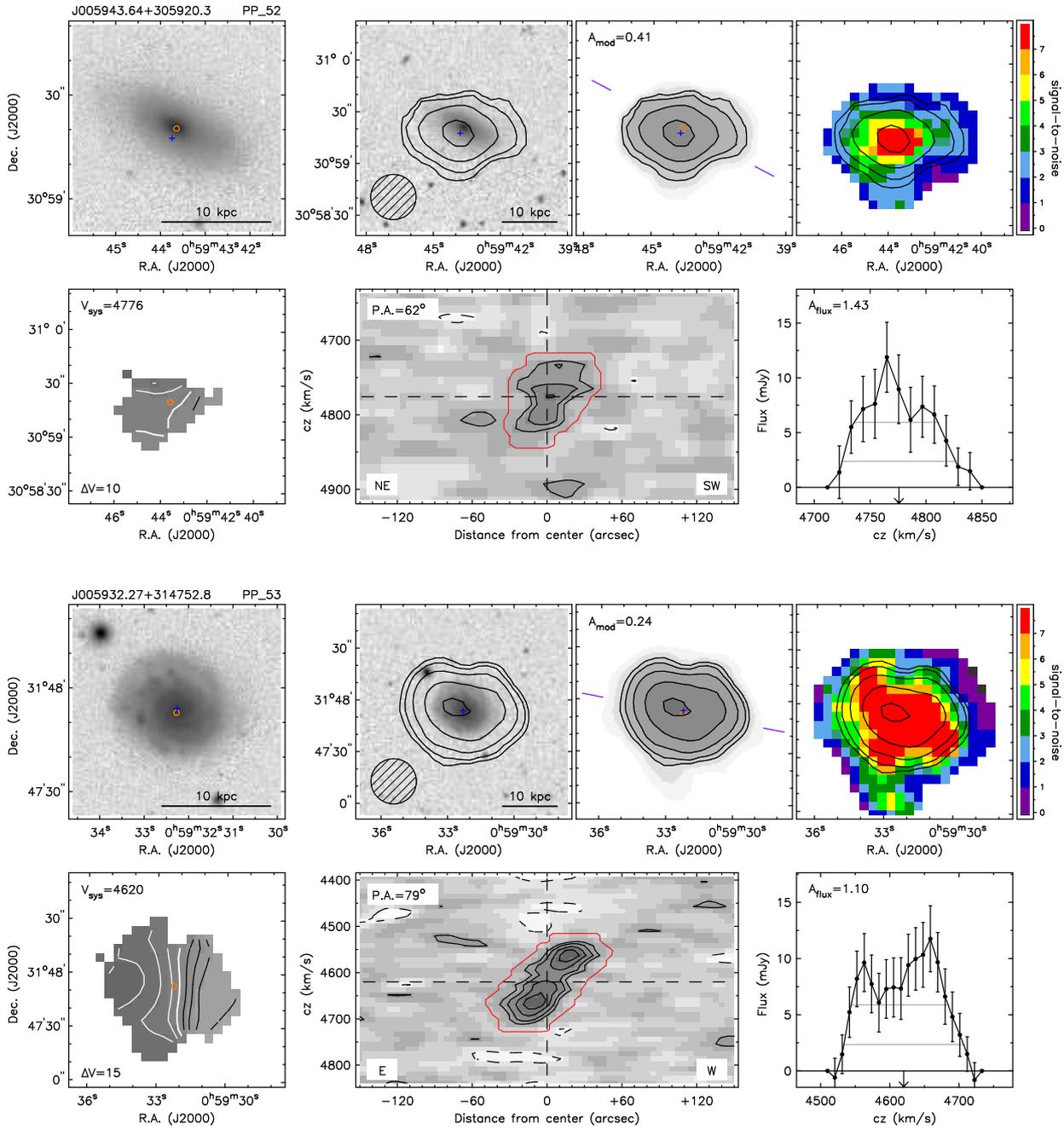

Atlas page 52 and 53

27

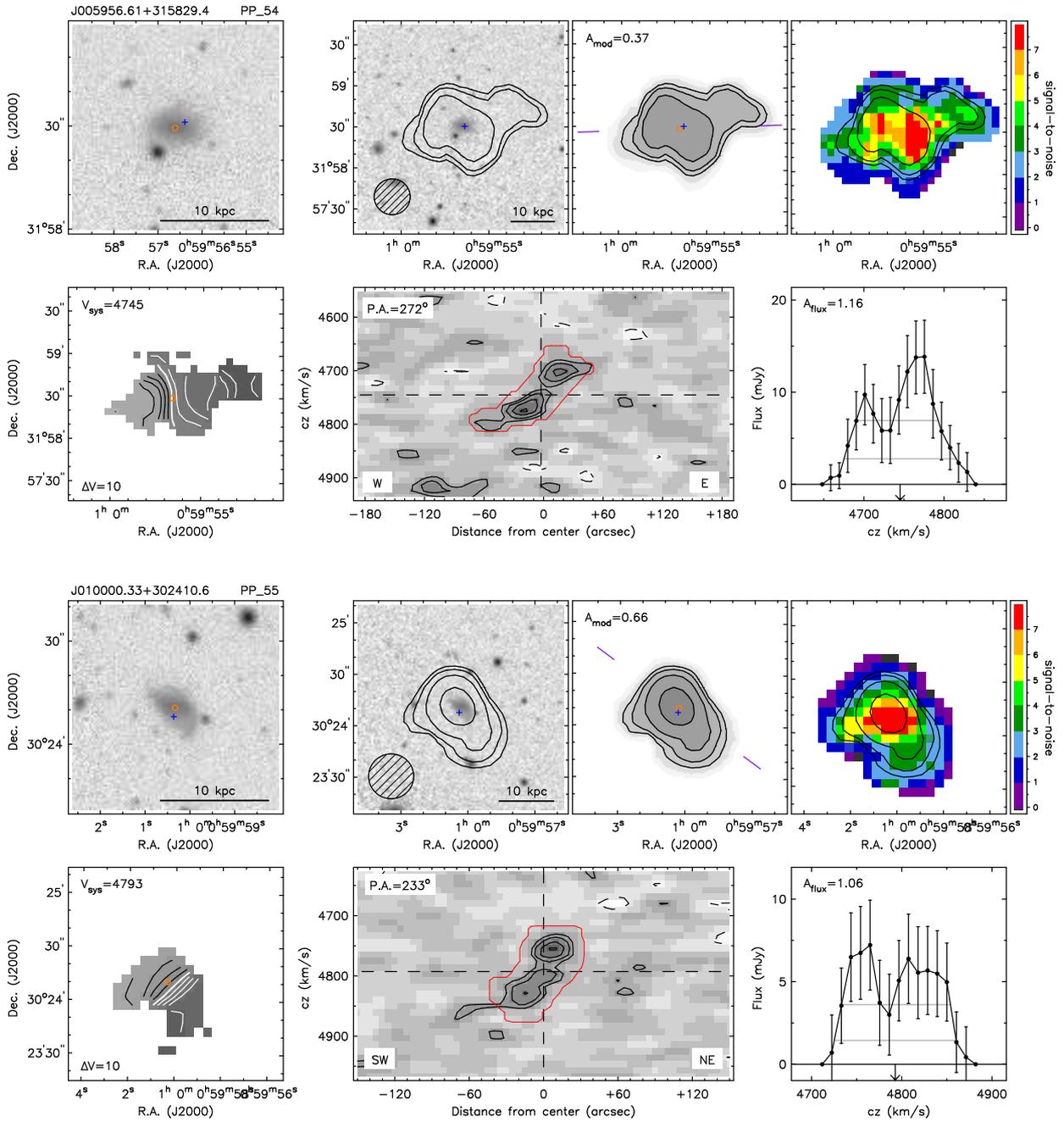

Atlas page 54 and 55

28

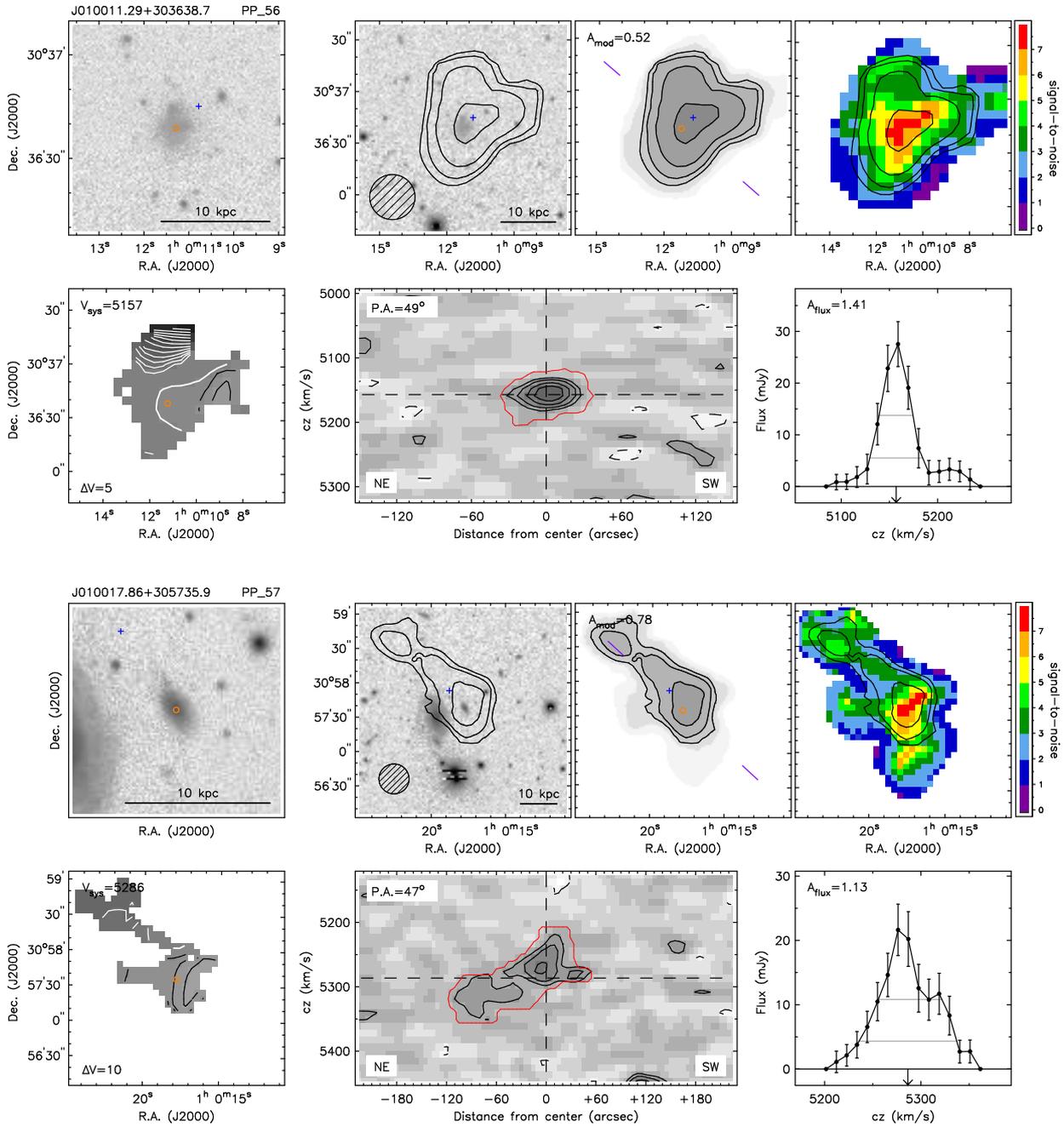

Atlas page 56 and 57

29

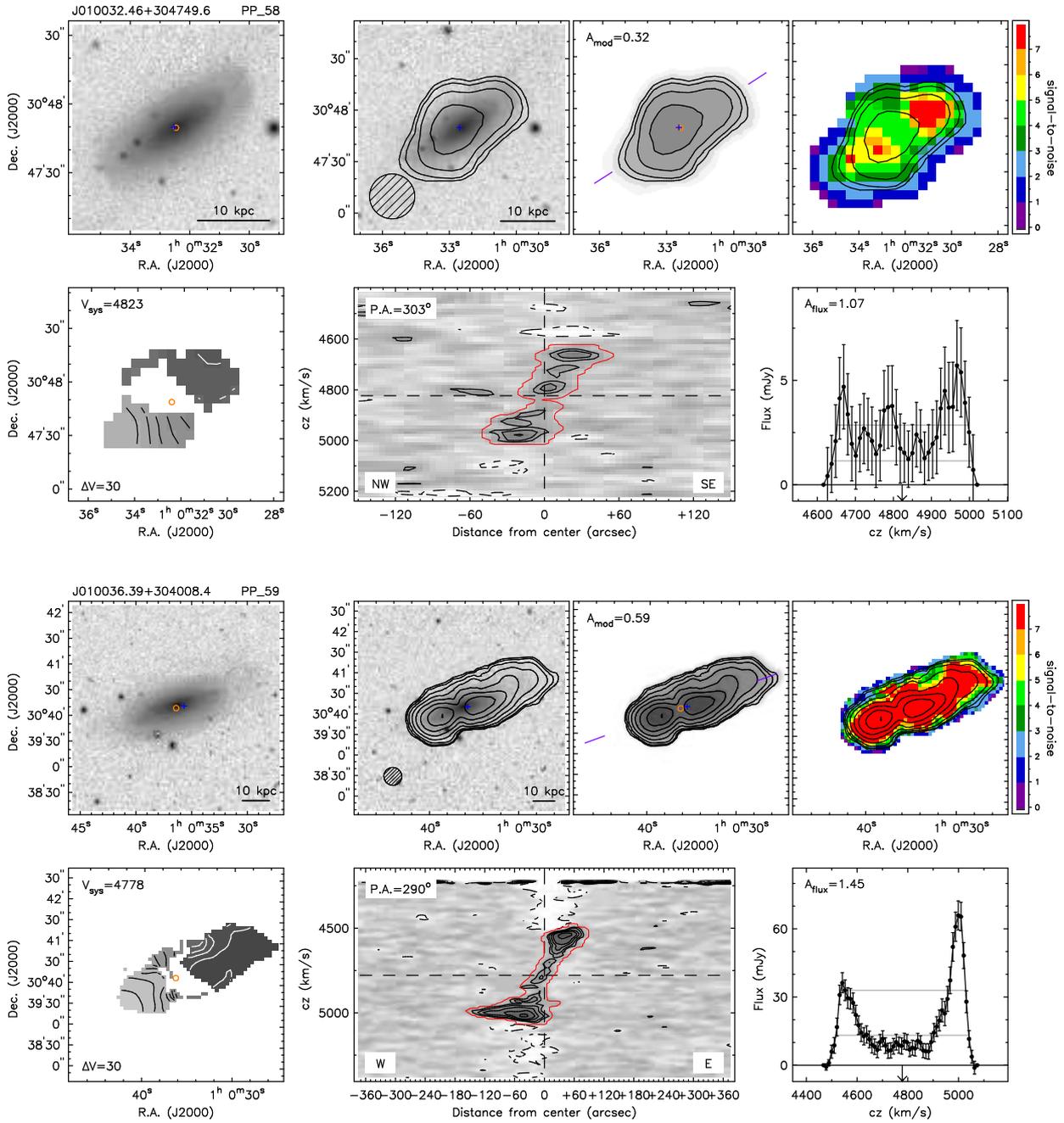

Atlas page 58 and 59

30

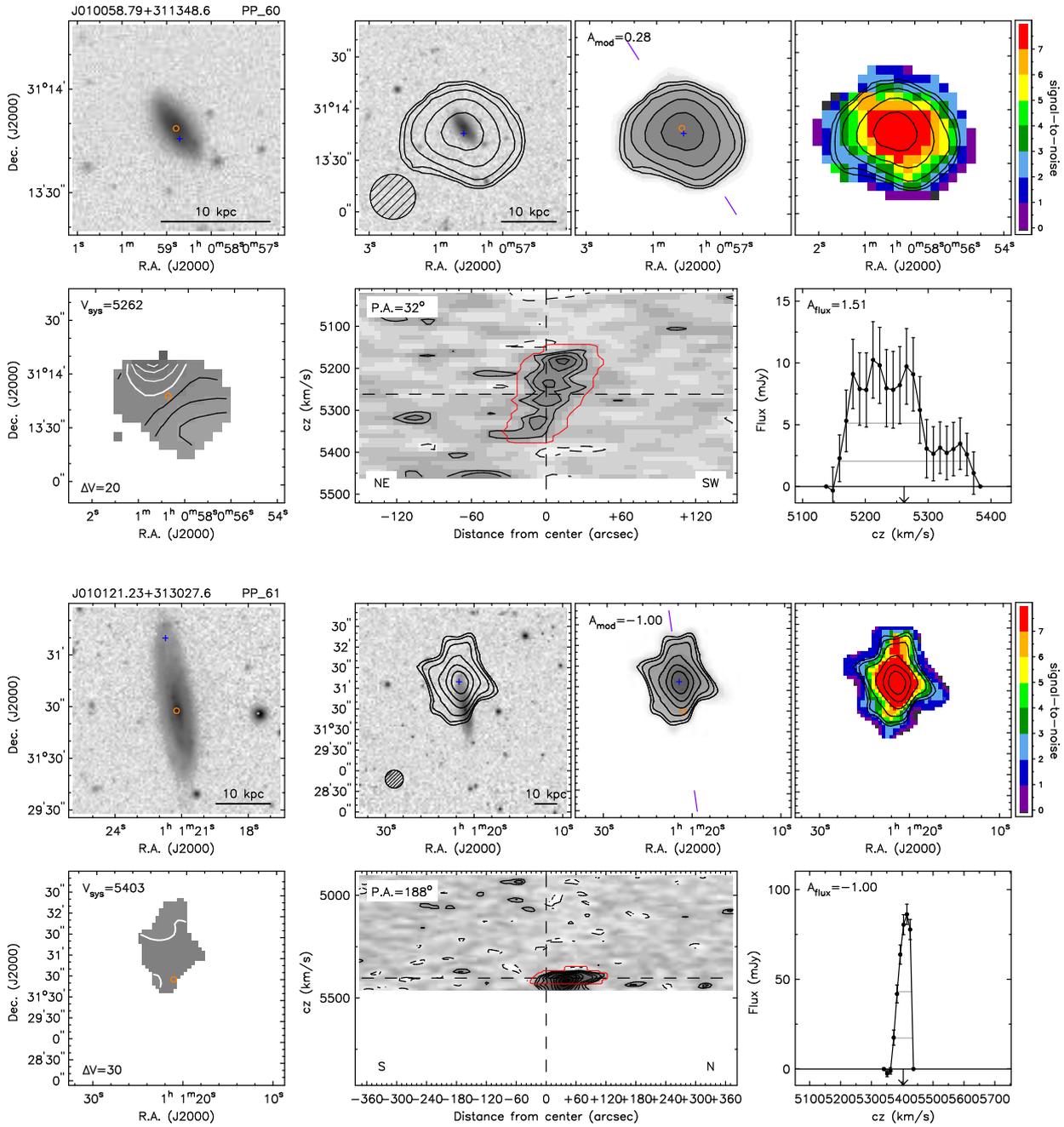

Atlas page 60 and 61

31

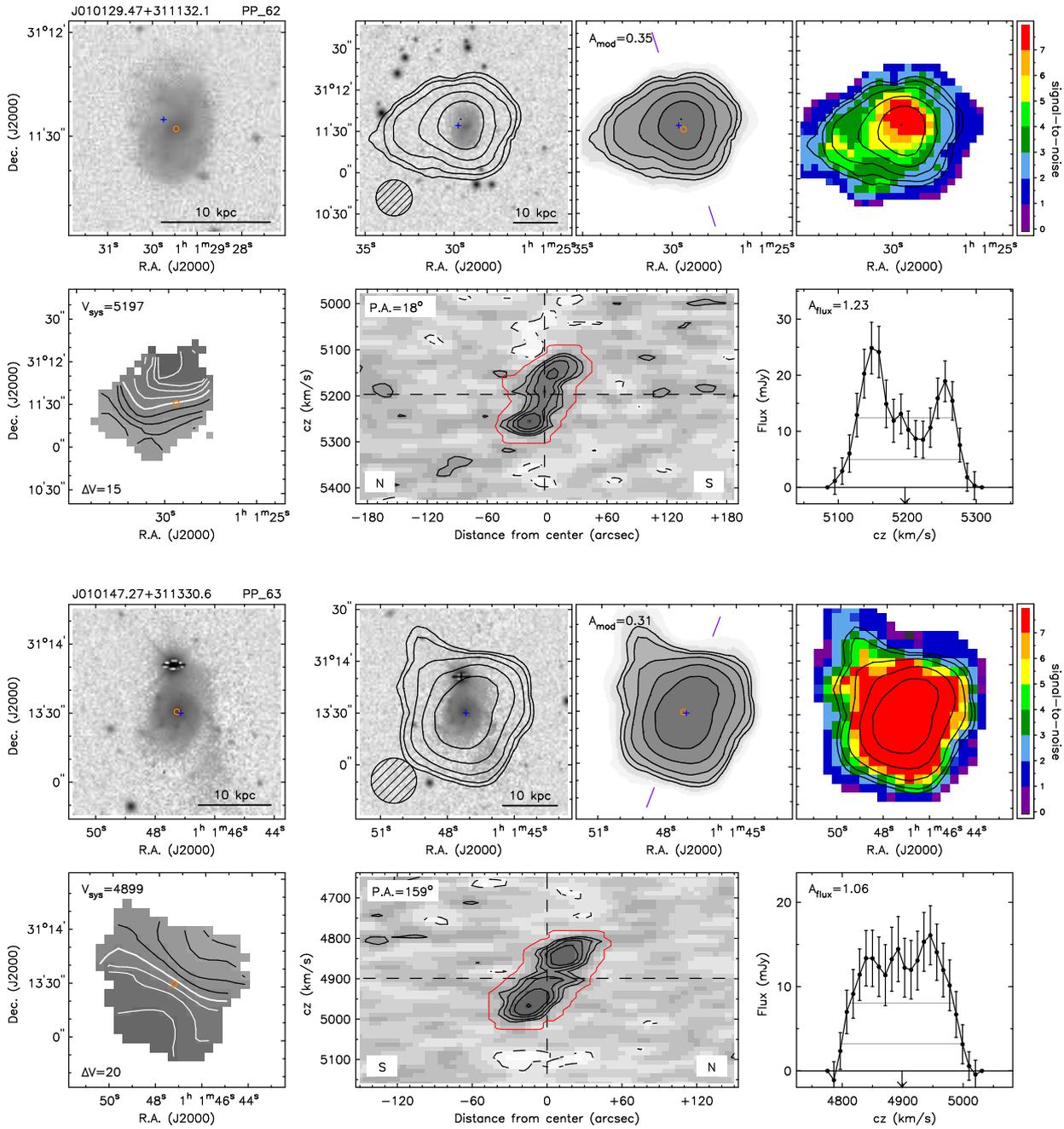

Atlas page 62 and 63

32

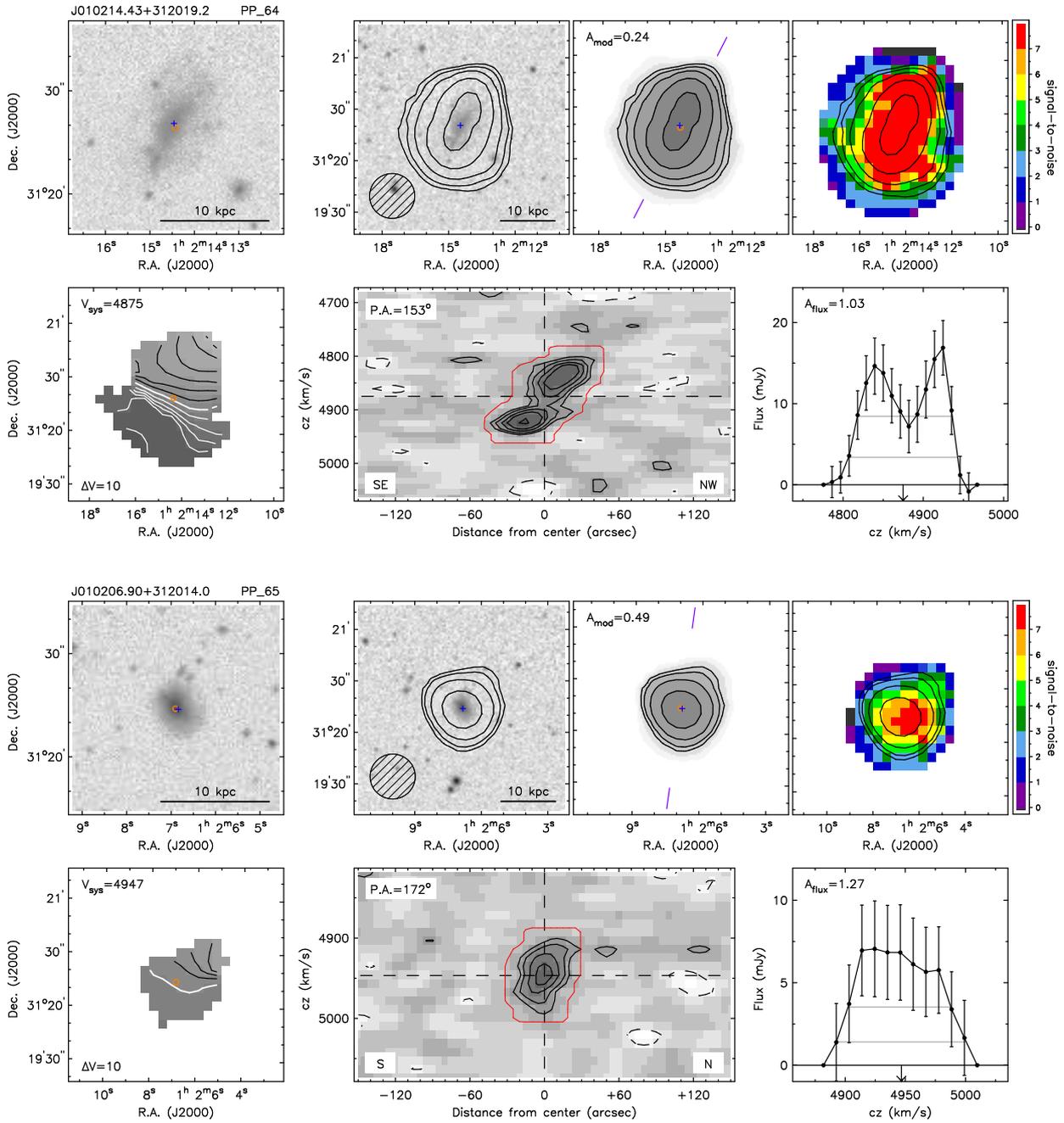

Atlas page 64 and 65

33

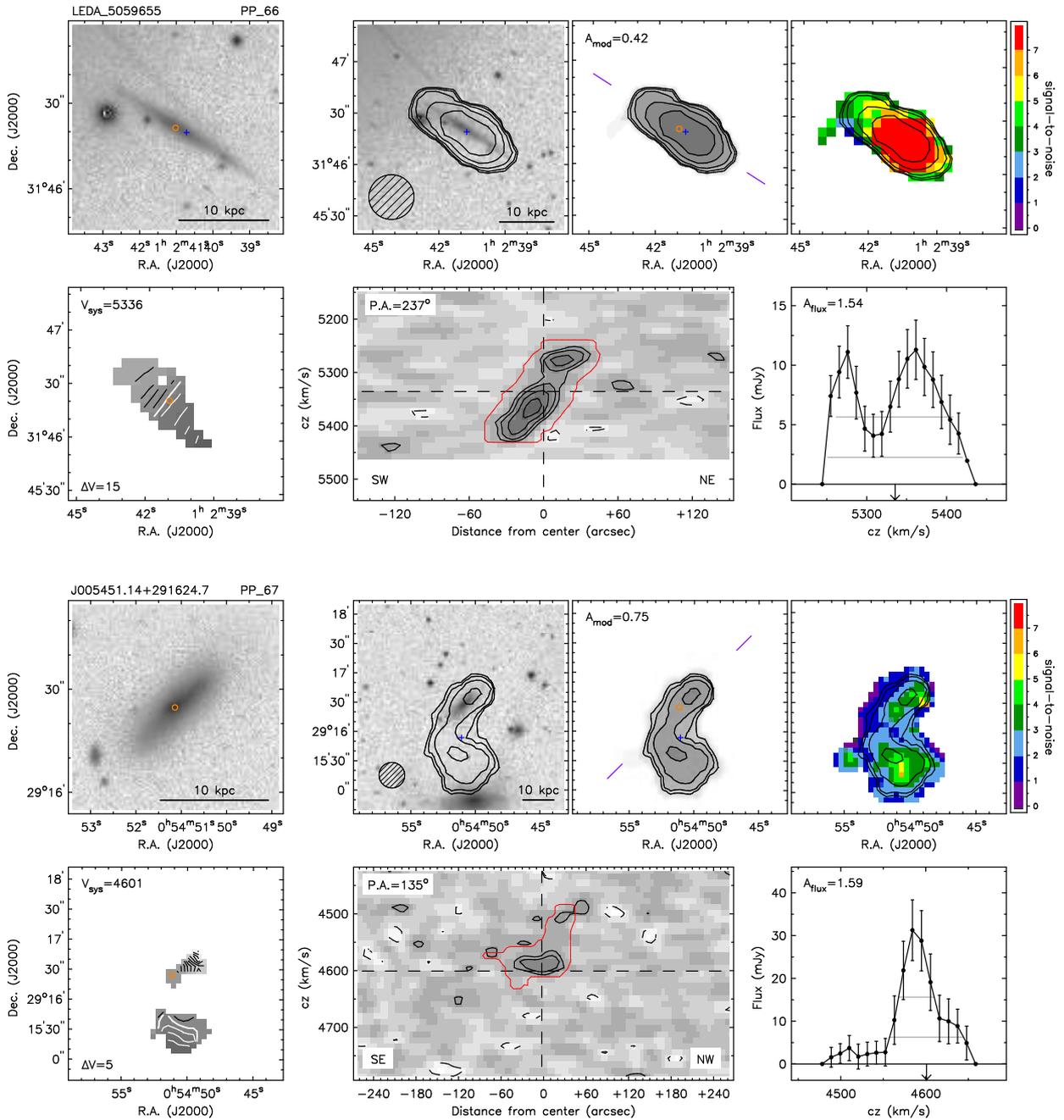

Atlas page 66 and 67

34



\begin{landscape}
\newpage
\section{Photometry of the UMa galaxies}\label{sec:PhotometryUMa}

\begin{small}
\begin{longtable}{c c c c c c c c c c|c c c c c c c c c c c} 

\caption{Parameters used for extinction correction calculation of the B- and R-band magnitudes of the UMa galaxies.}\\
\hline
& & & & & & & & & & & & & & & & &\\
\textbf{Name} & \textbf{b/a} & \textbf{P.A.} & $\mathbf{W_{20}}$ & \textbf{Incl.} & $\mathbf{W^i_r}$ & $\mathbf{A^i_B}$ & $\mathbf{A^i_R}$ & $\mathbf{A^b_B}$ & $\mathbf{A^b_R}$ & & \textbf{Name} & \textbf{b/a} & \textbf{P.A.} & $\mathbf{W_{20}}$ & \textbf{Incl.} & $\mathbf{W^i_r}$ & $\mathbf{A^i_B}$ & $\mathbf{A^i_R}$ & $\mathbf{A^b_B}$ & $\mathbf{A^b_R}$\\
& & (deg) & (\kms) & (deg) & & & & & & & & (deg) & (\kms) & (deg) & & & & &\\
& & & & & & & & & & & & & & & & &\\
\hline
\endfirsthead


\label{tab:PhotometryUMa_ECParams}
1135+48 & 0.31 & 114 & $-$ & 76 & $-$ & 0.00 & 0.00 & 0.10 & 0.06 & &NGC 4138 & 0.63 & 151 & 332 & 53 & 371 & 0.35 & 0.26 & 0.06 & 0.04  \\
1136+46 & 0.79 & 25 & $-$ & 39 & $-$ & 0.00 & 0.00 & 0.08 & 0.05 & &NGC 4143 & 0.55 & 143 & $-$ & 59 & $-$ & 0.00 & 0.00 & 0.06 & 0.03  \\
1137+46 & 0.74 & 117 & $-$ & 43 & $-$ & 0.00 & 0.00 & 0.08 & 0.05 & &NGC 4157 & 0.17 & 63 & 428 & 90 & 390 & 1.38 & 1.00 & 0.09 & 0.06  \\
1156+46 & 0.62 & 21 & $-$ & 54 & $-$ & 0.00 & 0.00 & 0.07 & 0.04 & &NGC 4183 & 0.14 & 166 & 250 & 90 & 212 & 0.94 & 0.71 & 0.07 & 0.04  \\
1203+43 & 0.69 & 101 & 40 & 47 & 16 & 0.00 & 0.00 & 0.06 & 0.04 & &NGC 4217 & 0.26 & 50 & 428 & 80 & 396 & 1.08 & 0.78 & 0.07 & 0.05  \\
MRK 1460 & 0.71 & 31 & 67 & 46 & 69 & 0.00 & 0.00 & 0.09 & 0.06 & &NGC 4218 & 0.60 & 136 & 138 & 55 & 135 & 0.12 & 0.10 & 0.07 & 0.04  \\
NGC 3718 & 0.42 & 15 & 493 & 68 & 492 & 0.78 & 0.56 & 0.06 & 0.04 & &NGC 4220 & 0.31 & 140 & 438 & 76 & 412 & 0.96 & 0.70 & 0.08 & 0.05  \\
NGC 3726 & 0.62 & 14 & 287 & 54 & 310 & 0.33 & 0.24 & 0.07 & 0.04 & &NGC 4346 & 0.33 & 98 & $-$ & 75 & $-$ & 0.00 & 0.00 & 0.06 & 0.04  \\
NGC 3729 & 0.68 & 164 & 271 & 48 & 312 & 0.26 & 0.19 & 0.05 & 0.03 & &NGC 4389 & 0.66 & 96 & 184 & 50 & 194 & 0.18 & 0.14 & 0.06 & 0.04  \\
NGC 3769 & 0.31 & 150 & 265 & 76 & 234 & 0.62 & 0.46 & 0.10 & 0.06 & &UGC 6399 & 0.28 & 140 & 188 & 79 & 156 & 0.41 & 0.32 & 0.07 & 0.04  \\
NGC 3782 & 0.76 & 12 & 132 & 42 & 156 & 0.09 & 0.07 & 0.08 & 0.05 & &UGC 6446 & 0.62 & 20 & 154 & 54 & 153 & 0.15 & 0.12 & 0.07 & 0.04  \\
NGC 3870 & 0.69 & 17 & $-$ & 47 & $-$ & 0.00 & 0.00 & 0.07 & 0.04 & &UGC 6628 & 0.94 & 42 & 52 & 20 & 94 & 0.00 & 0.00 & 0.12 & 0.08  \\
NGC 3877 & 0.23 & 36 & 373 & 84 & 337 & 1.07 & 0.78 & 0.10 & 0.06 & &UGC 6667 & 0.12 & 88 & 188 & 90 & 153 & 0.64 & 0.51 & 0.07 & 0.04  \\
NGC 3893 & 0.67 & 172 & 311 & 49 & 360 & 0.30 & 0.22 & 0.09 & 0.06 & &UGC 6713 & 0.83 & 124 & 104 & 35 & 144 & 0.05 & 0.04 & 0.08 & 0.05  \\
NGC 3896 & 0.67 & 128 & $-$ & 49 & $-$ & 0.00 & 0.00 & 0.09 & 0.06 & &UGC 6773 & 0.53 & 161 & 110 & 60 & 100 & 0.05 & 0.06 & 0.07 & 0.05  \\
NGC 3906 & 0.95 & 38 & 42 & 18 & 53 & 0.00 & 0.00 & 0.11 & 0.07 & &UGC 6805 & 0.79 & 120 & $-$ & 39 & $-$ & 0.00 & 0.00 & 0.09 & 0.06  \\
NGC 3913 & 0.93 & 165 & 52 & 23 & 86 & 0.00 & 0.00 & 0.06 & 0.03 & &UGC 6816 & 0.74 & 39 & 131 & 43 & 152 & 0.09 & 0.07 & 0.06 & 0.04  \\
NGC 3917 & 0.24 & 77 & 295 & 82 & 260 & 0.82 & 0.61 & 0.09 & 0.06 & &UGC 6818 & 0.28 & 77 & 167 & 79 & 137 & 0.32 & 0.26 & 0.09 & 0.06  \\
NGC 3924 & 0.83 & 105 & 115 & 35 & 159 & 0.06 & 0.05 & 0.10 & 0.06 & &UGC 6840 & 0.85 & 53 & 154 & 33 & 228 & 0.08 & 0.06 & 0.10 & 0.06  \\
NGC 3928 & 0.96 & 20 & $-$ & 16 & $-$ & 0.00 & 0.00 & 0.09 & 0.05 & &UGC 6894 & 0.16 & 89 & 142 & 90 & 113 & 0.27 & 0.25 & 0.06 & 0.04  \\
NGC 3931 & 0.81 & 155 & $-$ & 37 & $-$ & 0.00 & 0.00 & 0.10 & 0.06 & &UGC 6917 & 0.55 & 123 & 209 & 59 & 202 & 0.27 & 0.21 & 0.12 & 0.07  \\
NGC 3938 & 0.88 & 22 & 106 & 29 & 175 & 0.05 & 0.04 & 0.09 & 0.06 & &UGC 6922 & 0.84 & 71 & 140 & 34 & 200 & 0.08 & 0.06 & 0.11 & 0.07  \\
NGC 3949 & 0.62 & 117 & 287 & 54 & 310 & 0.33 & 0.24 & 0.09 & 0.06 & &UGC 6923 & 0.42 & 174 & 167 & 68 & 145 & 0.24 & 0.19 & 0.12 & 0.07  \\
NGC 3953 & 0.50 & 13 & 442 & 62 & 457 & 0.61 & 0.44 & 0.13 & 0.08 & &UGC 6930 & 0.86 & 39 & 137 & 32 & 207 & 0.07 & 0.05 & 0.13 & 0.08  \\
NGC 3972 & 0.28 & 118 & 281 & 79 & 248 & 0.72 & 0.53 & 0.06 & 0.04 & &UGC 6940 & 0.28 & 135 & 59 & 79 & 42 & 0.00 & 0.00 & 0.12 & 0.07  \\
NGC 3982 & 0.89 & 22 & 234 & 28 & 425 & 0.10 & 0.07 & 0.06 & 0.04 & &UGC 6956 & 0.72 & 124 & 60 & 45 & 60 & 0.00 & 0.00 & 0.09 & 0.06  \\
NGC 3985 & 0.63 & 70 & 160 & 53 & 162 & 0.15 & 0.12 & 0.11 & 0.07 & &UGC 6962 & 0.80 & 179 & 220 & 38 & 299 & 0.15 & 0.11 & 0.09 & 0.06  \\
NGC 3990 & 0.50 & 40 & $-$ & 62 & $-$ & 0.00 & 0.00 & 0.07 & 0.04 & &UGC 6969 & 0.31 & 150 & 132 & 76 & 108 & 0.15 & 0.14 & 0.12 & 0.08  \\
NGC 3992 & 0.56 & 68 & 479 & 58 & 521 & 0.55 & 0.39 & 0.13 & 0.08 & &UGC 6973 & 0.39 & 40 & 368 & 70 & 351 & 0.69 & 0.50 & 0.09 & 0.06  \\
NGC 3998 & 0.80 & 137 & & 38 & $-$ & 0.00 & 0.00 & 0.07 & 0.04 & &UGC 6983 & 0.66 & 90 & 188 & 50 & 199 & 0.19 & 0.14 & 0.12 & 0.07  \\
NGC 4010 & 0.12 & 65 & 278 & 90 & 240 & 1.13 & 0.85 & 0.11 & 0.07 & &UGC 6992 & 0.45 & 60 & $-$ & 65 & $-$ & 0.00 & 0.00 & 0.09 & 0.05  \\
NGC 4013 & 0.24 & 65 & 425 & 82 & 391 & 1.12 & 0.82 & 0.07 & 0.04 & &UGC 7089 & 0.19 & 35 & 157 & 90 & 126 & 0.34 & 0.29 & 0.07 & 0.04  \\
NGC 4026 & 0.26 & 177 & & 80 & $-$ & 0.00 & 0.00 & 0.10 & 0.06 & &UGC 7094 & 0.36 & 39 & 84 & 72 & 68 & 0.00 & 0.00 & 0.06 & 0.04  \\
NGC 4051 & 0.66 & 131 & 255 & 50 & 282 & 0.26 & 0.19 & 0.06 & 0.04 & &UGC 7129 & 0.69 & 72 & 130 & 47 & 140 & 0.09 & 0.08 & 0.06 & 0.04  \\
NGC 4085 & 0.24 & 75 & 277 & 82 & 242 & 0.77 & 0.57 & 0.08 & 0.05 & &UGC 7176 & 0.26 & 83 & 111 & 80 & 89 & 0.03 & 0.07 & 0.10 & 0.06  \\
NGC 4088 & 0.38 & 51 & 371 & 71 & 352 & 0.72 & 0.53 & 0.09 & 0.05 & &UGC 7218 & 0.53 & 172 & 107 & 60 & 98 & 0.05 & 0.05 & 0.11 & 0.07  \\
NGC 4100 & 0.29 & 164 & 402 & 77 & 373 & 0.94 & 0.69 & 0.10 & 0.06 & &UGC 7301 & 0.16 & 82 & 144 & 90 & 115 & 0.29 & 0.26 & 0.05 & 0.03  \\
NGC 4102 & 0.56 & 38 & 350 & 58 & 369 & 0.44 & 0.32 & 0.09 & 0.05 & &UGC 7401 & 0.59 & 16 & 57 & 56 & 47 & 0.00 & 0.00 & 0.06 & 0.04  \\
NGC 4111 & 0.23 & 150 & 244 & 84 & 208 & 0.00 & 0.00 & 0.06 & 0.04 \\
NGC 4117 & 0.44 & 21 & 289 & 67 & 274 & 0.50 & 0.37 & 0.06 & 0.04 \\
NGC 4118 & 0.60 & 151 & 49 & 55 & 35 & 0.00 & 0.00 & 0.06 & 0.04 \\
\hline
\end{longtable}
\end{small}
\end{landscape}

\begin{landscape}
\newpage
\begin{longtable}{c c c c c c c c c|c c c c c c c c c c} 
\caption{Magnitudes of the UMa galaxies from \cite{Tully1996} in the B- and R-bands, extinction corrected B- and R-band absolute magnitudes, and derived B- and R-band magnitudes derived from SDSS and DECaLS $g,r,z$ magnitudes.}\\
\hline
& & & & & & & & & & & & & & & & &\\
\textbf{Name} & $\mathbf{B_{T+96}}$ & $\mathbf{R_{T+96}}$ & $\mathbf{M^{b,i}_B}$ & $\mathbf{M^{b,i}_R}$ & $\mathbf{B^{SDSS}_{C+14}}$ & $\mathbf{R^{SDSS}_{C+14}}$ & $\mathbf{B^{LS}_{C+14}}$ & $\mathbf{R^{LS}_{C+14}}$ & & \textbf{Name} & $\mathbf{B_{T+96}}$ & $\mathbf{R_{T+96}}$ & $\mathbf{M^{b,i}_B}$ & $\mathbf{M^{b,i}_R}$ & $\mathbf{B^{SDSS}_{C+14}}$ & $\mathbf{R^{SDSS}_{C+14}}$ & $\mathbf{B^{LS}_{C+14}}$ & $\mathbf{R^{LS}_{C+14}}$\\
& & & & & & & & & & & & & & & & &\\
\hline
\endfirsthead
\hline
\endfoot

\label{tab:PhotometryUMa}
1135+48 & 14.95 & 14.05 & -16.32 & -17.18 & 15.31 & 14.47 & 15.10 & 14.23& &NGC 4118 & 15.85 & 14.82 & -15.37 & -16.38 & 15.97 & 15.01 & 15.94 & 14.90 \\
1136+46 & 16.53 & 15.50 & -14.71 & -15.71 & 16.79 & 15.69 & 16.50 & 15.35& &NGC 4138 & 12.27 & 10.72 & -19.31 & -20.74 & 12.25 & 10.86 & 12.22 & 10.90 \\
1137+46 & 16.54 & 15.74 & -14.70 & -15.47 & 24.17 & 24.39 & 16.63 & 15.82& &NGC 4143 & 12.06 & 10.55 & -19.16 & -20.65 & 12.20 & 10.80 & 12.16 & 10.80 \\
1156+46 & 16.24 & 15.05 & -14.99 & -16.16 & 22.40 & 21.63 & 16.38 & 15.53& &NGC 4157 & 12.12 & 10.60 & -20.52 & -21.63 & 13.11 & 11.53 & 12.25 & 10.76 \\
1203+43 & 16.65 & 15.79 & -14.57 & -15.41 & 21.88 & 21.16 & 17.11 & 16.23& &NGC 4183 & 12.96 & 11.99 & -19.21 & -19.92 & 13.80 & 12.73 & 13.11 & 12.01 \\
MRK 1460 & 16.98 & 16.12 & -14.28 & -15.10 & 16.61 & 15.90 & 17.00 & 16.24& &NGC 4217 & 12.15 & 10.62 & -20.17 & -21.37 & 13.40 & 11.27 & 12.05 & 10.62 \\
NGC 3718 & 11.28 & 9.95 & -20.73 & -21.82 & 13.99 & 11.86 & 11.72 & 10.33& &NGC 4218 & 13.69 & 12.83 & -17.67 & -18.48 & 13.77 & 12.91 & 13.79 & 12.87 \\
NGC 3726 & 11.00 & 9.97 & -20.56 & -21.48 & 13.29 & 12.09 & 11.20 & 10.03& &NGC 4220 & 12.34 & 10.79 & -19.86 & -21.12 & 12.75 & 11.37 & 12.42 & 11.03 \\
NGC 3729 & 12.31 & 10.94 & -19.16 & -20.44 & 12.84 & 11.52 & 12.19 & 10.93& &NGC 4346 & 12.14 & 10.69 & -19.08 & -20.51 & 12.38 & 11.07 & 12.24 & 10.95 \\
NGC 3769 & 12.80 & 11.56 & -19.08 & -20.13 & 13.04 & 11.92 & 12.78 & 11.65& &NGC 4389 & 12.56 & 11.33 & -18.85 & -20.01 & 12.86 & 11.69 & 12.74 & 11.50 \\
NGC 3782 & 13.22 & 12.51 & -18.11 & -18.77 & 13.79 & 12.86 & 13.52 & 12.58& &UGC 6399 & 14.33 & 13.31 & -17.31 & -18.22 & 14.59 & 13.63 & 14.35 & 13.33 \\
NGC 3870 & 13.67 & 12.71 & -17.56 & -18.50 & 13.68 & 12.80 & 13.80 & 12.77& &UGC 6446 & 13.52 & 12.81 & -17.86 & -18.51 & 15.15 & 14.09 & 13.61 & 12.69 \\
NGC 3877 & 11.91 & 10.46 & -20.42 & -21.54 & 13.05 & 11.56 & 12.09 & 10.73& &UGC 6628 & 13.17 & 12.26 & -18.12 & -18.99 & 13.84 & 12.86 & 13.51 & 12.51 \\
NGC 3893 & 11.20 & 10.19 & -20.36 & -21.25 & 11.97 & 10.91 & 11.42 & 10.33& &UGC 6667 & 14.33 & 13.11 & -17.55 & -18.61 & 14.54 & 13.43 & 14.33 & 13.16 \\
NGC 3896 & 13.75 & 12.96 & -17.51 & -18.26 & 14.14 & 13.41 & 14.31 & 13.48& &UGC 6713 & 14.90 & 13.88 & -16.40 & -17.38 & 15.20 & 14.26 & 14.80 & 13.78 \\
NGC 3906 & 13.66 & 12.62 & -17.62 & -18.61 & 14.05 & 13.00 & 13.64 & 12.58& &UGC 6773 & 14.42 & 13.61 & -16.87 & -17.66 & 20.56 & 19.80 & 14.55 & 13.65 \\
NGC 3913 & 13.27 & 12.29 & -17.95 & -18.91 & 13.80 & 12.74 & 13.30 & 12.26& &UGC 6805 & 14.90 & 13.75 & -16.36 & -17.47 & 18.72 & 17.66 & 14.95 & 13.80 \\
NGC 3917 & 12.66 & 11.42 & -19.42 & -20.41 & 13.43 & 12.13 & 12.72 & 11.46& &UGC 6816 & 14.31 & 13.62 & -17.01 & -17.65 & 20.67 & 19.92 & 14.38 & 13.54 \\
NGC 3924 & 14.83 & 13.88 & -16.49 & -17.39 & 15.67 & 14.69 & 14.85 & 13.85& &UGC 6818 & 14.43 & 13.62 & -17.15 & -17.86 & 14.68 & 13.73 & 14.56 & 13.60 \\
NGC 3928 & 13.29 & 11.93 & -17.96 & -19.29 & 13.25 & 12.09 & 13.20 & 12.00& &UGC 6840 & 14.39 & 13.35 & -16.96 & -17.94 & 22.94 & 22.32 & 14.66 & 13.65 \\
NGC 3931 & 14.19 & 12.84 & -17.07 & -18.38 & 13.99 & 12.74 & 14.14 & 12.93& &UGC 6894 & 15.27 & 14.31 & -16.23 & -17.14 & 15.18 & 14.25 & 15.20 & 14.24 \\
NGC 3938 & 10.98 & 9.98 & -20.32 & -21.28 & 12.79 & 11.54 & 11.12 & 9.99& &UGC 6917 & 13.15 & 12.16 & -18.40 & -19.28 & 27.87 & 17.37 & 13.16 & 12.09 \\
NGC 3949 & 11.55 & 10.69 & -20.03 & -20.77 & 11.93 & 10.92 & 11.72 & 10.75& &UGC 6922 & 14.52 & 13.65 & -16.83 & -17.64 & 15.21 & 14.10 & 14.70 & 13.59 \\
NGC 3953 & 11.03 & 9.66 & -20.87 & -22.02 & 12.27 & 10.83 & 11.00 & 9.64& &UGC 6923 & 13.91 & 12.97 & -17.61 & -18.46 & 14.17 & 13.27 & 14.09 & 13.13 \\
NGC 3972 & 13.09 & 11.90 & -18.85 & -19.83 & 13.52 & 12.28 & 13.15 & 11.94& &UGC 6930 & 12.70 & 11.71 & -18.66 & -19.59 & 17.45 & 15.49 & 12.80 & 11.75 \\
NGC 3982 & 12.26 & 11.20 & -19.06 & -20.07 & 12.48 & 11.30 & 12.38 & 11.29& &UGC 6940 & 16.45 & 15.65 & -14.83 & -15.59 & 0.22 & -0.27 & 16.29 & 15.52 \\
NGC 3985 & 13.25 & 12.26 & -18.18 & -19.10 & 13.87 & 12.86 & 13.52 & 12.43& &UGC 6956 & 15.12 & 13.83 & -16.14 & -17.39 & 21.97 & 21.06 & 15.43 & 14.35 \\
NGC 3990 & 13.53 & 12.08 & -17.70 & -19.13 & 13.50 & 12.20 & 13.45 & 12.15& &UGC 6962 & 12.88 & 11.88 & -18.52 & -19.45 & 13.16 & 12.09 & 13.04 & 11.93 \\
NGC 3992 & 10.86 & 9.55 & -20.98 & -22.09 & 0.22 & -0.27 & 10.81 & 9.46& &UGC 6969 & 15.12 & 14.32 & -16.32 & -17.06 & 15.73 & 14.64 & 15.12 & 14.22 \\
NGC 3998 & 11.73 & 9.55 & -19.66 & -21.34 & 12.37 & 11.32 & 11.57 & 10.22& &UGC 6973 & 12.94 & 11.26 & -19.01 & -20.47 & 13.11 & 11.55 & 13.14 & 11.51 \\
NGC 4010 & 13.36 & 12.14 & -19.05 & -19.94 & 20.06 & 20.28 & 17.38 & 16.13& &UGC 6983 & 13.10 & 12.27 & -18.37 & -19.11 & 25.43 & 24.60 & 13.29 & 12.27 \\
NGC 4013 & 12.44 & 10.79 & -19.92 & -21.23 & 14.94 & 12.81 & 12.69 & 11.19& &UGC 6992 & 14.77 & 13.51 & -16.48 & -17.71 & 14.96 & 13.80 & 14.82 & 13.64 \\
NGC 4026 & 11.71 & 10.25 & -19.55 & -20.97 & 12.05 & 10.68 & 11.86 & 10.52& &UGC 7089 & 13.73 & 12.77 & -17.84 & -18.72 & 19.60 & 19.01 & 13.89 & 12.88 \\
NGC 4051 & 10.98 & 9.88 & -20.50 & -21.51 & 0.22 & -0.27 & 11.87 & 10.60& &UGC 7094 & 14.74 & 13.70 & -16.48 & -17.50 & 21.29 & 21.02 & 14.89 & 13.81 \\
NGC 4085 & 13.09 & 11.87 & -18.92 & -19.92 & 13.76 & 12.39 & 13.23 & 11.95& &UGC 7129 & 14.13 & 12.80 & -17.19 & -18.48 & 14.28 & 13.10 & 14.19 & 12.94 \\
NGC 4088 & 11.23 & 10.00 & -20.74 & -21.75 & 12.99 & 11.54 & 11.58 & 10.31& &UGC 7176 & 16.20 & 15.61 & -15.10 & -15.68 & 16.42 & 15.65 & 20.33 & 19.41 \\
NGC 4100 & 11.91 & 10.62 & -20.30 & -21.29 & 12.78 & 11.31 & 12.11 & 10.80& &UGC 7218 & 14.88 & 13.99 & -16.44 & -17.29 & 22.74 & 23.15 & 15.12 & 14.17 \\
NGC 4102 & 12.04 & 10.54 & -19.65 & -21.00 & 12.88 & 11.21 & 12.25 & 10.80& &UGC 7301 & 15.54 & 14.57 & -15.96 & -16.88 & 15.49 & 14.53 & 15.40 & 14.39 \\
NGC 4111 & 11.40 & 9.95 & -19.83 & -21.25 & 11.61 & 10.87 & 11.74 & 10.48& &UGC 7401 & 15.74 & 14.83 & -15.49 & -16.37 & 23.01 & 21.95 & 17.18 & 16.28 \\
NGC 4117 & 14.05 & 12.47 & -17.68 & -19.10 & 13.93 & 12.63 & 13.91 & 12.57& & \\
\end{longtable}
\end{landscape}

\newpage
%
%

\newpage
\section{Photometry of the PP galaxies}\label{sec:PhotometryPP}
\begin{longtable}{c l c c c c c c c c c c} 
\caption{Parameters used for extinction correction calculation of the B- and R-band magnitudes of the PP galaxies.}\\
\hline
& & & & & & & & & & &\\
\textbf{\HI\ ID} & \textbf{Common name} & \textbf{b/a} & \textbf{P.A.} & $\mathbf{W_{20}}$ & \textbf{Incl.} & $\mathbf{W^i_r}$ & $\mathbf{A^i_B}$ & $\mathbf{A^i_R}$ & $\mathbf{A^b_B}$ & $\mathbf{A^b_R}$\\
& & & (deg) & (\kms) & (deg) & (\kms) & & & & \\
& & & & & & & & & & &\\
\hline
\endfirsthead
\caption{continued}\\

\hline
& & & & & & & & & & &\\
\textbf{\HI\ ID} & \textbf{Common name} & \textbf{b/a} & \textbf{P.A.} & $\mathbf{W_{20}}$ & \textbf{Incl.} & $\mathbf{W^i_r}$ & $\mathbf{A^i_B}$ & $\mathbf{A^i_R}$ & $\mathbf{A^b_B}$ & $\mathbf{A^b_R}$\\
& & & (deg) & (\kms) & (deg) & (\kms) & & & & \\
& & & & & & & & & & &\\
\hline
\endhead
\hline
\endfoot

\label{tab:PhotometryPP_ECParams}
PP 1 & GCG 500-081 & 0.76 & 172 & 97 & 42 & 115 & 0.04 & 0.04 & 0.24 & 0.15 \\
PP 2 & GALEXASC J004624.74+293926.8 & 0.84 & 115 & 127 & 33 & 184 & 0.07 & 0.05 & 0.23 & 0.14 \\
PP 3 & CGCG 500-083 & 0.74 & 20 & 105 & 44 & 120 & 0.05 & 0.05 & 0.23 & 0.14 \\
PP 4 & CGCG 500-085 & 0.69 & 99 & 171 & 48 & 187 & 0.15 & 0.12 & 0.24 & 0.15 \\
PP 5 & LEDA 1878721 & 0.56 & 129 & 188 & 58 & 182 & 0.23 & 0.17 & 0.24 & 0.15 \\
PP 6 & CGCG 500-090 & 0.54 & 111 & 202 & 59 & 193 & 0.26 & 0.2 & 0.27 & 0.17 \\
PP 7 & GALEXASC J004658.51+290734.7 & 0.2 & 100 & 83 & 90 & 64 & 0 & 0 & 0.27 & 0.17 \\
PP 8 & WISEA J004702.67+301244.0 & 0.26 & 98 & 140 & 80 & 113 & 0.2 & 0.18 & 0.26 & 0.16 \\
PP 9 & UGC 00485 & 0.17 & 179 & 377 & 90 & 339 & 1.28 & 0.93 & 0.29 & 0.18 \\
PP 10 & WISEA J004720.81+295249.9 & 0.8 & 137 & 46 & 38 & 39 & 0 & 0 & 0.24 & 0.15 \\
PP 11 & CGCG 501-016 & 0.9 & 43 & 121 & 27 & 212 & 0.05 & 0.04 & 0.27 & 0.17 \\
PP 12 & SDSS J004825.32+284535.4 & 0.59 & 127 & 115 & 55 & 111 & 0.07 & 0.07 & 0.28 & 0.17 \\
PP 13 & WISEA J004857.29+290345.5 & 0.95 & 32 & 76 & 18 & 192 & 0.02 & 0.01 & 0.24 & 0.15 \\
PP 14 & AGC 102806 & 0.29 & 118 & 166 & 77 & 137 & 0.3 & 0.25 & 0.27 & 0.17 \\
PP 15 & GALEXASC J005037.88+293029.7 & 0.84 & 40 & 105 & 34 & 150 & 0.05 & 0.04 & 0.24 & 0.15 \\
PP 16 & CGCG 501-024 & 0.39 & 137 & 120 & 70 & 101 & 0.08 & 0.09 & 0.29 & 0.18 \\
PP 17 & CGCG 501-026 & 0.87 & 49 & 210 & 30 & 343 & 0.1 & 0.07 & 0.26 & 0.16 \\
PP 18 & WISEA J005106.46+291435.1 & 0.68 & 143 & 102 & 49 & 107 & 0.05 & 0.04 & 0.27 & 0.17 \\
PP 19 & UGC 00525 & 0.63 & 153 & 227 & 53 & 239 & 0.25 & 0.19 & 0.26 & 0.16 \\
PP 20 & WISEA J005140.71+291235.6 & 0.38 & 128 & 83 & 71 & 68 & 0 & 0 & 0.25 & 0.16 \\
PP 21 & LEDA 1927673 & 0.73 & 32 & 112 & 44 & 127 & 0.07 & 0.06 & 0.28 & 0.17 \\
PP 22 & UGC 00536 & 0.54 & 48 & 226 & 59 & 221 & 0.3 & 0.23 & 0.23 & 0.14 \\
PP 23 & WISEA J005255.31+292419.2 & 0.44 & 86 & 172 & 66 & 152 & 0.25 & 0.19 & 0.29 & 0.18 \\
PP 24 & WISEA J005254.86+293351.6 & 0.52 & 92 & 105 & 61 & 95 & 0.04 & 0.05 & 0.25 & 0.15 \\
PP 25 & UGC 00540 & 0.63 & 133 & 270 & 52 & 293 & 0.3 & 0.22 & 0.23 & 0.14 \\
PP 26 & WISEA J005319.55+311417.3 & 0.33 & 32 & 206 & 74 & 177 & 0.42 & 0.32 & 0.25 & 0.15 \\
PP 27 & WISEA J005321.76+300505.4 & 0.55 & 61 & 110 & 58 & 102 & 0.06 & 0.06 & 0.28 & 0.18 \\
PP 28 & UGC 00542 & 0.22 & 158 & 385 & 85 & 349 & 1.11 & 0.81 & 0.24 & 0.15 \\
PP 29 & WISEA J005337.40+313425.5 & 0.58 & 7 & 150 & 56 & 144 & 0.15 & 0.12 & 0.24 & 0.15 \\
PP 30 & WISEA J005352.96+301338.5 & 0.7 & 79 & 128 & 47 & 139 & 0.09 & 0.07 & 0.27 & 0.17 \\
PP 31 & WISEA J005434.66+291335.4 & 1 & 0 & 64 & 3 & 1019 & 0 & 0 & 0.25 & 0.15 \\
PP 32 & UGC 00557 & 0.46 & 37 & 276 & 65 & 263 & 0.46 & 0.34 & 0.26 & 0.16 \\
PP 33 & UGC 00556 & 0.3 & 92 & 369 & 77 & 340 & 0.87 & 0.64 & 0.25 & 0.15 \\
PP 34 & UGC 00561 & 0.64 & 171 & 381 & 52 & 436 & 0.38 & 0.28 & 0.25 & 0.15 \\
PP 35 & NGC 0296 & 0.46 & 163 & 261 & 65 & 246 & 0.43 & 0.32 & 0.27 & 0.17 \\
PP 36 & CGCG 501-043 & 0.44 & 75 & 180 & 66 & 160 & 0.27 & 0.21 & 0.28 & 0.17 \\
PP 37 & AGC 102972 & 0.28 & 119 & 138 & 79 & 112 & 0.18 & 0.17 & 0.27 & 0.17 \\
PP 38 & WISEA J005607.74+302335.6 & 0.55 & 138 & 167 & 59 & 157 & 0.19 & 0.15 & 0.27 & 0.16 \\
PP 39 & UGC 00575 & 0.25 & 159 & 284 & 81 & 249 & 0.77 & 0.58 & 0.26 & 0.16 \\
PP 40 & WISEA J005634.35+305331.2 & 0.74 & 83 & 177 & 43 & 210 & 0.14 & 0.11 & 0.27 & 0.17 \\
PP 41 & WISEA J005642.77+302656.4 & 0.72 & 178 & 120 & 45 & 133 & 0.08 & 0.06 & 0.26 & 0.16 \\
PP 42 & AGC 102822 & 0.51 & 151 & 105 & 61 & 94 & 0.04 & 0.05 & 0.26 & 0.16 \\
PP 43 & WISEA J005735.85+310826.8 & 0.62 & 90 & 123 & 53 & 122 & 0.09 & 0.08 & 0.28 & 0.17 \\
PP 44 & WISEA J005757.47+315440.7 & 0.31 & 139 & 161 & 76 & 133 & 0.27 & 0.22 & 0.25 & 0.15 \\
PP 45 & CGCG 501-055 & 0.59 & 76 & 200 & 56 & 199 & 0.24 & 0.18 & 0.28 & 0.17 \\
PP 46 & WISEA J005834.23+305425.3 & 0.95 & 44 & 135 & 18 & 347 & 0.04 & 0.03 & 0.26 & 0.16 \\
PP 47 & GALEXASC J005839.60+312149.5 & 0.56 & 13 & 84 & 58 & 77 & 0 & 0 & 0.29 & 0.18 \\
PP 48 & WISEA J005902.14+321737.4 & 0.45 & 63 & — & 66 & — & 0 & 0 & 0.29 & 0.18 \\
PP 49 & WISEA J005903.15+310959.7 & 0.83 & 35 & 108 & 35 & 149 & 0.05 & 0.04 & 0.29 & 0.18 \\
PP 50 & AGC 102859 & 0.43 & 91 & 100 & 67 & 86 & 0 & 0.03 & 0.28 & 0.17 \\
PP 51 & MRK 0352 & 0.75 & 102 & 249 & 42 & 313 & 0.19 & 0.14 & 0.27 & 0.16 \\
PP 52 & WISEA J005943.66+305920.6 & 0.47 & 64 & 102 & 64 & 89 & 0.02 & 0.04 & 0.28 & 0.17 \\
PP 53 & CGCG 501-056 & 0.85 & 69 & 173 & 33 & 257 & 0.1 & 0.07 & 0.28 & 0.17 \\
PP 54 & AGC 102830 & 0.63 & 88 & 140 & 53 & 140 & 0.12 & 0.1 & 0.26 & 0.16 \\
PP 55 & WISEA J010000.30+302410.6 & 0.68 & 49 & 135 & 49 & 143 & 0.1 & 0.08 & 0.25 & 0.16 \\
PP 56 & WISEA J010011.31+303638.6 & 0.71 & 174 & 55 & 46 & 51 & 0 & 0 & 0.25 & 0.15 \\
PP 57 & WISEA J010019.81+305730.7 & 0.64 & 11 & 102 & 51 & 103 & 0.04 & 0.04 & 0.27 & 0.16 \\
PP 58 & IC 0066 & 0.54 & 124 & 368 & 59 & 384 & 0.48 & 0.35 & 0.26 & 0.16 \\
PP 59 & NGC 0338 & 0.44 & 110 & 525 & 67 & 531 & 0.78 & 0.56 & 0.24 & 0.15 \\
PP 60 & WISEA J010058.70+311346.9 & 0.6 & 34 & 208 & 55 & 211 & 0.24 & 0.18 & 0.26 & 0.16 \\
PP 61 & UGC 00633 & 0.28 & 8 & 404 & 79 & 373 & 0.99 & 0.72 & 0.28 & 0.17 \\
PP 62 & WISEA J010129.47+311132.1 & 0.64 & 179 & 168 & 52 & 173 & 0.16 & 0.13 & 0.25 & 0.16 \\
PP 63 & CGCG 501-068 & 0.66 & 155 & 200 & 50 & 214 & 0.2 & 0.15 & 0.25 & 0.16 \\
PP 64 & WISEA J010214.50+312020.2 & 0.59 & 158 & 136 & 56 & 131 & 0.12 & 0.1 & 0.24 & 0.15 \\
PP 65 & WISEA J010206.90+312014.9 & 0.77 & 31 & 108 & 40 & 132 & 0.06 & 0.05 & 0.24 & 0.15 \\
PP 66 & AGC 113898 & 0.2 & 57 & 187 & 90 & 152 & 0.49 & 0.39 & 0.26 & 0.16 \\
PP 67 & WISEA J005451.16+291624.3 & 0.46 & 135 & 87 & 65 & 75 & 0 & 0 & 0.25 & 0.15 \\
PP 68 & WISEA J010020.60+305856.3 & 0.48 & 71 & 60 & 64 & 47 & 0 & 0 & 0.27 & 0.17 \\
\end{longtable}

\begin{landscape}
\newpage
\begin{longtable}{c c c c c c c c|c c c c c c c c c} 
\caption{Magnitudes of the PP galaxies in the $g-,r-,z-$ bands from the DECaLS survey, derived B- and R-band magnitudes derived from the DECaLS magnitudes, and extinction corrected B- and R-band absolute magnitudes}\\
\hline
 & & & & & & & & & & & & & & & &\\
\textbf{\HI\ ID} & \textbf{$g$} & \textbf{$r$} & \textbf{$z$} & $\mathbf{B^{LS}_{ C+14}}$ & $\mathbf{R^{LS}_{ C+14}}$ & $\mathbf{M^{b,i}_{ B}}$ & $\mathbf{M^{b,i}_{ B}}$ & & \textbf{\HI\ ID} & \textbf{$g$} & \textbf{$r$} & \textbf{$z$} & $\mathbf{B^{LS}_{ C+14}}$ & $\mathbf{R^{LS}_{ C+14}}$ & $\mathbf{M^{b,i}_{ B}}$ & $\mathbf{M^{b,i}_{ B}}$\\
& & & & & & & & & & & & & & & &\\
\hline
\endfirsthead
\hline
\endfoot

\label{tab:PhotometryPP}
PP 1 & 15.84 & 15.33 & 15.03 & 16.19 & 15.17 & -18.19 & -19.12& &PP 35 & 12.96 & 12.07 & 11.33 & 13.41 & 11.99 & -21.39 & -22.59 \\
PP 2 & 17.68 & 17.17 & 16.84 & 18.04 & 17.01 & -16.36 & -17.29& &PP 36 & 15.09 & 14.31 & 13.69 & 15.51 & 14.21 & -19.13 & -20.27 \\
PP 3 & 14.88 & 14.17 & 13.65 & 15.29 & 14.05 & -19.09 & -20.23& &PP 37 & 16.39 & 15.76 & 15.17 & 16.77 & 15.62 & -17.78 & -18.81 \\
PP 4 & 14.89 & 14.32 & 13.9 & 15.27 & 14.17 & -19.23 & -20.19& &PP 38 & 16.93 & 16.39 & 16.01 & 17.29 & 16.23 & -17.26 & -18.18 \\
PP 5 & 17.1 & 16.56 & 16.2 & 17.46 & 16.4 & -17.1 & -18.01& &PP 39 & 15.34 & 14.6 & 13.98 & 15.76 & 14.48 & -19.37 & -20.35 \\
PP 6 & 15.74 & 15.11 & 14.61 & 16.13 & 14.97 & -18.51 & -19.49& &PP 40 & 15.2 & 14.43 & 13.91 & 15.63 & 14.32 & -18.88 & -20.05 \\
PP 7 & 20.07 & 19.76 & 19.54 & 20.38 & 19.55 & -13.99 & -14.71& &PP 41 & 16.51 & 15.98 & 15.63 & 16.87 & 15.83 & -17.56 & -18.50 \\
PP 8 & 17.81 & 17.37 & 17.1 & 18.15 & 17.19 & -16.41 & -17.25& &PP 42 & 21.84 & 21.11 & 20.55 & 22.26 & 20.99 & -12.13 & -13.32 \\
PP 9 & 14.7 & 13.89 & 13.12 & 15.14 & 13.79 & -20.52 & -21.42& &PP 43 & 17.62 & 17.13 & 16.82 & 17.97 & 16.96 & -16.49 & -17.38 \\
PP 10 & 17.69 & 17.18 & 16.85 & 18.04 & 17.01 & -16.3 & -17.23& &PP 44 & 17.3 & 16.83 & 16.52 & 17.64 & 16.66 & -16.97 & -17.82 \\
PP 11 & 14.85 & 14.16 & 13.61 & 15.26 & 14.03 & -19.16 & -20.27& &PP 45 & 15.38 & 14.83 & 14.41 & 15.75 & 14.68 & -18.86 & -19.77 \\
PP 12 & 15.69 & 15.35 & 15.17 & 16 & 15.15 & -18.44 & -19.18& &PP 46 & 16.01 & 15.54 & 15.22 & 16.35 & 15.37 & -18.04 & -18.92 \\
PP 13 & 16.47 & 16.11 & 15.89 & 16.78 & 15.91 & -17.58 & -18.35& &PP 47 & 18.77 & 18.37 & 18.13 & 19.1 & 18.18 & -15.29 & -16.09 \\
PP 14 & 17.14 & 16.7 & 16.39 & 17.48 & 16.52 & -17.2 & -17.99& &PP 48 & 19.82 & 19.48 & 19.22 & 20.12 & 19.28 & -14.26 & -14.99 \\
PP 15 & 16.79 & 16.34 & 16.08 & 17.13 & 16.17 & -17.26 & -18.12& &PP 49 & 16.76 & 16.33 & 16.06 & 17.09 & 16.15 & -17.35 & -18.17 \\
PP 16 & 15.5 & 14.95 & 14.53 & 15.87 & 14.8 & -18.6 & -19.57& &PP 50 & 18.67 & 18.32 & 18.06 & 18.98 & 18.12 & -15.39 & -16.17 \\
PP 17 & 14.46 & 13.69 & 13.11 & 14.88 & 13.59 & -19.58 & -20.75& &PP 51 & 14.75 & 14.02 & 13.52 & 15.16 & 13.91 & -19.39 & -20.50 \\
PP 18 & 18.1 & 17.7 & 17.48 & 18.42 & 17.52 & -15.99 & -16.79& &PP 52 & 15.87 & 15.26 & 14.85 & 16.25 & 15.12 & -18.15 & -19.19 \\
PP 19 & 14.4 & 13.71 & 13.12 & 14.8 & 13.59 & -19.81 & -20.86& &PP 53 & 14.95 & 14.42 & 13.99 & 15.31 & 14.26 & -19.16 & -20.08 \\
PP 20 & 17.77 & 17.47 & 17.32 & 18.07 & 17.26 & -16.28 & -16.99& &PP 54 & 18 & 17.52 & 17.2 & 18.35 & 17.35 & -16.13 & -17.01 \\
PP 21 & 16.4 & 15.92 & 15.6 & 16.75 & 15.75 & -17.69 & -18.57& &PP 55 & 18.14 & 17.78 & 17.56 & 18.45 & 17.59 & -16 & -16.75 \\
PP 22 & 15.2 & 14.51 & 14.01 & 15.6 & 14.39 & -19.03 & -20.08& &PP 56 & 18.79 & 18.41 & 18.16 & 19.1 & 18.22 & -15.24 & -16.04 \\
PP 23 & 17.93 & 17.44 & 17.14 & 18.27 & 17.27 & -16.36 & -17.20& &PP 57 & 15.94 & 15.27 & 14.75 & 16.33 & 15.14 & -18.07 & -19.16 \\
PP 24 & 17.18 & 16.84 & 16.65 & 17.48 & 16.64 & -16.9 & -17.66& &PP 58 & 14.55 & 13.75 & 13.08 & 14.98 & 13.65 & -19.86 & -20.96 \\
PP 25 & 14.01 & 13.52 & 13.17 & 14.36 & 13.35 & -20.27 & -21.10& &PP 59 & 13.05 & 12.22 & 11.52 & 13.49 & 12.12 & -21.63 & -22.68 \\
PP 26 & 16 & 15.39 & 14.91 & 16.38 & 15.25 & -18.38 & -19.33& &PP 60 & 16.04 & 15.52 & 15.13 & 16.4 & 15.36 & -18.2 & -19.08 \\
PP 27 & 19.15 & 18.61 & 18.24 & 19.51 & 18.45 & -14.92 & -15.88& &PP 61 & 14.26 & 13.54 & 12.88 & 14.67 & 13.42 & -20.69 & -21.57 \\
PP 28 & 14.07 & 13.19 & 12.41 & 14.53 & 13.11 & -20.92 & -21.95& &PP 62 & 16.31 & 15.81 & 15.48 & 16.66 & 15.65 & -17.85 & -18.73 \\
PP 29 & 16.33 & 15.83 & 15.49 & 16.68 & 15.66 & -17.81 & -18.70& &PP 63 & 15.09 & 14.6 & 14.25 & 15.43 & 14.43 & -19.12 & -19.97 \\
PP 30 & 17.39 & 16.84 & 16.48 & 17.75 & 16.68 & -16.7 & -17.66& &PP 64 & 17.73 & 17.34 & 17.12 & 18.06 & 17.16 & -16.4 & -17.19 \\
PP 31 & 16.73 & 16.09 & 15.85 & 17.12 & 15.96 & -17.22 & -18.30& &PP 65 & 17.62 & 17.24 & 16.98 & 17.94 & 17.05 & -16.45 & -17.25 \\
PP 32 & 14.72 & 14.23 & 13.84 & 15.07 & 14.06 & -19.75 & -20.54& &PP 66 & 20.67 & 21.2 & 20.98 & 20.74 & 20.82 & -14.11 & -13.83 \\
PP 33 & 14.76 & 13.78 & 12.95 & 15.23 & 13.72 & -19.98 & -21.17& &PP 67 & 16.18 & 15.56 & 15.08 & 16.56 & 15.42 & -17.78 & -18.83 \\
PP 34 & 15.35 & 14.46 & 13.78 & 15.8 & 14.38 & -18.92 & -20.15& &PP 68 & 19.55 & 19.29 & 19.25 & 19.83 & 19.07 & -14.53 & -15.19 \\ 
\end{longtable}
\end{landscape}

\newpage
%
%

\begin{landscape}
\newpage

\section{Asymmetry values of UMa galaxies}
\begin{longtable}{c c c c c c c c|c c c c c c c c c}
\caption{Morphological and global profile asymmetry values of the UMa galaxies}\\
\hline
& & & & & & & & & & & & & & &\\
\textbf{Name} & \textbf{\logMHI} & \textbf{A$^{^\mathrm{15}}_\mathrm{mod}$} & \textbf{N$^{^\mathrm{15}}_\mathrm{beams}$} & \textbf{A$^{^\mathrm{5}}_\mathrm{mod}$} & \textbf{N$^{^\mathrm{5}}_\mathrm{beams}$} & \textbf{A$_\mathrm{flux}$} & \textbf{S/N} & & \textbf{Name} & \textbf{\logMHI} & \textbf{A$^{^\mathrm{15}}_\mathrm{mod}$} & \textbf{N$^{^\mathrm{15}}_\mathrm{beams}$} & \textbf{A$^{^\mathrm{5}}_\mathrm{mod}$} & \textbf{N$^{^\mathrm{5}}_\mathrm{beams}$} & \textbf{A$_\mathrm{flux}$} & \textbf{S/N} \\
& & & & & & & & & & & & & & & &\\
\hline
\endfirsthead
\hline
\endfoot

\label{UMaAsyTable}
NGC 3718 & 9.93 & 0.27 & 10.9 & 0.50 & 14.4 & 1.10 & 10.50& &UGC 6973 & 9.14 & 0.36 & 3.8 & 0.37 & 5.2 & 1.09 & 5.16 \\
NGC 3726 & 9.82 & 0.34 & 9.1 & 0.36 & 11.0 & 1.08 & 0.06& &UGC 6983 & 9.39 & 0.17 & 4.8 & 0.19 & 5.7 & 1.10 & 6.96 \\
NGC 3729 & 8.38 & 0.30 & 1.7 & 0.49 & 2.7 & 2.26 & 1.39& &UGC 7176 & 8.47 & 0.30 & 1.9 & 0.36 & 2.8 & 1.30 & 2.41 \\
NGC 3769 & 9.60 & 0.56 & 6.1 & 0.65 & 10.7 & 1.44 & 7.75& &UGC 7301 & 8.54 & 0.34 & 2.1 & 0.30 & 3.1 & 1.04 & 2.14 \\
NGC 3782 & 9.24 & 0.34 & 3.9 & 0.44 & 4.8 & 1.04 & 6.26& &UM 01 & 7.92 & $-$ & $-$ & 0.49 & 1.5 & 1.29 & 3.35 \\
NGC 3893 & 9.78 & 0.69 & 7.7 & 0.76 & 12.0 & 1.04 & 6.89& &UM 02 & 6.54 & $-$ & $-$ & $-$ & $-$ & 1.74 & 1.03 \\
NGC 3917 & 9.22 & 0.29 & 4.2 & 0.30 & 4.9 & 1.45 & 4.87& &UM 03 & 7.84 & $-$ & $-$ & 0.52 & 1.6 & 1.13 & 7.23 \\
NGC 3938 & 9.07 & 0.27 & 7.0 & 0.27 & 9.4 & 1.00 & 2.95& &MRK1460 & 6.97 & $-$ & $-$ & $-$ & $-$ & 1.05 & 1.31 \\
NGC 3953 & 9.40 & 0.18 & 5.1 & 0.28 & 6.4 & 1.03 & 3.45& &UM 04 & 7.08 & $-$ & $-$ & $-$ & $-$ & 1.30 & 1.46 \\
NGC 3972 & 8.95 & 0.44 & 3.1 & 0.50 & 3.7 & 1.04 & 2.49& &UM 05 & 7.08 & $-$ & $-$ & $-$ & $-$ & $-$ & $-$ \\
NGC 3982 & 9.11 & 0.38 & 3.7 & 0.42 & 5.2 & 1.04 & 5.60& &UM 06 & 7.17 & $-$ & $-$ & $-$ & $-$ & 1.08 & 1.70 \\
NGC 3992 & 9.68 & 0.29 & 7.8 & 0.35 & 9.9 & 1.18 & 3.88& &UM 08 & 8.17 & 0.35 & 1.0 & 0.21 & 2.1 & 1.27 & 6.12 \\
NGC 4010 & 9.39 & 0.23 & 4.6 & 0.30 & 5.8 & 1.06 & 6.94& &UM 09 & 7.09 & $-$ & $-$ & $-$ & $-$ & 1.12 & 1.57 \\
NGC 4051 & 9.41 & 0.31 & 4.7 & 0.35 & 5.4 & 1.02 & 11.00& &UM 12 & 7.62 & $-$ & $-$ & 0.97 & 0.8 & 1.29 & 3.19 \\
NGC 4085 & 9.03 & 0.19 & 3.0 & 0.28 & 3.4 & 1.70 & 3.06& &UM 13 & 7.49 & $-$ & $-$ & $-$ & $-$ & 1.03 & 3.55 \\
NGC 4088 & 9.85 & 0.35 & 7.1 & 0.46 & 9.8 & 1.01 & 6.36& &UM 14 & 8.09 & $-$ & $-$ & 0.67 & 1.0 & 1.09 & 3.17 \\
NGC 4100 & 9.43 & 0.16 & 5.5 & 0.27 & 6.7 & 1.01 & 2.64& &NGC 4111 & 8.87 & $-$ & $-$ & 0.85 & 9.3 & 1.09 & 5.98 \\
NGC 4102 & 8.59 & 0.45 & 2.1 & 0.44 & 3.4 & 1.41 & 1.13& &NGC 3877 & 9.12 & 0.27 & 3.6 & 0.42 & 4.3 & 1.24 & 18.05 \\
NGC 4138 & 8.97 & 0.52 & 4.2 & 0.56 & 5.7 & 1.03 & 2.30& &NGC 3906 & 8.47 & 0.54 & 1.8 & 0.43 & 2.6 & 1.10 & 15.72 \\
NGC 4157 & 9.86 & 0.28 & 8.1 & 0.33 & 9.0 & 1.07 & 8.65& &NGC 3913 & 8.98 & 0.33 & 3.2 & 0.35 & 4.1 & 1.27 & 28.74 \\
NGC 4183 & 9.54 & 0.26 & 5.6 & 0.28 & 6.5 & 1.12 & 17.63& &UGC 6840 & 9.19 & 0.33 & 3.8 & 0.52 & 6.6 & 1.16 & 16.73 \\
NGC 4217 & 9.33 & 0.20 & 4.5 & 0.25 & 5.3 & 1.11 & 5.32& &NGC 3949 & 9.46 & 0.19 & 4.3 & 0.33 & 6.7 & 1.00 & 32.68 \\
NGC 4218 & 8.63 & 0.44 & 2.2 & 0.47 & 3.0 & 1.04 & 2.29& &UGC 6917 & 9.27 & 0.26 & 3.9 & 0.27 & 4.8 & 1.17 & 21.15 \\
NGC 4220 & 8.71 & 0.75 & 2.4 & 0.69 & 3.4 & 1.08 & 7.00& &NGC 3985 & 9.05 & 0.33 & 3.0 & 0.35 & 5.1 & 1.16 & 10.86 \\
NGC 4389 & 8.64 & 0.20 & 2.2 & 0.23 & 3.3 & 1.19 & 1.50& &UGC 6922 & 8.88 & 0.06 & 2.8 & 0.12 & 4.0 & 1.15 & 11.39 \\
UGC 6399 & 8.76 & 0.33 & 2.7 & 0.35 & 3.5 & 1.02 & 2.71& &UGC 6930 & 9.45 & 0.27 & 4.9 & 0.29 & 6.4 & 1.05 & 31.14 \\
UGC 6446 & 9.41 & 0.10 & 4.8 & 0.12 & 5.6 & 1.02 & 7.20& &UGC 6956 & 9.00 & 0.42 & 4.0 & 0.57 & 5.8 & 1.35 & 33.61 \\
UGC 6667 & 8.78 & 0.19 & 2.8 & 0.19 & 3.4 & 1.01 & 3.15& &UGC 7089 & 9.06 & 0.51 & 3.1 & 0.54 & 4.0 & 1.66 & 17.14 \\
UGC 6773 & 8.46 & 0.16 & 1.9 & 0.18 & 2.6 & 1.26 & 1.61& &1203+43 & 7.67 & $-$ & $-$ & 0.45 & 1.1 & 1.09 & 5.10 \\
UGC 6818 & 8.93 & 0.31 & 3.0 & 0.44 & 4.3 & 1.24 & 3.91& &UGC 7094 & 8.26 & 0.24 & 1.0 & 0.26 & 2.3 & 1.09 & 6.95 \\
UGC 6894 & 8.39 & 0.19 & 1.7 & 0.25 & 2.4 & 1.03 & 2.64& &NGC 4117 & 8.34 & 0.79 & 1.0 & 0.45 & 2.8 & 1.36 & 3.99 \\
UGC 6923 & 8.81 & 0.14 & 2.5 & 0.45 & 3.7 & 1.07 & 2.32& &UGC 7129 & 7.80 & $-$ & $-$ & 0.36 & 1.1 & 2.13 & 1.96 \\
UGC 6940 & 7.92 & 0.24 & 0.8 & 0.39 & 1.8 & 1.33 & 0.77& &NGC 4026 & & $-$ & $-$ & 1.0 & & & \\
UGC 6962 & 8.83 & 0.47 & 2.9 & 0.55 & 4.2 & 1.09 & 2.34& &NGC 3998 & & 0.97 & & 0.76 & & & \\
UGC 6969 & 8.56 & 0.30 & 2.1 & 0.37 & 2.8 & 1.48 & 1.73& & \\
\end{longtable}
\end{landscape}

\begin{landscape}
\newpage

\section{Asymmetry values of the PP galaxies}
\begin{longtable}{c c c c c c |c c c c c c c}
\caption{Morphological and global profile asymmetry values of the PP galaxies}\\
\hline
& & & & & & & & & & & &\\
\textbf{\HI\ ID} & \textbf{\logMHI} & \textbf{A$^{^\mathrm{15}}_\mathrm{mod}$} & \textbf{N$^{^\mathrm{15}}_\mathrm{beams}$} & \textbf{A$_\mathrm{flux}$} & \textbf{S/N} & & \textbf{\HI\ ID} & \textbf{\logMHI} & \textbf{A$^{^\mathrm{15}}_\mathrm{mod}$} & \textbf{N$^{^\mathrm{15}}_\mathrm{beams}$} & \textbf{A$_\mathrm{flux}$} & \textbf{S/N}\\
 & & & & & & & & & & &\\
\hline
\endfirsthead
\hline
\endfoot

\label{PPAsyTable}
PP 1 & 8.94 & 0.48 & 2.9 & 2.87 & 1.63 & &PP 35 & 9.53 & 0.98 & -100 & -100 & -100  \\
PP 2 & 8.87 & 0.28 & 2.9 & 1.32 & 1.61 & &PP 36 & 8.99 & 0.57 & 3.1 & 1.36 & 2.12  \\
PP 3 & 8.84 & 0.65 & 3.1 & 1.32 & 1.69 & &PP 37 & 8.98 & 0.75 & 3 & 1.39 & 2.26  \\
PP 4 & 8.96 & 0.52 & 3.6 & 1.83 & 1.45 & &PP 38 & 9.01 & 0.55 & 3.4 & 1.53 & 2.21  \\
PP 5 & 9.5 & 0.6 & 5.8 & 1.43 & 2.72 & &PP 39 & 9.26 & 0.62 & 5.2 & 2.07 & 2.03  \\
PP 6 & 9.08 & 0.65 & 4.2 & 1.1 & 1.77 & &PP 40 & 8.55 & 0.73 & 1.9 & 1.22 & 1.21  \\
PP 7 & 8.52 & 0.46 & 1.9 & 1.16 & 1.76 & &PP 41 & 9.08 & 0.34 & 3.5 & 1.34 & 3.01  \\
PP 8 & 9.06 & 0.41 & 3.3 & 1.12 & 2.33 & &PP 42 & 9.55 & 0.22 & 5.7 & 1.21 & 5.13  \\
PP 9 & 10.15 & 0.62 & 10.4 & 2.07 & 6.21 & &PP 43 & 8.99 & 0.27 & 3.1 & 1.61 & 2  \\
PP 10 & 8.88 & 0.3 & 3 & 1.1 & 2.43 & &PP 44 & 9.07 & 0.7 & 3.7 & 2.23 & 2.57  \\
PP 11 & 9.98 & 0.38 & 10.8 & 1.21 & 8.25 & &PP 45 & 9.05 & 0.44 & 3.7 & 1.14 & 2.15  \\
PP 12 & 8.74 & 0.55 & 2.6 & 1.02 & 1.75 & &PP 46 & 8.96 & 0.67 & 3.6 & 1.07 & 2.1  \\
PP 13 & 9.4 & 0.43 & 4.7 & 1.52 & 4.63 & &PP 47 & 8.69 & 0.21 & 2 & 1.2 & 1.61  \\
PP 14 & 9.2 & 0.31 & 4.2 & 1.47 & 3.47 & &PP 48 & 9.39 & 0.48 & 2 & 1.14 & 0.98  \\
PP 15 & 9.28 & 0.5 & 4.4 & 1.21 & 3.57 & &PP 49 & 8.8 & 0.86 & 5.2 & 1 & 1.86  \\
PP 16 & 9.71 & 0.47 & 8.2 & 1.11 & 4.01 & &PP 50 & 8.62 & 0.35 & 2.1 & 1.33 & 1.91  \\
PP 17 & 9.6 & 0.66 & 8.5 & 2.86 & 3.91 & &PP 51 & 9.39 & 0.31 & 6.6 & 1.12 & 3.83  \\
PP 18 & 8.56 & 0.66 & 1.8 & 1.53 & 1.55 & &PP 52 & 8.86 & 0.41 & 3.3 & 1.43 & 2.11  \\
PP 19 & 9.43 & 0.49 & 6 & 1.35 & 3.37 & &PP 53 & 9.09 & 0.24 & 3.5 & 1.1 & 2.76  \\
PP 20 & 8.79 & 0.61 & 2.3 & 1.02 & 2.1 & &PP 54 & 9.06 & 0.37 & 4.5 & 1.16 & 1.98  \\
PP 21 & 9.01 & 0.26 & 3.5 & 1.46 & 2.29 & &PP 55 & 8.84 & 0.66 & 3 & 1.06 & 1.72  \\
PP 22 & 9 & 0.41 & 3.7 & 1.99 & 1.83 & &PP 56 & 9.12 & 0.52 & 3.7 & 1.41 & 2.18  \\
PP 23 & 9.01 & 0.26 & 3.5 & 1.33 & 2.07 & &PP 57 & 9.22 & 0.78 & 5.7 & 1.13 & 2.91  \\
PP 24 & 9.05 & 0.43 & 4 & 1.05 & 2.75 & &PP 58 & 9.01 & 0.32 & 3.4 & 1.07 & 1.45  \\
PP 25 & 9.62 & 0.55 & 5.3 & 2.55 & 2.92 & &PP 59 & 10.02 & 0.59 & 9.6 & 1.45 & 4.11  \\
PP 26 & 9.09 & 0.49 & 4.1 & 1.05 & 2.28 & &PP 60 & 9.19 & 0.28 & 3.3 & 1.51 & 2.15  \\
PP 27 & 8.82 & 0.56 & 2.6 & 1.14 & 2.09 & &PP 61 & 9.68 & 0.96 & -100 & -100 & -100  \\
PP 28 & 10.04 & 0.44 & 17.1 & 1 & 7.3 & &PP 62 & 9.45 & 0.35 & 4.8 & 1.23 & 3.07  \\
PP 29 & 9.28 & 0.34 & 3.8 & 1.05 & 1.24 & &PP 63 & 9.38 & 0.31 & 4.4 & 1.06 & 3.23  \\
PP 30 & 8.58 & 0.68 & 2.6 & 1.05 & 1.51 & &PP 64 & 9.19 & 0.24 & 3.7 & 1.03 & 2.76  \\
PP 31 & 8.86 & 0.54 & 3.3 & 1.04 & 1.95 & &PP 65 & 8.79 & 0.49 & 2.5 & 1.27 & 1.87  \\
PP 32 & 9.5 & 0.52 & 4.8 & 1.13 & 3.28 & &PP 66 & 9.05 & 0.42 & 2.9 & 1.54 & 2.44  \\
PP 33 & 9.3 & 0.69 & 6.2 & 2.14 & 1.4 & &PP 67 & 9.2 & 0.75 & 5.3 & 1.59 & 1.88  \\
PP 34 & 9.94 & 0.63 & 11.1 & 2.52 & 4.6 & &PP 68 & 8.52 & 0.69 & 1.6 & 1.17 & 1.73  \\
\end{longtable}
\end{landscape}

\newpage
\section{Optical asymmetries in UMa galaxies}
\begin{figure}[h!]
    \centering
    
    \includegraphics[width=0.8\textwidth]{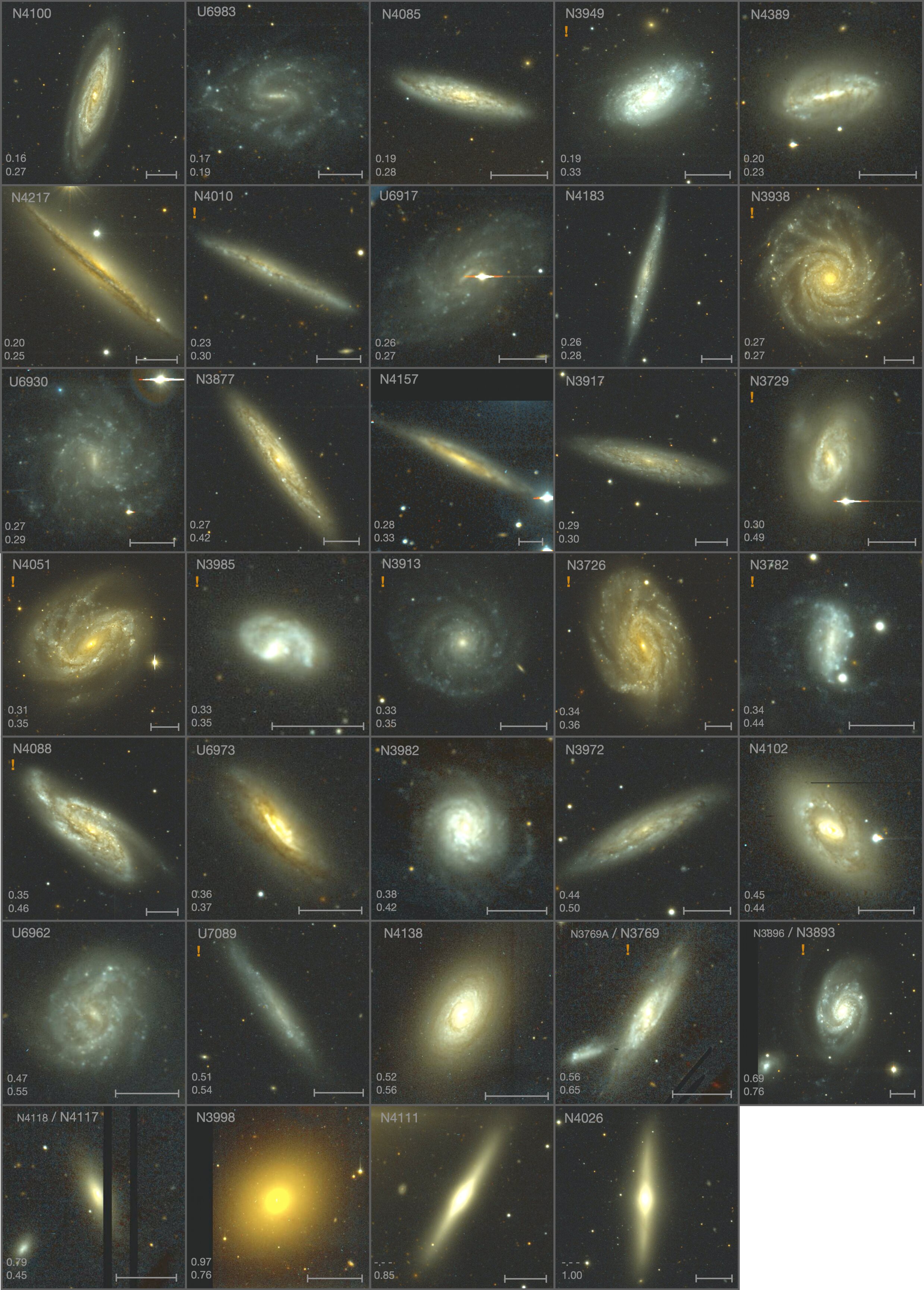}
    \caption{Colour images of galaxies with \logMHI$>8.42$ in the Ursa Major volume, created from photometric imaging data by \cite{Tully1996}, arranged in order of increasing \AmodH\ values, which is shown at the bottom left corner of the panel. Their \AmodL\ values are also shown underneath the \AmodH\ values. Galaxies with optical disturbances are identified with an orange exclamation mark in the top left corner, underneath the name of the galaxy. A scale bar, of 5 kpc, is shown at the bottom right corner of each panel as the images are not shown on the same scale.}
    \label{fig:UMaOptMosaic1}

\end{figure}

\newpage
\begin{figure}[h!]
    \centering
    \includegraphics[width=0.85\textwidth]{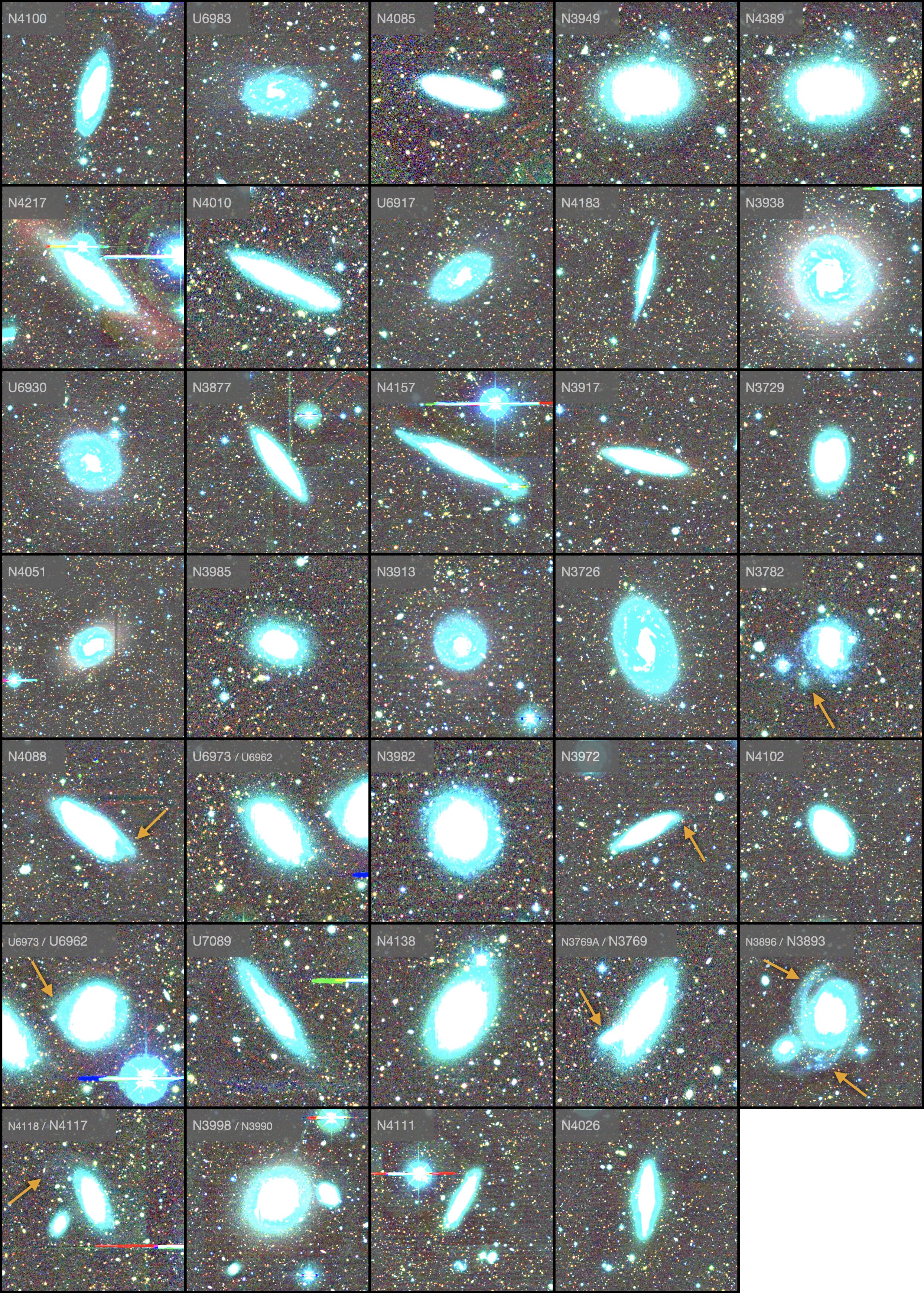}
    \caption{A compilation of the DECaLS $g,r,z$ band images of the UMa galaxies but with increased contrast. The galaxies in this figure have maximum rotational velocities in the range of 80\,\kms$<V_\mathrm{max}<$200\,\kms, arranged in order of increasing \AmodH\ values. Optical disturbances in galaxies are identified with an orange arrow in the panel. A scale bar, of 5 kpc, is shown at the bottom right corner of each panel as the images are not shown on the same scale.}
    \label{fig:UMaOptMosaic15_HC1}
\end{figure}

\newpage
\section{Optical asymmetries in PP galaxies}
\begin{figure}[h!]
    \centering
    
    \includegraphics[width=0.85\textwidth]{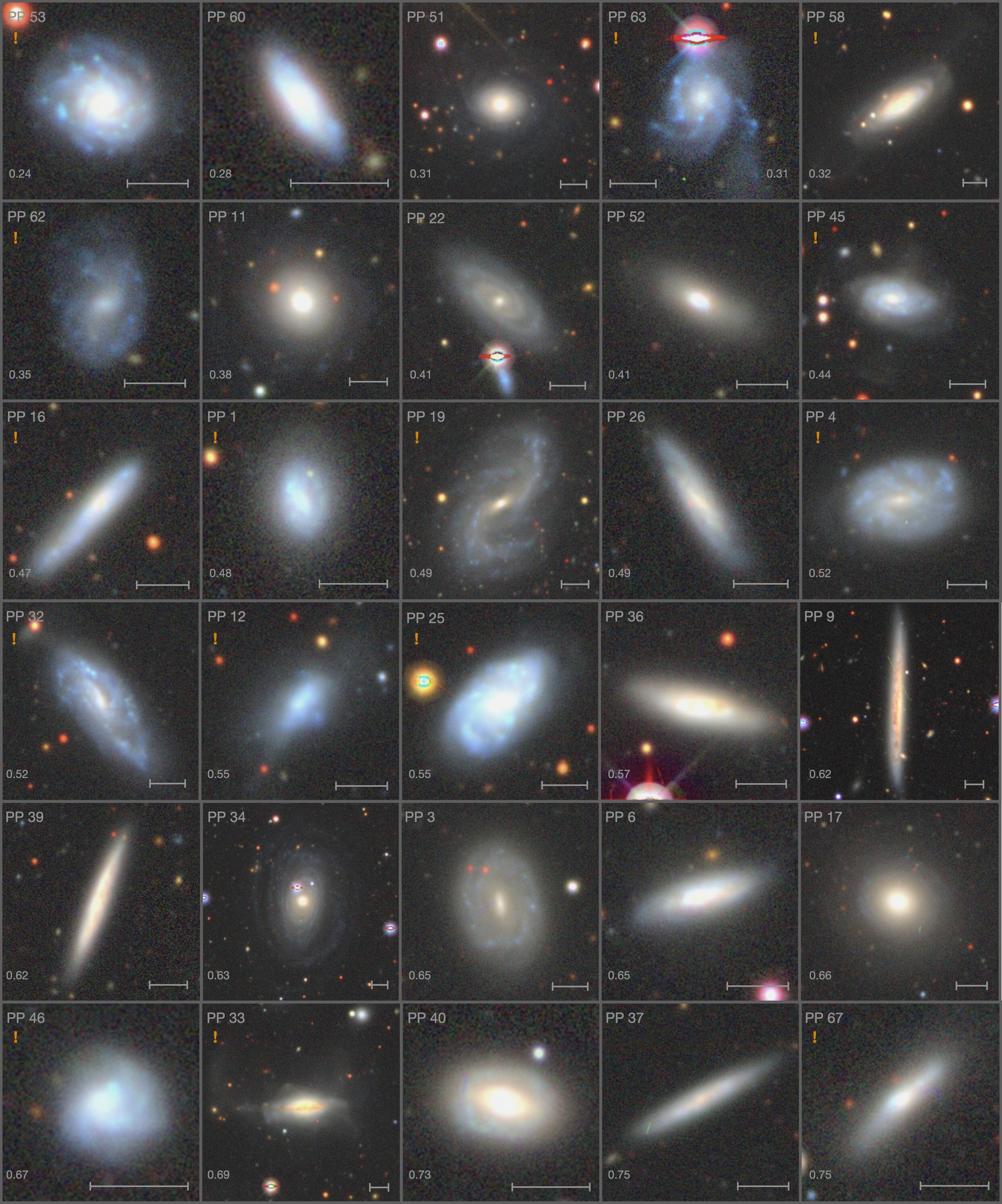}
    \caption{A compilation of the DECaLS $g,r,z$ band images of the PP galaxies that have maximum rotational velocities in the range of 80\,\kms$<V_\mathrm{max}<$200\,\kms, arranged in order of increasing \AmodH\ values. Galaxies with optical disturbances are identified with an orange exclamation mark in the top left corner, underneath the name of the galaxy. A scale bar, of 5 kpc, is shown at the bottom right corner of each panel as the images are not shown on the same scale.}
    \label{fig:PPOptMosaic1}
    
\end{figure}

\newpage
\begin{figure}[h!]
    \centering
    
    \includegraphics[width=0.85\textwidth]{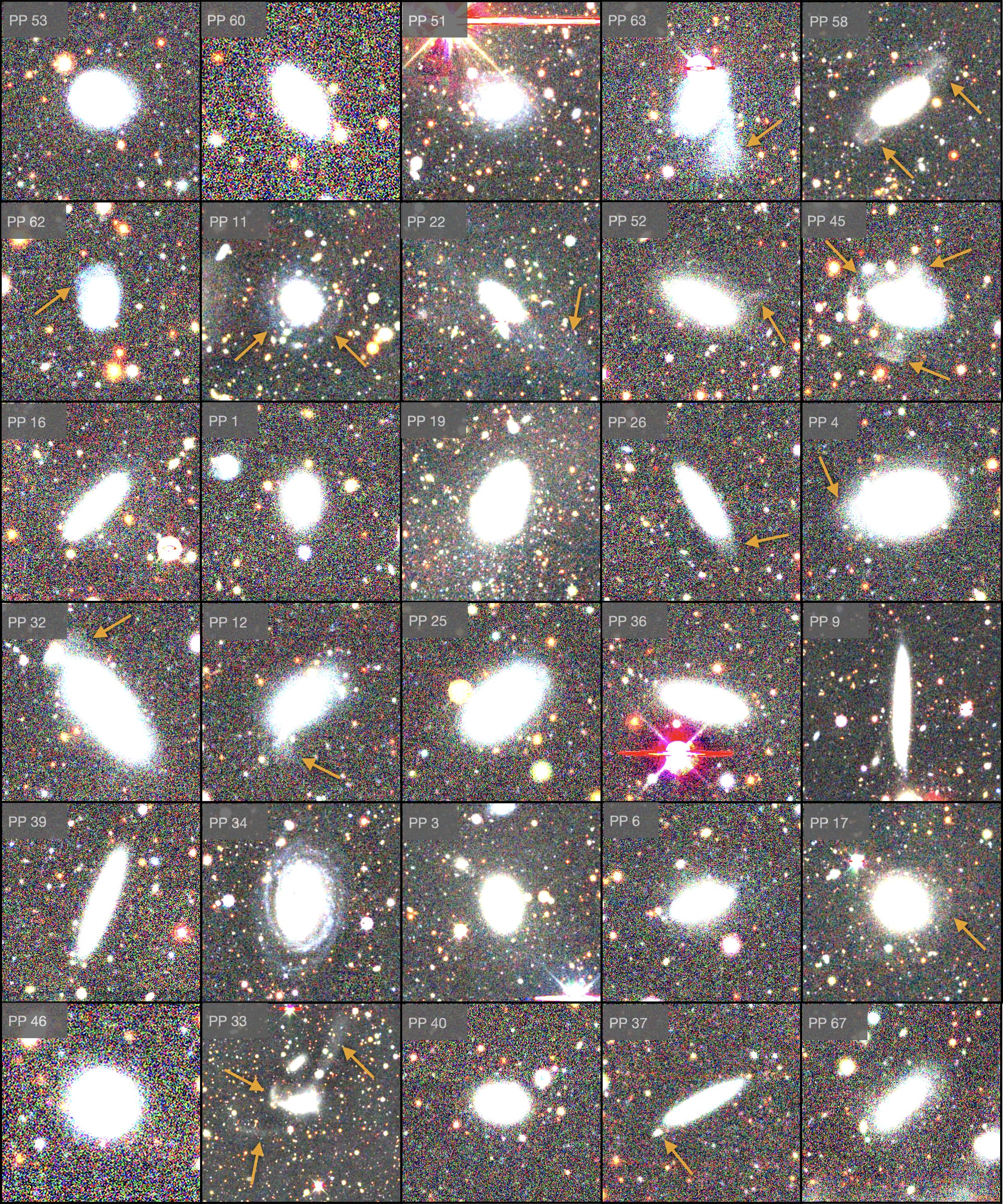}
    \caption{A compilation of the DECaLS $g,r,z$ band images of the PP galaxies but with increased contrast. The galaxies in this figure have maximum rotational velocities in the range of 80\,\kms$<V_\mathrm{max}<$200\,\kms, arranged in order of increasing \AmodH\ values. Optical disturbances in galaxies are identified with an orange arrow in the panel. A scale bar, of 5 kpc, is shown at the bottom right corner of each panel as the images are not shown on the same scale.}
    \label{fig:PPOptMosaic15_HC1}
    
\end{figure}

\end{appendix}
%

\end{document}